\documentclass[11pt]{article}
\pdfoutput=1
\usepackage{amsmath,array,bm,amssymb,epsfig,amsfonts}
\usepackage{graphicx}
\usepackage{tabularx}
\usepackage{subfig}
\usepackage[usenames, dvipsnames]{color}
\usepackage[mathscr]{euscript} 

\usepackage[dvipsnames]{xcolor}
\usepackage{cite}
\usepackage{multirow}
\usepackage{slashed}
\usepackage{verbatim} 
\usepackage{enumitem}
\usepackage{rotating}
\usepackage{xcolor}
\usepackage{multirow}
\usepackage{bm}
\usepackage[normalem]{ulem}
\usepackage{caption}
\usepackage{booktabs} 
\usepackage{tikz-cd} 

\usepackage{makecell}
\usepackage[fleqn,tbtags]{mathtools}
\usepackage{tikz}
\usepackage{ytableau}

\usepackage{tikz}
\usepackage{diagbox}

\usetikzlibrary{positioning}
\usetikzlibrary{calc}
\usetikzlibrary{decorations.pathreplacing,calligraphy}

\usetikzlibrary{arrows.meta}

\usepackage{xstring}
\usetikzlibrary{decorations.pathmorphing} 
\usetikzlibrary{decorations.markings} 
\usetikzlibrary{arrows} 
\usetikzlibrary{shapes} 
\usetikzlibrary{matrix} 
\usetikzlibrary{positioning} 
\usepackage[english]{babel} 
\usepackage[autostyle]{csquotes}

\usepackage{CJK}
\usepackage{tcolorbox}

\usepackage{titlesec}
\usepackage{footmisc}

\usepackage{hyperref}

\usepackage{cleveref}

\titlespacing{\paragraph}{%
  0pt}{
  0.2\baselineskip}{
  0.5em}

\makeatletter

\newcommand{\be}{\begin{equation}}
\newcommand{\ee}{\end{equation}}
\newcommand{\ba}{\begin{aligned}}
\newcommand{\ea}{\end{aligned}}

\newcommand{\bea}{\begin{eqnarray}}
\newcommand{\eea}{\end{eqnarray}}

\renewcommand{\dim}{\text{dim}}

\def\diag{\mathop{\mathrm{diag}}\nolimits}

\def\tr{\mathop{\mathrm{tr}}\nolimits}

\def\ket#1{{|{#1}\rangle}}

\def\unit{{1\kern-.65ex {\rm l}}}
\def\1{{1\kern-.65ex {\rm l}}}

\newcount\hour \newcount\minute
\hour=\time \divide \hour by 60
\minute=\time
\def\now{%
\ifnum \hour<13
  \ifnum \hour=0 \advance \hour by 12 \number\hour:\else \number\hour:\fi%
     \ifnum \minute<10 0\fi%
     \number\minute%
\ A.M.%
\else \advance \hour by -12 \number\hour:%
  \ifnum \minute<10 0\fi%
  \number\minute%
  \ P.M.%
\fi%
}

\makeatother

\begin{document}

\baselineskip=18pt  
\numberwithin{equation}{section}  
\allowdisplaybreaks  


%
%


\thispagestyle{empty}

\vspace*{1.2cm} 
\begin{center}
{{\LARGE \bf  S-Duality for Non-Abelian Monopoles }}

 \vspace*{1.5cm}
Shan Hu\\

\vspace*{1.0cm} 
{\it   Department of Physics, Hubei University \\
Wuhan 430062, China}\\

\vspace{6mm}

{\small \tt hushan@hubu.edu.cn} \\

\vspace*{0.8cm}
\end{center}

\vspace*{.5cm}

\noindent

In $\mathcal{N}=4$ super-Yang-Mills theory with gauge group $G$ spontaneously broken to a subgroup $H$, S-duality requires that the BPS monopole spectrum organizes into the same representation as W-bosons in the dual theory, where $G^{\vee}$ is broken to $H^{\vee}$. The expectation has been extensively verified in the maximally broken phase $G\to U(1)^r$. Here we address the non-Abelian regime in which $H$ contains a semisimple factor $H^{s}$. Using the stratified description of monopole moduli space, we give a general proof of this matching for any simple gauge group $G$. Each BPS monopole state is naturally labeled by a weight of the relevant $W$-boson representation of $(H^{\vee})^{s}$. We construct non-Abelian magnetic gauge transformation operators implementing the $(H^{\vee})^{s}$-action on the monopole Hilbert space, which commute with the electric $H^{s}$-transformations and thereby realize the $H^{s}\times (H^{\vee})^{s}$ symmetry at the level of monopole quantum mechanics.

\newpage

\tableofcontents

\section{Introduction}

In \cite{GNO}, Goddard, Nuyts and Olive showed that magnetic charges of monopoles in a theory with unbroken gauge group $H$ are classified by the weight lattice of the GNO dual group $H^{\vee}$, and conjectured that $H$-monopoles transform as $H^{\vee}$-multiplets, with the true symmetry group given by $H \times H^{\vee}$. In the context of $\mathcal{N}=4$ super-Yang-Mills theory, this kinematical GNO duality is embedded into the Montonen-Olive S-duality, which relates a theory with gauge group $G$ to a dual theory with gauge group $G^{\vee}$ and inverted coupling \cite{Montonen:1977,Osborn:1979}. Under the symmetry breaking $G \to H$ and $G^{\vee} \to H^{\vee}$, the duality identifies monopoles in one theory with W-bosons in the dual theory. Including the $\theta$-angle, the S-duality group of $\mathcal{N}=4$ SYM is $SL(2,\mathbb{Z})$, under which monopoles and dyons are permuted and, in particular, are mapped to the dual W-bosons \cite{Sen:1994,Dorey:1996}.

In $\mathcal{N}=4$ SYM, the low-energy dynamics of BPS monopoles is governed by a supersymmetric quantum mechanics on the monopole moduli space, in which each bosonic modulus is paired with two fermionic partners \cite{Gauntlett, Blum}. Quantization of the center-of-mass sector produces the required $16$-fold degeneracy of the massive vector multiplet, while BPS ground states on the relative moduli space are realized as (normalizable) harmonic forms \cite{Witten:1982, Sen:1994}. In the maximally broken phase $G \to U(1)^{r}$, there are $r$ fundamental monopoles (one for each simple coroot), which, together with their BPS bound states, reproduce the W-boson spectrum in the dual theory \cite{GauntlettLowe1996, LWY:SU3, LWY:ManyBPS, Gibbons:1996}. With the electric charge turned on, dyonic bound states built from these monopoles realize the $SL(2,\mathbb{Z})$ S-duality orbit of the dual W-bosons \cite{Sen:1994, SegalSelby1996}.

When $G$ is broken to a non-Abelian subgroup $H=H^{s} \times U(1)^{t}$ with semisimple factor $H^{s}$, as in the GNO setting, the resulting $H$-monopoles are often referred to as non-Abelian monopoles. For related literature, see, e.g., \cite{Balachandran1971, Balachandran1984a, Balachandran1984b, Abouelsaood1983a, Abouelsaood1984b, NelsonManohar1983, Dorey1995a, Weinberg2, Weinberg1, Weinberg1982, Abouelsaood1984a, 1, BaisSchroers1998, SchroersBais1998, Bais2009, Bais2010, BolognesiKonishi2002, Konishi2002, Auzzi2004b, Auzzi2004a, Konishi2004, Bolognesi2005, Konishi2008, Konishi2009}. Classically, for each $H^{\vee}$-multiplet of W-bosons in the dual theory, there is a corresponding multiplet of $SU(2)$-embedded monopole solutions in the original theory, with the same mass and the same multiplicity \cite{Auzzi2004a}. Quantum mechanically, however, the semiclassical analysis of non-Abelian monopoles is obstructed by subtle issues associated with their non-Abelian gauge-orientation modes.

At large $r$, the Higgs field in a monopole solution has the expansion
\begin{equation}
\Phi(\vec{r}) = \Phi_{0}(\hat{r}) - \frac{G_{0}(\hat{r})}{2r} + \cdots,
\end{equation}
where the vacuum expectation value $\Phi_{0}(\hat r)$ defines a map 
$\mathbb{S}^{2}_{\infty} \to G/H$, representing a class in $\pi_{2}(G/H) \cong \mathbb{Z}^{t}$ characterized by a $ t $-component topological charge, and $G_{0}$ is the $ r $-component magnetic charge (up to an angle-dependent gauge rotation). The non-Abelian gauge-orientation modes associated with $H^{s}$ are generated at $r=\infty$ by sections $\{T_{i}(\hat r)|i=1,\ldots,\dim H^{s}\}$ of the adjoint $H^{s}$-bundle over $\mathbb{S}^{2}_{\infty}$. It is natural to require
\begin{equation}\label{vev1}
[T_{i}(\hat r),\Phi_{0}(\hat r)] = 0
\qquad \text{on the large sphere } \mathbb{S}^{2}_{\infty},
\end{equation}
so that the deformation preserves the asymptotic vacuum configuration. However, for many choices of topological charge, the adjoint $H^{s}$-bundle over $\mathbb{S}^{2}_{\infty}$ is twisted, so not all of $T_i(\hat r)$ can be chosen globally \cite{Balachandran1971, Balachandran1984a, Balachandran1984b, Abouelsaood1984b, NelsonManohar1983}. In these cases, (\ref{vev1}) cannot be imposed everywhere on $\mathbb{S}^{2}_{\infty}$; the variation $\delta\Phi$ does not decay at infinity, and the norm develops an $\int dr\, r^{2}$ divergence. Moreover, if
\begin{equation}
[T_{i}(\hat r),G_{0}(\hat r)] \neq 0,
\end{equation}
the variation of the $1/r$ term induces a $\delta\Phi \sim 1/r$ falloff, leading to an additional $\int dr$ divergence. These considerations are only schematic, since the physical inner product on zero modes must be computed in background gauge. A careful treatment shows that the non-Abelian modes which do not commute with the magnetic charge cannot be brought into background gauge by any local gauge transformation; enforcing the background gauge may remove them away \cite{Abouelsaood1983a}. Finally, with the non-Abelian zero modes taken into account, the moduli space has a dimension that is not always divisible by $4$ and thus need not be hyper-K\"ahler. On the other hand, if these modes are simply discarded, the unbroken non-Abelian symmetry $H^{s}$ has no manifestation on the monopole moduli space. It is only when the magnetic charge is invariant under $H^{s}$ that all these difficulties disappear, and the corresponding moduli space is as well defined as in the maximally broken case \cite{1}.  Such ``neutral'' non-Abelian monopoles, however, are not generic.

Despite all these problems, Dorey et al. \cite{Dorey1995a} found that, in non-degenerate situations,\footnote{In degenerate situations, for a fixed topological charge, monopole solutions can carry gauge-inequivalent magnetic charges, and there are solutions interpolating between them; see, e.g., \cite{Weinberg1982}. In the language of the stratification of monopole moduli space \cite{Murray1989, MurraySinger2003}, the degenerate case corresponds to a moduli space composed of multiple strata, whereas in the non-degenerate case the moduli space consists of a single stratum.} once the non-Abelian modes are included, the relative moduli space of a monopole with topological charge $1$ is a coset space (of infinite volume) whose Euler characteristic, which counts the number of BPS ground states, equals the dimension of the dual W-boson multiplet of unit $U(1)$ charge. These results provide strong evidence for GNO conjecture in $\mathcal{N}=4$ SYM.

In this paper, based on the stratified formulation of monopole moduli space \cite{Murray1989, MurraySinger2003}, which is a suitable framework for studying non-Abelian monopoles, we give a general proof of the matching between the monopole spectrum and the W-boson spectrum under the symmetry breakings $G \rightarrow H^{s} \times U(1)^{t}$ and $G^{\vee} \rightarrow (H^{\vee})^{s} \times U(1)^{t}$ for any simple gauge group $G$. In this picture, the monopole moduli space admits a natural geometric realization of the dual W-boson representation. The discussion also extends to dyonic states. In our analysis, it is assumed that, in the maximally broken case $G \rightarrow U(1)^{r}$, there is a one-to-one correspondence between monopoles and the dual W-bosons, an expectation that has been extensively tested, though not yet when the $U(1)^{r}$ charge vector of the W-bosons has components larger than one \cite{Yi:1996}.

Our approach may be viewed as a smooth-monopole analogue of the Kapustin-Witten construction for geometric Langlands program \cite{KW}, where 't~Hooft operators in the topologically twisted $\mathcal N=4$ theory are defined by imposing supersymmetric singular monopole behavior, and the resulting magnetic data are encoded in the geometry of Hecke modifications. In this framework, a 't~Hooft (Hecke) operator $T(w^{\vee})$ is labeled by a dominant weight $w^{\vee}$ of the Langlands dual group $G^{\vee}$, and the moduli space of singular monopoles inserted at a single point is identified with the space $\mathscr{Y}(w^{\vee})$ of Hecke modifications of type $w^{\vee}$ for a $G$-bundle. Its natural compactification $\overline{\mathscr{Y}}(w^{\vee})$ is a Schubert variety in the affine Grassmannian. The space of physical states $\mathcal{H}(w^{\vee})$ is given by the (intersection) cohomology of $\overline{\mathscr{Y}}(w^{\vee})$ and is isomorphic to $R(w^{\vee})$, the irreducible $G^{\vee}$-representation of highest weight $w^{\vee}$. Here we consider smooth monopoles arising from symmetry breaking, but the underlying geometric picture is similar. In particular, compactifying the Hecke modification space $\mathscr{Y}(w^{\vee})$ by adding all associated lower-weight Hecke modifications (which share the same topological type determined by $w^{\vee}$) mirrors the stratification of the monopole moduli space by allowing gauge-inequivalent magnetic charges at fixed topological charge.

We will return to the conceptual issues in the semiclassical quantization of non-Abelian monopoles in Section~\ref{qm}. In the stratified formulation of the monopole moduli space, the condition~(\ref{vev1}) is relaxed to 
\begin{equation}\label{vev2}
[T_{i}(\hat r_{0}),\Phi_{0}(\hat r_{0})] = 0
\qquad \text{for a fixed direction $\hat r_{0}$ on the large sphere } \mathbb{S}^{2}_{\infty}.
\end{equation}
For example, when $SU(N+1)$ breaks to $U(N)$, there are $N$ $SU(2)$-embedded monopole solutions of unit topological charge, related by global $SU(N)$ Weyl transformations \cite{Auzzi2004a}. They are the classical counterparts of dual W-bosons in the fundamental of $SU(N)$. With~(\ref{vev1}) relaxed to~(\ref{vev2}), these solutions can be incorporated into a single connected moduli space labeled by the topological charge. The low-energy effective action for monopoles is obtained by a collective-coordinate expansion of $\mathcal N=4$ SYM. Starting from an arbitrary BPS trajectory, integrating out $A_{0}$ automatically enforces the background-gauge condition on the moduli derivatives. We will argue that, even with the background gauge imposed, the non-Abelian modes that do not commute with $G_{0}$ still have divergent norm. This divergence is also required for the stratified structure of the moduli space. For instance, in a relative moduli space consisting of an open stratum $\mathbb{C}^{2N}$ and a closed stratum $\mathbb{CP}^{2N-1}$, $\mathbb{CP}^{2N-1}$ must sit at the asymptotic boundary of $\mathbb{C}^{2N}$ and thus inherits a divergent metric. With the non-Abelian modes included, the moduli space is not necessarily hyper-K\"ahler. To be self-contained, we will derive the effective action via a collective-coordinate expansion without relying on hyper-K\"ahler structure\footnote{The hyper-K\"ahler structure is needed only for the fermionic zero-mode counting, where it guarantees that the number of fermionic zero modes is twice the number of bosonic moduli. Here we take this relation as an input. In fact, with the non-normalizable modes included, the index calculation \cite{Weinberg2, Weinberg1} yields a moduli-space dimension that agrees with the dimension in the stratified formulation \cite{Murray1989, MurraySinger2003}. Although this dimension is not necessarily divisible by $4$, the index still gives a $2:1$ ratio of fermionic to bosonic modes for $\mathcal N=4$ SYM \cite{Index, Callias1978}.}. The resulting action takes the same form as in the maximally broken case, but the associated supersymmetry is $\mathcal N=2$ (enhanced to $\mathcal N=4$ when the moduli space is K\"ahler), rather than the $\mathcal N=8$ available for a hyper-K\"ahler target \cite{Bagger:1984ge}. This reduction does not affect the physical conclusions: the $16$-fold degeneracy of the $\mathcal N=4$ massive vector multiplet is already saturated by the eight fermionic zero modes associated with the center-of-mass motion, while the ground states in the relative sector are still given by harmonic forms on the moduli space.

Aside from the numerical match, the monopole ground states, which furnish the same $(H^{\vee})^{s}$-representation as the dual W-bosons, also carry definite weights of that representation (up to Weyl transformations). This allows us to construct magnetic operators generating the $(H^{\vee})^{s}$-action on the monopole Hilbert space. While the $H^{s}$-action is realized geometrically by isometric diffeomorphisms of the relative moduli space, the $(H^{\vee})^{s}$-action is realized algebraically on differential forms by wedge and contraction operations. In \cite{Dorey1995a}, it was noted that the harmonic forms on the coset space, which represent monopole ground states in that case, are $H^{s}$-invariant. Here we show that the ground states are $H^{s}$-invariant in general. Consequently, the $(H^{\vee})^{s}$-action may be implemented by operators commuting with both $H^{s}$ and the Hamiltonian. For monopole moduli spaces with a single stratum, we give an explicit construction of the resulting $H^{s}\times (H^{\vee})^{s}$-representation on the Hilbert space, thereby realizing the $H\times H^{\vee}$ GNO conjecture at the level of monopole quantum mechanics.

The rest of the paper is organized as follows. Section~\ref{str} reviews the stratified structure of the monopole moduli space. Section~\ref{pr} presents a general proof of the matching between the monopole (and dyon) spectrum and the dual W-boson spectrum when the unbroken gauge group is non-Abelian. In Section~\ref{qm} we study the supersymmetric quantum mechanics on the monopole moduli space with the non-Abelian degrees of freedom included. In Section~\ref{st} we construct the magnetic gauge transformation generators and give an explicit realization of the $H^{s}\times (H^{\vee})^{s}$-representation on the monopole Hilbert space. We end with a discussion in Section~\ref{cda}.

\section{Stratification of the monopole moduli space}\label{str}

In this section, we review the stratified formulation of monopole moduli space, following \cite{Murray1989, MurraySinger2003} and \cite{Bais2009}. For a gauge group $G$ with Lie algebra $\mathfrak{g}$, a monopole on $\mathbb{R}^{3}$ is a pair $(A,\Phi)$, where $ A $ is a connection on the trivial $G$-bundle and $ \Phi $ is a section of the adjoint bundle. $(A,\Phi)$ satisfies the Bogomolny equations
\begin{equation}
D_{i}\Phi=B_{i},\qquad i=1,2,3.
\end{equation}
At infinity, $ \Phi $ gives a map
\begin{equation}\label{frf}
\Phi^{\infty}: \mathbb{S}^{2} \rightarrow C[\Phi_{0}]=\{g\Phi_{0} g^{-1}\, | \,  \forall \; g \in G\} \cong G/H,
\end{equation}
where $\Phi_{0}\in  \mathfrak{g}  $ and $ H \subset G$ is the centralizer of $\Phi_{0} $. The homotopy class in $  \pi_{2}(G/H)\cong \mathbb{Z}^{t}$ is specified by $t$ nonnegative integers, the topological charge $m=(m_{1},\ldots,m_{t})$. For a fixed $ \Phi_{0} $, we impose a framing along the positive $x_{3}$-axis by requiring the asymptotic expansion
\begin{equation}\label{fr}
\Phi(0,0,x_{3})=\Phi_{0}-\frac{G_{0}}{2 x_{3}}+\mathcal{O}\left( \frac{1}{(x_{3})^{2}}\right) .
\end{equation}
Since $ [\Phi_{0}, G_{0}]=0 $, $G_{0} \in \mathfrak{h}=  \text{Lie} (H)$. $ \exp \{ 2 \pi i \,G_{0}\}=I $. $G_{0}  $ is the magnetic charge of the monopole.

Let $ \mathcal{M} (m,\Phi_{0})$ denote the moduli space of framed monopoles with asymptotic Higgs field in $C[\Phi_{0}]$ and the topological charge $m$. Monopoles with the same topological charge can be continuously deformed into each other, so $ \mathcal{M} (m,\Phi_{0})  $ is a connected manifold. (\ref{fr}) induces a map 
\begin{equation}
e: \;  \mathcal{M} (m,\Phi_{0})  \rightarrow \mathfrak{h},
\end{equation}
which assigns $G_{0}$ to $(A,\Phi)$. Its image $\mathcal{K} \subset \mathfrak{h} $ is a disjoint union of $H$-orbits
\begin{equation}\label{ck}
\mathcal{K}=\bigcup_{i=1}^{n}C(k_{i}) ,\qquad  C(k_{i})=\{h k_{i} h^{-1} | \forall \; h \in H\} \cong H/Z_{H}(k_{i})
\end{equation}
for integral elements $ k_{i} $ with $\exp \{2 \pi i k_{i} \}=I $. Here $Z_{H}(k_{i})$ is the centralizer of $ k_{i} $ in $H$. The BPS mass 
\begin{equation}
 M=\frac{2\pi}{g} \tr(\Phi_{0} G_{0}) 
\end{equation}
is a constant on $\mathcal{M} (m,\Phi_{0})$, $ \forall \;G_{0} \in \mathcal{K}$.

For each orbit $  C(k_{i})$, define the $i$-th stratum 
\begin{equation}
\mathcal{M}_{i} (m,\Phi_{0})\equiv e^{-1} \left(  C(k_{i}) \right) ,
\end{equation}
then 
\begin{equation}
\mathcal{M}(m,\Phi_{0})
=\bigcup_{i=1}^{n}\mathcal{M}_{i}(m,\Phi_{0})
\end{equation}
with
\begin{equation}
\dim \mathcal{M}(m,\Phi_{0})
=\max_{1\le i\le n}\,\dim \mathcal{M}_{i}(m,\Phi_{0}).
\end{equation}
For fixed $G_{0} \in \mathcal{K}  $, define
\begin{equation}\label{ckk}
\mathcal{M} (m,\Phi_{0}, G_{0})=\{(A,\Phi) \in \mathcal{M} (m,\Phi_{0}), e(A,\Phi)=G_{0}\},
\end{equation}
which is the moduli space of framed monopoles of type $(\Phi_{0}, G_{0})$. $ \mathcal{M} (m,\Phi_{0}, G_{0}) $ is a well-defined hyper-K\"ahler manifold with the dimension divisible by $4$. $\forall \;G_{0} \in C(k_{i})$, $ \mathcal{M} (m,\Phi_{0}, G_{0})$ is isometric to $ \mathcal{M} (m,\Phi_{0}, k_{i})$, because the corresponding $(A,\Phi)$ are related by a global gauge transformation. As a result, 
\begin{equation}
\mathcal{M}_{i} (m,\Phi_{0})=\{(A,\Phi) \in \mathcal{M} (m,\Phi_{0}), e(A,\Phi)=G_{0}\in C(k_{i})\}
\end{equation}
forms a fiber bundle over $C(k_i)$ with fiber $\mathcal M(m,\Phi_0,k_i)$.
\begin{equation}
\dim \mathcal M_i (m,\Phi_{0})=\dim \mathcal M(m,\Phi_0,k_i)+\dim C(k_i).
\end{equation}
By construction, $ \mathcal M_i (m,\Phi_{0}) $ is $H$-stable and inherits an $H$-invariant metric (which is divergent along $C(k_i)$). Since each stratum is $H$-invariant, the full moduli space $\mathcal{M}(m,\Phi_{0})  $ is also invariant under the unbroken symmetry $H$.

To compute the dimension of the moduli space, we work in a Cartan subalgebra $ \mathfrak{h}_{0}\subset  \mathfrak{g}$ containing $\Phi_{0}$. Let $ T_{1},\ldots,T_{r} $ be a basis of $\mathfrak{h}_{0}$ with $\tr (T_{a}T_{b})=\delta_{ab}$, where $r = \text{rank}\,G$, and write
\begin{equation}
 \Phi_{0}=\sum^{r}_{a=1}h_{a}T_{a} \equiv \mathbf{h}\cdot\mathbf{T}.
\end{equation}
For the given $ \mathbf{h}$, choose simple roots $ \boldsymbol{\alpha}_{1},\ldots, \boldsymbol{\alpha}_{r}$ such that
\begin{equation}\label{110}
\boldsymbol{\alpha}_{1}\cdot \mathbf{h} >0,\qquad\ldots\qquad\boldsymbol{\alpha}_{t}\cdot \mathbf{h} >0,\qquad \text{and} \qquad \boldsymbol{\alpha}_{t+1}\cdot \mathbf{h} =0,\qquad\ldots\qquad\boldsymbol{\alpha}_{r}\cdot \mathbf{h} =0.
\end{equation}
Let $\boldsymbol{\alpha}_{a}^{\vee}=2 \boldsymbol{\alpha}_{a}/(\boldsymbol{\alpha}_{a} \cdot  \boldsymbol{\alpha}_{a})$ be the coroots. The unbroken gauge group is $ H= H^{s} \times U(1)^{t} $, where $ H^{s} $ is semisimple with simple roots $ \boldsymbol{\alpha}_{t+1},\ldots,\boldsymbol{\alpha}_{r} $.

In (\ref{ck}), we may take $ k_{i} =\mathbf{k}_{i} \cdot \mathbf{T}\in \mathfrak{h}_{0}$. The integrality condition $ \exp \{2 \pi i k_{i} \}=I $ implies
\begin{equation}\label{ki}
\mathbf{k}_{i}  = \sum^{r}_{a=1}  n_{a}\boldsymbol{\alpha}_{a}^{\vee},\qquad n_{a} \in \mathbb{Z}, 
\end{equation}
where the integers $ n_{a} $ can be split into the topological charges
\begin{equation}\label{113}
m_{1}= n_{1},\qquad\ldots\qquad m_{t}= n_{t}
\end{equation}
and the holomorphic charges
\begin{equation}
q_{1}= n_{t+1},\qquad \ldots\qquad q_{r-t}= n_{r}.
\end{equation}
$m_{a}$ are gauge invariant, while $ q_{a} $ transform under the Weyl group of $H$. For each $1 \leq i \leq n$, the intersection $ \mathfrak{h}_{0} \cap C(k_{i}) $ is a Weyl ($H$)-orbit. We can choose a unique representative by imposing the anti-dominant condition
\begin{equation}\label{ad}
\boldsymbol{\alpha}_{t+1}\cdot \mathbf{k}_{i} \leq 0,\qquad \ldots\qquad \boldsymbol{\alpha}_{r}\cdot \mathbf{k}_{i}  \leq 0.
\end{equation}
Given $m_{1},\ldots,m_{t}$ and the constraints $q_{1} ,\ldots,q_{r-t} \geq 0$, the condition \eqref{ad} selects $ n $ admissible sets of $ q_{a} $, which in turn determine $k_{1},\ldots,k_{n}$ in (\ref{ck}).

With this choice, $\dim \mathcal{M} (m,\Phi_{0}, k_{i})$ can be computed from $n_{a}$ exactly as in the maximally broken case. $\forall\, G_{0} \in C(k_{i}) $, $ \mathcal{M} (m,\Phi_{0}, G_{0}) $ and $\mathcal{M} (m,\Phi_{0}, k_{i})$ are isometric, so
\begin{equation}
 \dim \mathcal{M} (m,\Phi_{0}, G_{0})=\dim \mathcal{M} (m,\Phi_{0}, k_{i})=4 \sum^{r}_{a=1}n_{a} =4 \left(   \sum^{t}_{a=1}m_{a} +\sum^{r-t}_{a=1}q_{a} \right) .
\end{equation}
Consequently, the dimension of the $i$-th stratum is
\begin{equation}\label{m1}
\dim \mathcal{M}_{i}(m,\Phi_{0})=4 \left(   \sum^{t}_{a=1}m_{a} +\sum^{r-t}_{a=1}q_{a} \right) +\dim C(k_{i}).
\end{equation}
Because of the last term, $\dim \mathcal{M}_{i}(m,\Phi_{0})$ is not always divisible by $4$, and $\mathcal{M}_{i}(m,\Phi_{0})$ need not be hyper-K\"ahler. 
By contrast, the fiber $\mathcal{M}(m,\Phi_{0},G_{0})$ is a well-defined hyper-K\"ahler manifold, and the problematic non-normalizable directions in $\mathcal{M}_{i}(m,\Phi_{0})$ lie along the base $C(k_{i})$. (\ref{m1}) coincides with the index calculation of the moduli space dimension when the non-normalizable modes are also included \cite{Weinberg2, Weinberg1}.

It is convenient to decompose the full moduli space $ \mathcal{M} (m,\Phi_{0}) $ into a free center-of-mass factor and a relative part $\mathcal{M}_{\mathrm{rel}} (m,\Phi_{0})$, which inherits the stratification:
\begin{equation}
\mathcal{M}_{\mathrm{rel}} (m,\Phi_{0})=\bigcup_{i=1}^{n} \mathcal{M}_{\mathrm{rel},i}(m,\Phi_{0}),
\end{equation}
where each $\mathcal{M}_{\mathrm{rel},i}(m,\Phi_{0})   $ is a fiber bundle over $C(k_i)$ with fiber $ \mathcal{M}_{\mathrm{rel}} (m,\Phi_{0}, k_{i})$.
\begin{equation}
\dim \mathcal{M}_{\mathrm{rel}} (m,\Phi_{0}) = \max_{1\le i\le n}\,\dim \mathcal{M}_{\mathrm{rel},i}(m,\Phi_{0})
\end{equation}
with
\begin{equation}\label{dim}
\dim \mathcal{M}_{\mathrm{rel},i}(m,\Phi_{0})=4 \left(   \sum^{t}_{a=1}m_{a} +\sum^{r-t}_{a=1}q_{a} \right) +\dim C(k_{i})-4.
\end{equation}

\

\paragraph{Example: $G=SU(3)$ and $\Phi_{0}=\diag (2v,-v,-v)$. }\label{su31}
\leavevmode\par\vspace{0.8em}
\noindent
The unbroken group is $H=U(2)$, $\pi_{2}(SU(3)/U(2))\cong \mathbb{Z}$, and $t=1$.

\begin{itemize}
  \item $m=1$: $\mathcal{K}=C(k_{1})$ with $k_{1}=\diag(1,-1,0)$,
  $M=6\pi v/g$. $\mathcal{M}(1,\Phi_{0})=\mathcal{M}_{1}(1,\Phi_{0})$. The orbit is 
  \begin{equation}
  C(k_{1})\cong SU(2)/U(1)\cong  \mathbb{CP}^{1}.
  \end{equation}
$\dim \mathcal{M}(1,\Phi_{0},k_{1})=4$, so $\dim \mathcal{M}(1,\Phi_{0})=6$, and
\begin{equation}
\mathcal{M}_{\mathrm{rel}}(1,\Phi_{0})=\mathcal{M}_{\mathrm{rel},1}(1,\Phi_{0}) =    C(k_{1}) \cong \mathbb{CP}^{1}.
\end{equation}

\item $m=2$: In this case $\mathcal{K}=C(k_{1})\cup C(k_{2})$ with
\begin{equation}
k_{1}=\diag(2,-2,0),
\qquad
k_{2}=\diag(2,-1,-1).
\end{equation}
$M=12\pi v/g$. $\mathcal{M}(2,\Phi_{0})= \mathcal{M}_{1}(2,\Phi_{0}) \cup  \mathcal{M}_{2}(2,\Phi_{0}) $. The orbits are  
\begin{equation}
C(k_{1})\cong \mathbb{CP}^{1},   \qquad C(k_{2})=\{k_{2}\}.
\end{equation}
Dimensions of the fibers are 
\begin{equation}
\dim \mathcal{M}(2,\Phi_{0},k_{1})=8 , \qquad \dim \mathcal{M}(2,\Phi_{0},k_{2})=12, 
\end{equation}
so 
\begin{equation}
\dim \mathcal{M}_{1}(2,\Phi_{0})=10 , \qquad \dim \mathcal{M}_{2}(2,\Phi_{0})=12 , \qquad \dim \mathcal{M}(2,\Phi_{0})=12. 
\end{equation}
For the relative spaces,
\begin{equation}
\dim \mathcal{M}_{\mathrm{rel},1}(2,\Phi_{0})=6,
\qquad
\dim \mathcal{M}_{\mathrm{rel},2}(2,\Phi_{0})=8,
\qquad
\dim \mathcal{M}_{\mathrm{rel}}(2,\Phi_{0})=8.
\end{equation}

The stratum $\mathcal{M}_{\mathrm{rel},2}(2,\Phi_{0})$
  contains a parameter $a\in[0,\infty)$ describing the size of the non-Abelian cloud. The corresponding monopole solutions and the moduli space metric can be found in \cite{Dancer1992, DancerLeese1997, Irwin1997}. When $a\to\infty$, $G_{0}$ jumps from $k_{2}$ to $C(k_{1})$, consistent with the
  connectedness of $\mathcal{M}_{\mathrm{rel}}(2,\Phi_{0})$. Note that the boundary of
  $\mathcal{M}_{\mathrm{rel},2}(2,\Phi_{0})$ would be $7$-dimensional, while $\mathcal{M}_{\mathrm{rel},1}(2,\Phi_{0})$ is $6$-dimensional, so $\mathcal{M}_{\mathrm{rel}}(2,\Phi_{0})$
  is not obtained by attaching a boundary to $\mathcal{M}_{\mathrm{rel},2}(2,\Phi_{0})$, but
  rather by compactifying $\mathcal{M}_{\mathrm{rel},2}(2,\Phi_{0})$ with
  $\mathcal{M}_{\mathrm{rel},1}(2,\Phi_{0})$ placed at infinity, analogous to compactifying
  $\mathbb{R}^{2}$ to $\mathbb{S}^{2}$ by adding a point at infinity.
\end{itemize}

\paragraph{Example: $G=SO(5)$ and $H= U(2)$ }
\leavevmode\par\vspace{0.8em}
\noindent
Simple roots of $SO(5)$ may be chosen as $\boldsymbol{\gamma}=\boldsymbol{e}_{1}-\boldsymbol{e}_{2}$ and $\boldsymbol{\mu}=\boldsymbol{e}_{2}$. Removing the node $\boldsymbol{\mu}$ in Dynkin diagram breaks $SO(5)$ to $H=U(2)$. $\pi_2(SO(5)/U(2))\cong\mathbb Z$, $t=1$. For a fundamental monopole with $m=1$:

\begin{itemize}
  \item $\mathcal{K}=C(k_{1})\cup C(k_{2})$ with
  $ \mathbf{k}_{1} =\boldsymbol{\mu}^{\vee}$ and $\mathbf{k}_{2} =\boldsymbol{\mu}^{\vee}+\boldsymbol{\gamma}^{\vee}$. Here $ C(k_{1})\cong SU(2)/U(1)\cong \mathbb{CP}^{1} $, while $C(k_{2})=\{k_{2}\}$ is a point.

  \item The strata are
  \begin{equation}
  \mathcal{M}_{1}(1,\Phi_{0}) \cong \mathbb{R}^{3} \times\mathbb{S}^{1}  \times  \mathbb{CP}^{1} ,\qquad \mathcal{M}_{2}(1,\Phi_{0}) \cong \mathbb{R}^{3} \times\mathbb{S}^{1}  \times  \mathbb{R}^{4} .
  \end{equation}
The eight-parameter family of solutions for $ \mathcal{M}_{2}(1,\Phi_{0})$ is constructed in \cite{Weinberg1982},
  and the corresponding moduli-space metric is computed in \cite{Abouelsaood1984a, 1}. $ \mathcal{M}_{2}(1,\Phi_{0})$ contains a modulus $a\in[0,\infty)$ with $2\sqrt{a}$ the radial coordinate on the $\mathbb{R}^{4}$ factor. As $a\to\infty$, $G_{0}$ jumps from $k_2$ onto the orbit $C(k_1)$, and the two strata glue along this boundary. Since $\mathbb{R}^{4}\cong\mathbb{C}^{2}$ compactifies to $\mathbb{CP}^{2}$ by adjoining $\mathbb{CP}^{1}$ at infinity, it follows that
\begin{equation}
 \mathcal{M}(1,\Phi_{0})
 =\mathcal{M}_{1}(1,\Phi_{0})\cup \mathcal{M}_{2}(1,\Phi_{0})
= \mathbb{R}^{3}\times \mathbb{S}^{1}\times \mathcal{M}_{\mathrm{rel}}(1,\Phi_{0})
 \approx \mathbb{R}^{3}\times \mathbb{S}^{1}\times \mathbb{CP}^{2},
\end{equation}
where $\approx$ denotes a homeomorphism and $\mathcal{M}_{\mathrm{rel}}(1,\Phi_{0})\approx \mathbb{CP}^{2}$.
\end{itemize}
For more details, see Appendix \ref{reg}, where explicit monopole solutions for both strata are presented and the $a\to\infty$ transition is exhibited.

\section{Spectrum matching of monopoles and W-bosons}\label{pr}

In $\mathcal N=4$ SYM with gauge group $G$ broken to $H^{s}\times U(1)^{t}$, S-duality requires that the spectrum of BPS monopole states matches the spectrum of W-bosons in the dual theory, where $G^{\vee}$ is broken to $(H^{\vee})^{s}\times U(1)^{t}$. In the semiclassical description, BPS monopole states are realized as (normalizable) harmonic forms on the relative moduli space \cite{Witten:1982}. In our setting, the relevant moduli space is $\mathcal{M}_{\mathrm{rel}}(m,\Phi_{0})$, but the same conclusion holds as will be shown in Section~\ref{qm}.

In the single-stratum case, $\mathcal{M}_{\mathrm{rel}}(1,\Phi_{0})=C(k_{1})$ is a compact equal-rank homogeneous space with vanishing odd cohomology, so the number of harmonic forms is given by the Euler characteristic $ \chi [C(k_{1}) ] $. In Subsection~\ref{si}, we collect representative examples (most of which already appeared in \cite{Dorey1995a}) illustrating that $ \chi [C(k_{1}) ] $ equals the dimension of the dual W-boson representation. In Subsection~\ref{mut}, we present two typical examples in which $\mathcal{M}_{\mathrm{rel}} (1,\Phi_{0})=C(k_{1})\cup \mathcal{M}_{\mathrm{rel},2}(1,\Phi_{0}) $ to show that the correspondence persists in the multi-stratum situation. Subsection~\ref{sur} gives a general proof for simple $G$, including the dyonic sector.

We begin with the simplest setup, in which $H=H^{s}\times U(1)$ and $\pi_{2}(G/H)\cong \mathbb{Z}$, so the topological charge is a single integer $m$. Here $G$ is simple and $H$ is obtained by removing a single node from the Dynkin diagram. If $\boldsymbol{\alpha}_{a}$ is the removed simple root of $G$, then $\{\boldsymbol{\alpha}_{b}\}_{b\neq a}$ are the simple roots of $H^{s}$. $ \Phi_{0} = \mathbf{h} \cdot  \mathbf{T}$ with 
\begin{equation}
\mathbf{h}=v \boldsymbol{\omega}^{\vee}_{a},
\end{equation}
where $v$ is a constant and the fundamental coweight $\boldsymbol{\omega}^{\vee}_{a}$ is characterized by $ \boldsymbol{\omega}^{\vee}_{a} \cdot \boldsymbol{\alpha}_{b} = \delta_{a,b} $. For each stratum, $ k_{i}=\mathbf{k}_{i} \cdot \mathbf{T} $ with 
\begin{equation}
\mathbf{k}_{i}  =m \boldsymbol{\alpha}_{a}^{\vee}+\sum_{b \neq a} q_{b} \boldsymbol{\alpha}_{b}^{\vee}.
\end{equation}
The holomorphic charges $  q_{b}$ obey
\begin{equation}
\boldsymbol{\alpha}_{c}\cdot \mathbf{k}_{i} \leq 0,\qquad  q_{b}  \in \mathbb{Z}_{\geq 0}\qquad ( c \neq a),
\end{equation}
which can be explicitly written as 
\begin{equation}\label{asa}
m A_{ac}+\sum_{b \neq a} q_{b}A_{bc} \leq 0, \qquad q_{b}  \in \mathbb{Z}_{\geq 0}  \qquad (c \neq a)
\end{equation}
with Cartan matrix $ A_{ab}=\boldsymbol{\alpha}_{a}^{\vee}\cdot \boldsymbol{\alpha}_{b} $. If (\ref{asa}) admits $n$ solutions, the relative moduli space decomposes as
\begin{equation}\label{3.5a}
 \mathcal{M}_{\mathrm{rel}} (m,\Phi_{0})=\bigcup_{i=1}^{n}\mathcal{M}_{\mathrm{rel},i}(m,\Phi_{0}).
\end{equation}
There is always a trivial solution $  q_{b}=0$, which gives $\mathbf{k}_{1}   =m \boldsymbol{\alpha}_{a}^{\vee} $. So $ \mathcal{M}_{\mathrm{rel}} (m,\Phi_{0})  $ always contains a stratum $\mathcal{M}_{\mathrm{rel},1}(m,\Phi_{0})$. We call $\mathbf{k}_{1}   =m \boldsymbol{\alpha}_{a}^{\vee} $ the principal magnetic charge. For each solution $k_{i}$, using (\ref{dim}),
\begin{equation}\label{dimf}
\dim \mathcal{M}_{\mathrm{rel},i}(m,\Phi_{0})=4(\sum_{b \neq a }q_{b}+m-1) +\dim C(k_{i}).
\end{equation}

When $m=1$, $ \dim \mathcal{M}_{\mathrm{rel},1}(1,\Phi_{0})=\dim C(k_{1}) $, so 
\begin{equation}
\mathcal{M}_{\mathrm{rel},1}(1,\Phi_{0})=C(k_{1}),
\end{equation}
and 
\begin{equation}
\mathcal{M}_{\mathrm{rel}} (1,\Phi_{0})=C(k_{1}) \cup  \mathcal{M}_{\mathrm{rel},2}(1,\Phi_{0})     \cup \cdots   \cup   \mathcal{M}_{\mathrm{rel},n}(1,\Phi_{0}) .
\end{equation}
As a topological space, \(\mathcal{M}_{\mathrm{rel}}(1,\Phi_{0})\) is compact, even though the metric on each $C(k_{i})  $ is divergent. Consequently, the BPS ground states are harmonic forms counted topologically by the Betti numbers of \(\mathcal{M}_{\mathrm{rel}}(1,\Phi_{0})\).

\subsection{$\mathcal{M}_{\mathrm{rel}} (1,\Phi_{0})=C(k_{1}) $}\label{si}
\medskip
\begin{enumerate}[label=\textbf{(\arabic*)}, leftmargin=2em, itemsep=1.2em]

\item \textit{$SU(N+1)\rightarrow U(N) \Longleftrightarrow SU(N+1)/\mathbb{Z}_{N+1}\rightarrow U(N) $}

\medskip

\begin{center}
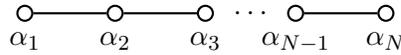

\begin{tikzpicture}[scale=1.2, thick]

\node[circle, draw=black, fill=white, inner sep=2pt, label=below:{\(\alpha_1\)}] (a1) at (0,0) {};
\node[circle, draw=black, fill=white, inner sep=2pt, label=below:{\(\alpha_2\)}] (a2) at (1,0) {};
\node[circle, draw=black, fill=white, inner sep=2pt, label=below:{\(\alpha_3\)}] (a3) at (2,0) {};
\node[circle, draw=black, fill=white, inner sep=2pt, label=below:{\(\alpha_{N-1}\)}] (aNminus1) at (3,0) {};
\node[circle, draw=black, fill=white, inner sep=2pt, label=below:{\(\alpha_N\)}] (aN) at (4,0) {};

\draw (a1) -- (a2) -- (a3);
\node at (2.5,0) {\(\cdots\)};
\draw (aNminus1) -- (aN);

\end{tikzpicture}

\captionof{figure}{Dynkin diagram of \(SU(N+1)\) with simple roots \(\alpha_1,\dots,\alpha_N\).}
\label{fig1}
\end{center}

In the standard orthonormal basis $\{\boldsymbol{e}_{i}\}^{N+1}_{i=1}  $, the simple roots of $SU(N+1)$ are 
\begin{equation}
\boldsymbol{\alpha}_{1}= \boldsymbol{e}_{1}-\boldsymbol{e}_{2}, \qquad \boldsymbol{\alpha}_{2}=\boldsymbol{e}_{2}-\boldsymbol{e}_{3},\qquad \ldots \qquad \boldsymbol{\alpha}_{N}=\boldsymbol{e}_{N}-\boldsymbol{e}_{N+1},
\end{equation}
with $\boldsymbol{\alpha}^{\vee}_{a}=\boldsymbol{\alpha}_{a}$. Removing $\boldsymbol{\alpha}_{N}$ breaks $SU(N+1)$ to $U(N)$ with the semisimple part $ H^{s} =SU(N) $. $ \mathbf{k}_{1}  = \boldsymbol{\alpha}_{N}^{\vee} = \boldsymbol{e}_{N}-\boldsymbol{e}_{N+1} $. The centralizer of $k_{1}$ in $SU(N)$ is $Z_{SU(N) }(k_{1})=S[ U(N-1)\times U(1)  ]$, so the relative moduli space is
\begin{equation}
\mathcal{M}_{\mathrm{rel}} (1,\Phi_{0})= C(k_{1})\cong SU(N)/S[ U(N-1)\times U(1)  ] \cong \mathbb{CP}^{N-1}
\end{equation}
with Euler characteristic
\begin{equation}
 \chi[ \mathcal{M}_{\mathrm{rel}} (1,\Phi_{0})]=N.
\end{equation}
In the dual theory where $SU(N+1)/\mathbb{Z}_{N+1}$ is broken to $U(N)$, the massive gauge bosons transform as
\begin{equation}
(\mathbf{N})_{+1}\ \oplus\ (\overline{\mathbf{N}})_{-1}
\end{equation}
where $\mathbf{N}$ denotes the $N$-dimensional fundamental representation of $SU(N)$ and the subscripts indicate the $U(1)$ charge.

\item \textit{$ SO(2N+2)\rightarrow SO(2N) \times U(1) \Longleftrightarrow  SO(2N+2)\rightarrow SO(2N) \times U(1) $}

\medskip

\begin{center}
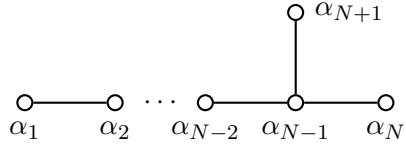

\begin{tikzpicture}[scale=1.2, thick]

\node[circle, draw=black, fill=white, inner sep=2pt, label=below:{\(\alpha_N\)}] (a3) at (4,0) {};
\node[circle, draw=black, fill=white, inner sep=2pt, label=below:{\(\alpha_{N-1}\)}] (a4) at (3,0) {};
\node[circle, draw=black, fill=white, inner sep=2pt, label=below:{\(\alpha_{N-2}\)}] (a5) at (2,0) {};
\node[circle, draw=black, fill=white, inner sep=2pt, label=below:{\(\alpha_2\)}] (a6) at (1,0) {};
\node[circle, draw=black, fill=white, inner sep=2pt, label=below:{\(\alpha_1\)}] (a7) at (0,0) {};

\node[circle, draw=black, fill=white, inner sep=2pt, label=right:{\(\alpha_{N+1}\)}] (a8) at (3,1) {};

\draw (a3) -- (a4) -- (a5);
\draw (a6) -- (a7);
\draw (a4) -- (a8);

\node at (1.5,0) {\(\cdots\)};

\end{tikzpicture}
\captionof{figure}{Dynkin diagram of \(SO(2N+2)\) with simple roots \(\alpha_1,\dots,\alpha_{N+1}\).}
\label{fig2}
\end{center}

Simple roots of $SO(2N+2)$ are 
\begin{equation}
\boldsymbol{\alpha}_{1}= \boldsymbol{e}_{1}-\boldsymbol{e}_{2} ,\qquad \boldsymbol{\alpha}_{2}=\boldsymbol{e}_{2}-\boldsymbol{e}_{3},\qquad \ldots\qquad \boldsymbol{\alpha}_{N}=\boldsymbol{e}_{N}-\boldsymbol{e}_{N+1},\qquad \boldsymbol{\alpha}_{N+1}=\boldsymbol{e}_{N}+\boldsymbol{e}_{N+1}
\end{equation}
with $\boldsymbol{\alpha}^{\vee}_{a}=\boldsymbol{\alpha}_{a}$. Removing $\boldsymbol{\alpha}_{1}$ breaks $SO(2N+2)$ to $SO(2N) \times U(1) $ with $H^{s} =SO(2N)$. $ \mathbf{k}_{1}  = \boldsymbol{\alpha}_{1}^{\vee} = \boldsymbol{e}_{1}-\boldsymbol{e}_{2}  $. The centralizer of $k_{1}$ in $SO(2N)$ is $Z_{SO(2N)}(k_{1})= SO(2N-2)\times U(1)  $, so
\begin{equation}
\mathcal{M}_{\mathrm{rel}} (1,\Phi_{0})= C(k_{1})\cong SO(2N)/[SO(2N-2) \times U(1) ]
\end{equation}
with Euler characteristic
\begin{equation}
\chi[\mathcal{M}_{\mathrm{rel}} (1,\Phi_{0})]= 2N.
\end{equation}
In the dual theory where $SO(2N+2)$ is broken to $SO(2N) \times U(1) $, the massive gauge bosons transform as
\begin{equation}
(\mathbf{2N})_{+1}\ \oplus\ (\mathbf{2N})_{-1}.
\end{equation}

\item \textit{$ SO(2N+2)\rightarrow U(N+1) \Longleftrightarrow  SO(2N+2)\rightarrow U(N+1) $}

\medskip

If $\boldsymbol{\alpha}_{N+1}$ is removed instead, the symmetry breaks to $U(N+1)$ with semisimple part $H^{s}=SU(N+1)$. $ \mathbf{k}_{1}  = \boldsymbol{\alpha}_{N+1}^{\vee} = \boldsymbol{e}_{N}+\boldsymbol{e}_{N+1}  $. The centralizer of $k_{1}$ in $SU(N+1)$ is $ S[U(N-1)\times  U(2) ]  $, so
\begin{equation}
\mathcal{M}_{\mathrm{rel}} (1,\Phi_{0})= C(k_{1})\cong SU(N+1)/S[U(N-1)\times  U(2) ]\cong \mathrm{Gr}(2, N+1)
\end{equation}
with Euler characteristic
\begin{equation}
 \chi[\mathcal{M}_{\mathrm{rel}} (1,\Phi_{0})]=\binom{N+1}{2}=\frac{N(N+1)}{2}.
\end{equation}
In the dual theory where $SO(2N+2)$ is broken to $U(N+1)$, the massive gauge bosons transform as\footnote{We normalize the $U(1)$ generator so that the $U(1)$ charge of W-bosons are integers with unit spacing.} 
\begin{equation}
(\Lambda^{2}\,\mathbf{N{+}1})_{+1}\ \oplus\ (\Lambda^{2}\,\overline{\mathbf{N{+}1}})_{-1},
\end{equation}
where $\Lambda^{2}\,\mathbf{N{+}1}$ denotes the $2$-index antisymmetric representation of $SU(N+1)$.

\item \textit{$SO(2N+3)\rightarrow SO(2N+1) \times U(1)\Longleftrightarrow USp(2N+2)\rightarrow USp(2N) \times U(1)$}

\medskip

\begin{center}
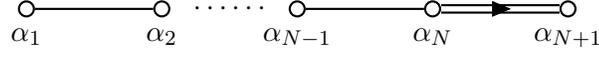

\begin{tikzpicture}[scale=1.2, thick, >=stealth,
  decorate double/.style={
    double,
    double distance=2pt,
    decoration={
      markings,
      mark=at position 0.63 with {\arrow[scale=0.5]{triangle 45}}
    },
    postaction={decorate}
  }
]

  \node[circle,draw,fill=white,inner sep=2pt,label=below:{\(\alpha_{1}\)}] (a1) at (-1.5,0) {};
  \node[circle,draw,fill=white,inner sep=2pt,label=below:{\(\alpha_{2}\)}] (a2) at (0,0) {};
  \node[circle,draw,fill=white,inner sep=2pt,label=below:{\(\alpha_{N-1}\)}] (a3) at (1.5,0) {};
  \node[circle,draw,fill=white,inner sep=2pt,label=below:{\(\alpha_{N}\)}] (a4) at (3.0,0) {};
  \node[circle,draw,fill=white,inner sep=2pt,label=below:{\(\alpha_{N+1}\)}] (a5) at (4.5,0) {};

  \draw (a1) -- (a2);
  \node at (0.75, 0) {\(\cdots\cdots\)};
  \draw (a3) -- (a4);
  \draw[decorate double] (a4) -- (a5); 

\end{tikzpicture}
\captionof{figure}{Dynkin diagram of \(SO(2N+3)\) with simple roots \(\alpha_1,\dots,\alpha_{N+1}\).}
\label{fig3}
\end{center}

Simple roots of $SO(2N+3)$ are 
\begin{equation}\label{so1}
\boldsymbol{\alpha}_{1}= \boldsymbol{e}_{1}-\boldsymbol{e}_{2} ,\qquad \boldsymbol{\alpha}_{2}=\boldsymbol{e}_{2}-\boldsymbol{e}_{3},\qquad \ldots \qquad \boldsymbol{\alpha}_{N}=\boldsymbol{e}_{N}-\boldsymbol{e}_{N+1},\qquad \boldsymbol{\alpha}_{N+1}=\boldsymbol{e}_{N+1},
\end{equation}
with $\boldsymbol{\alpha}^{\vee}_{a}=\boldsymbol{\alpha}_{a}$ for $1 \leq a \leq N$ and $\boldsymbol{\alpha}^{\vee}_{N+1}=2 \boldsymbol{e}_{N+1}$. Removing $\boldsymbol{\alpha}_{1}$ breaks $SO(2N+3)$ to $H^{s}\times U(1) =SO(2N+1) \times U(1) $. $ \mathbf{k}_{1}  = \boldsymbol{\alpha}_{1}^{\vee} = \boldsymbol{e}_{1}-\boldsymbol{e}_{2}  $. The centralizer of $k_{1}$ in $SO(2N+1)$ is $ SO(2N-1)\times U(1)  $, so
\begin{equation}
\mathcal{M}_{\mathrm{rel}} (1,\Phi_{0})= C(k_{1})\cong SO(2N+1)/[SO(2N-1) \times U(1) ]
\end{equation}
with 
\begin{equation}
\chi[\mathcal{M}_{\mathrm{rel}} (1,\Phi_{0})]= 2N.
\end{equation}
In the dual theory, where $USp(2N+2)$ is broken to $USp(2N) \times U(1)$, the massive gauge bosons transform as 
\begin{equation}
(\mathbf{2N})_{+1}\ \oplus\ (\mathbf{2N})_{-1}\ \oplus\ (\mathbf{1})_{+2}\ \oplus\ (\mathbf{1})_{-2}.
\end{equation}
We will identify the monopole counterparts of $(\mathbf{1})_{\pm 2}$ in Subsection \ref{sur}.

\item \textit{$USp(2N+2) \rightarrow U(N+1)    \Longleftrightarrow  SO(2N+3)\rightarrow U(N+1)  $}

\medskip

\begin{center}
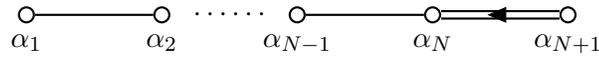

\begin{tikzpicture}[scale=1.2, thick, >=stealth,
  decorate double/.style={
    double,
    double distance=2pt,
    decoration={
      markings,
      mark=at position 0.65 with {\arrow[scale=0.5]{triangle 45}}
    },
    postaction={decorate}
  }
]

  \node[circle,draw,fill=white,inner sep=2pt,label=below:{\(\alpha_{1}\)}] (a1) at (-1.5,0) {};
  \node[circle,draw,fill=white,inner sep=2pt,label=below:{\(\alpha_{2}\)}] (a2) at (0,0) {};
  \node[circle,draw,fill=white,inner sep=2pt,label=below:{\(\alpha_{N-1}\)}] (a3) at (1.5,0) {};
  \node[circle,draw,fill=white,inner sep=2pt,label=below:{\(\alpha_{N}\)}] (a4) at (3.0,0) {};
  \node[circle,draw,fill=white,inner sep=2pt,label=below:{\(\alpha_{N+1}\)}] (a5) at (4.5,0) {};

  \draw (a1) -- (a2);
  \node at (0.75, 0) {\(\cdots\cdots\)};
  \draw (a3) -- (a4);
  \draw[decorate double] (a5) -- (a4); 

\end{tikzpicture}
\captionof{figure}{Dynkin diagram of \(USp(2N+2)\) with simple roots \(\alpha_1,\dots,\alpha_{N+1}\).}
\label{fig4}
\end{center}

Simple roots of $USp(2N+2)$ are 
\begin{equation}\label{usp1}
\boldsymbol{\alpha}_{1}= \boldsymbol{e}_{1}-\boldsymbol{e}_{2} ,\qquad \boldsymbol{\alpha}_{2}=\boldsymbol{e}_{2}-\boldsymbol{e}_{3},\qquad \ldots\qquad \boldsymbol{\alpha}_{N}=\boldsymbol{e}_{N}-\boldsymbol{e}_{N+1},\qquad \boldsymbol{\alpha}_{N+1}=2\boldsymbol{e}_{N+1}
\end{equation}
with $\boldsymbol{\alpha}^{\vee}_{a}=\boldsymbol{\alpha}_{a}$ for $1 \leq a \leq N$ and $\boldsymbol{\alpha}^{\vee}_{N+1}= \boldsymbol{e}_{N+1}$. Removing $\boldsymbol{\alpha}_{N+1}$ breaks $USp(2N+2)$ to $U(N+1) $ with $H^{s}=SU(N+1)$. $ \mathbf{k}_{1}  = \boldsymbol{\alpha}_{N+1}^{\vee} = \boldsymbol{e}_{N+1} $. The centralizer of $k_{1}$ in $SU(N+1)$ is $S[ U(N)\times U(1)  ]$, so
\begin{equation}
\mathcal{M}_{\mathrm{rel}} (1,\Phi_{0})= C(k_{1})\cong SU(N+1)/S[U(N)\times U(1) ]\cong \mathbb{CP}^{N} 
\end{equation}
with 
\begin{equation}
\chi[\mathcal{M}_{\mathrm{rel}} (1,\Phi_{0})]= N+1.
\end{equation}
In the dual theory, where $SO(2N+3)$ is broken to $U(N+1)$, the massive gauge bosons transform as 
\begin{equation}
(\mathbf{N{+}1})_{+1}\ \oplus\
(\overline{\mathbf{N{+}1}})_{-1}      \ \oplus\
(\Lambda^{2}\,\mathbf{N{+}1})_{+2}\ \oplus\
(\Lambda^{2}\,\overline{\mathbf{N{+}1}})_{-2}   .
\end{equation}
We will discuss the monopole counterparts of $(\Lambda^{2}\,\mathbf{N{+}1})_{+2}$ and $(\Lambda^{2}\,\overline{\mathbf{N{+}1}})_{-2}$ in Subsection \ref{sur}.

\end{enumerate}

\subsection{$ \mathcal{M}_{\mathrm{rel}} (1,\Phi_{0})=C(k_{1})\cup \mathcal{M}_{\mathrm{rel},2}(1,\Phi_{0}) $}\label{mut}

\medskip
\begin{enumerate}[label=\textbf{(\arabic*)}, leftmargin=2em, itemsep=1.2em]

\item \textit{$USp(2N+2)\rightarrow USp(2N)\times U(1)  \Longleftrightarrow   SO(2N+3)\rightarrow SO(2N+1) \times U(1) $}\label{mut1}

\medskip

In Figure \ref{fig4}, removing $\boldsymbol{\alpha}_{1}$ breaks $USp(2N+2)$ to $ H^{s} \times U(1) = USp(2N)\times U(1)$. For $m=1$, (\ref{asa}) admits two solutions,
\begin{equation}
 \mathbf{k}_{1}  = \boldsymbol{\alpha}_{1}^{\vee}=\boldsymbol{e}_{1}-\boldsymbol{e}_{2} ,\qquad \mathbf{k}_{2}  = \boldsymbol{\alpha}_{1}^{\vee}+ \cdots +\boldsymbol{\alpha}_{N+1}^{\vee}=\boldsymbol{e}_{1}. 
\end{equation}
$  \mathcal{K}=C(k_{1}) \cup C(k_{2})  $. The centralizers in $USp(2N)$ are 
\begin{equation}
Z_{USp(2N)}(k_{1})=USp(2N-2) \times U(1) ,\qquad Z_{USp(2N)}(k_{2})=USp(2N),
\end{equation}
so
\begin{equation}
C(k_{1})\cong USp(2N) /[USp(2N-2) \times U(1)]\cong  \mathbb{CP}^{2N-1}  ,\qquad C(k_{2})=\{k_{2}\}.
\end{equation}
From (\ref{dimf}), 
\begin{equation}
 \dim \mathcal{M}_{\mathrm{rel},1}(1,\Phi_{0})=\dim C(k_{1})= 4N-2,\qquad  \dim \mathcal{M}_{\mathrm{rel},2}(1,\Phi_{0})=4N +\dim C(k_{2}) =4N.
\end{equation}
The closed stratum is $\mathcal{M}_{\mathrm{rel},1}(1,\Phi_{0})=\mathbb{CP}^{2N-1}$, while the open stratum $\mathcal{M}_{\mathrm{rel},2}(1,\Phi_{0})=\mathcal{M}_{\mathrm{rel}}(1,\Phi_{0},k_{2})$ is the relative moduli space for monopoles of type $(\Phi_{0},k_{2})$. As shown in \cite{1}, this moduli space carries a flat metric with
\begin{equation}
\mathcal{M}_{\mathrm{rel}}(1,\Phi_{0},k_{2})\cong \mathbb{R}^{4N}\cong \mathbb{C}^{2N}.
\end{equation}
Consequently,
\begin{equation}
  \mathcal{M}_{\mathrm{rel}} (1,\Phi_{0}) \cong\mathbb{CP}^{2N-1} \cup  \mathbb{C}^{2N}.
\end{equation}
Compactifying $ \mathbb{C}^{2N} $ by adjoining $ \mathbb{CP}^{2N-1} $ at infinity gives $  \mathbb{CP}^{2N}  $ \cite{GriffithsHarris}. Hence, as a topological space,
\begin{equation}\label{3.34a}
\mathcal{M}_{\mathrm{rel}} (1,\Phi_{0}) \approx \mathbb{CP}^{2N}.
\end{equation}
As Riemannian manifolds, however, the natural metric on $\mathcal{M}_{\mathrm{rel}}(1,\Phi_{0})$ admits a $USp(2N)$ symmetry coming from the unbroken group, whereas the standard Fubini-Study metric on $\mathbb{CP}^{2N}$ has isometry group $SU(2N+1)$. The Euler characteristic is
\begin{equation}
\chi[\mathcal{M}_{\mathrm{rel}} (1,\Phi_{0})]= 2N+1.
\end{equation}
In the dual theory, where $SO(2N+3)$ is broken to $SO(2N+1) \times U(1) $, the massive gauge bosons transform as 
\begin{equation}
(\mathbf{2N{+}1})_{+1}\ \oplus\
(\mathbf{2N{+}1})_{-1}.
\end{equation}

For $N=1$, explicit $USp(4) \rightarrow USp(2)\times U(1) $ (equivalent to $SO(5)\rightarrow U(2)$) solutions of type $ (\Phi_{0}, k_{2}) $ are reviewed in Appendix \ref{reg}. The solutions are parametrized by eight moduli: three translations, one $U(1)$ electric phase, three $USp(2)$ orientations, and a size $a\in[0,\infty)$ of the ``non-Abelian cloud'' (see \cite{1}). For finite $a$, the charge is $k_{2}$ and the $USp(2)$ modes are normalizable. At $a=\infty$, the solutions are gauge equivalent to monopoles of type $(\Phi_{0},G_{0})$ with $G_{0}\in C(k_{1})$, for which the $USp(2)$ orientation metric diverges, in agreement with the boundary metric on $\mathbb{S}^{3}_{\infty}$. Moreover, at $a=\infty$, a $U(1)\subset USp(2)$ acts as the electric $U(1)$ and must be quotiented out in the relative moduli space. For instance, in the $SU(2)$ embedding with generators
\begin{equation}
S_{1}=\frac{1}{2}\begin{pmatrix}
 0& 1&0& 0\\ 1 &0&0& 0\\ 0 &0&0& -1\\ 0 & 0&-1& 0
\end{pmatrix}\qquad S_{2}=\frac{1}{2}\begin{pmatrix}
 0& i&0& 0\\- i &0&0& 0\\ 0 &0&0&- i\\ 0 & 0&i& 0
\end{pmatrix}\qquad S_{3}=\frac{1}{2}\begin{pmatrix}
 -1& 0&0& 0\\ 0 &1&0& 0\\ 0 &0&1& 0\\ 0 & 0&0& -1
\end{pmatrix},
\end{equation}
the solution is invariant under $T=\operatorname{diag}(1,1,-1,-1)$, while the electric $U(1)$ is generated by $T'=\operatorname{diag}(0,1,0,-1)$; the combination $ T''=\operatorname{diag}(1,0,-1,0)=T-T' $ is a $U(1) \subset USp(2)$ that acts purely electrically and should be quotiented out. As a result, the boundary $\mathbb{S}^{3}_{\infty}$ projects to $\mathbb{CP}^{1}$ via the Hopf fibration.

For $N>1$, embedding the $USp(4)$ solution into $USp(2N+2)$ leaves a $USp(2N-2)$ subgroup unbroken, and the gauge orientations are parametrized by 
\begin{equation}
USp(2N)/ USp(2N-2) \cong \mathbb{S}^{4N-1}.
\end{equation}
At the boundary of $\mathbb{R}^{4N}$, $k_{2}\to C(k_{1})$, and a $U(1)\subset USp(2N)$ becomes an electric gauge symmetry. Quotienting by this $U(1)$ projects $ \mathbb{S}^{4N-1}$ onto $ \mathbb{CP}^{2N-1}$ via the Hopf fibration.

\item \textit{$SO(2N+3)\rightarrow U(N+1) \Longleftrightarrow  USp(2N+2)\rightarrow U(N+1)  $}\label{mut2}

In Figure \ref{fig3}, removing $\boldsymbol{\alpha}_{N+1}$ breaks $SO(2N+3)$ to $U(N+1)$ with $H^{s}=SU(N+1)$. For $m=1$, (\ref{asa}) admits two solutions, 
\begin{equation}
 \mathbf{k}_{1}  = \boldsymbol{\alpha}_{N+1}^{\vee}=2\boldsymbol{e}_{N+1},\qquad  \mathbf{k}_{2} = \boldsymbol{\alpha}_{N}^{\vee}+  \boldsymbol{\alpha}_{N+1}^{\vee}=\boldsymbol{e}_{N}+\boldsymbol{e}_{N+1},
\end{equation}
whose centralizers in $SU(N+1)$ are 
\begin{equation}
Z_{SU(N+1)}(k_{1})=S[U(N)\times U(1)] ,\qquad Z_{SU(N+1)}(k_{2})=S[U(N-1) \times U(2) ].
\end{equation}
Accordingly,
\begin{eqnarray}
 && C(k_{1})\cong SU(N+1)/S[U(N)\times U(1)]\cong  \mathbb{CP}^{N}  ,\\ &&  C(k_{2})\cong SU(N+1)/S[U(N-1) \times U(2) ]\cong\mathrm{Gr}(2,N+1).
\end{eqnarray} 
From (\ref{dimf}),
\begin{equation}
 \dim \mathcal{M}_{\mathrm{rel},1}(1,\Phi_{0})=\dim  C(k_{1})= 2N,\qquad  \dim \mathcal{M}_{\mathrm{rel},2}(1,\Phi_{0})= 4 +\dim C(k_{2}) =4N,
\end{equation}
with $\dim \mathcal{M}_{\mathrm{rel}} (1,\Phi_{0}, k_{2}) =4  $. The closed stratum is $\mathcal{M}_{\mathrm{rel},1} (1,\Phi_{0})=\mathbb{CP}^{N}$. The open stratum $\mathcal{M}_{\mathrm{rel},2}(1,\Phi_{0})   $ is a fiber bundle over $C(k_2)$ with fiber $ \mathcal{M}_{\mathrm{rel}} (1,\Phi_{0}, k_{2})$. 
Since $\mathcal{M}_{\mathrm{rel}} (1,\Phi_{0}, k_{2}) \cong \mathbb{C}^{2}$\footnote{When $N=1$, the symmetry breaks from $SO(5)$ to $ U(2)$ and $ \mathcal{M}_{\mathrm{rel}} (1,\Phi_{0}, k_{2})=\mathbb{R}^{4}\cong \mathbb{C}^{2}$ \cite{1}. When $N>1$, $  \mathcal{M}_{\mathrm{rel}} (1,\Phi_{0}, k_{2})$ remains the same.}, $\mathcal{M}_{\mathrm{rel},2}(1,\Phi_{0})     $ is the total space of a rank-$2$ holomorphic vector bundle over $\mathrm{Gr}(2,N+1)$:
\begin{equation}
\mathcal{M}_{\mathrm{rel},2}(1,\Phi_{0})\cong \mathrm{Tot}(S^{\vee}),
\qquad
\pi: \mathrm{Tot}(S^{\vee})\to \mathrm{Gr}(2,N{+}1),
\end{equation}
where $S$ is the tautological rank-$2$ subbundle on $\mathrm{Gr}(2,N{+}1)$ and $\pi$ is the bundle projection. Gluing this open bundle to the closed Schubert variety at infinity yields a compactification \cite{GriffithsHarris}
\begin{equation}
\mathcal{M}_{\mathrm{rel}}(1,\Phi_{0})
\;\cong\;
\mathbb{CP}^{N}\,\cup\, \mathrm{Tot}(S^{\vee})
\;\approx\;
\mathrm{Gr}(2,N{+}2).
\end{equation}
$\mathcal{M}_{\mathrm{rel}}(1,\Phi_{0})$ is homeomorphic to $\mathrm{Gr}(2,N{+}2)$. In particular, when $N=1$,
\begin{equation}
  \mathcal{M}_{\mathrm{rel}} (1,\Phi_{0}) \cong\mathbb{CP}^{1} \cup  \mathbb{C}^{2} \approx \mathrm{Gr}(2,3) =\mathbb{CP}^{2}.
\end{equation}
The Euler characteristic is
\begin{equation}
 \chi[\mathcal{M}_{\mathrm{rel}} (1,\Phi_{0})]=\binom{N+2}{2}=\frac{(N+1)(N+2)}{2}.
\end{equation}
In the dual theory, where $USp(2N+2)$ is broken to $U(N+1) $, the massive gauge bosons transform as 
\begin{equation}
( \mathrm{Sym}^2 \, \mathbf{N{+}1} )_{+1} \; \oplus \; ( \mathrm{Sym}^2 \, \overline{\mathbf{N{+}1}} )_{-1}, 
\end{equation}
where $\mathrm{Sym}^2 \, \mathbf{N{+}1} $ denotes the symmetric two-tensor of $SU(N+1)$.

\end{enumerate}

\subsection{General proof}\label{sur}

In this subsection, we give a general proof of the matching between the BPS monopole spectrum and the W-boson spectrum for the symmetry breakings $G\to H$ and $G^{\vee}\to H^{\vee}$. Before proceeding, we collect a few facts that will be used below.

For a compact, connected semisimple group $\mathcal{G}$, let $ \lambda $ be an anti-dominant coweight of $\mathcal{G}$ (equivalently, a weight of $\mathcal{G}^{\vee}$), and let $V_{\lambda}$ be the irreducible $\mathcal{G}^{\vee}$-representation with lowest weight $ \lambda $\footnote{Representations are usually labeled by a dominant highest weight; here we use the equivalent convention of an anti-dominant lowest weight to match our setting.}. The set of weights in $V_{\lambda}$ decomposes into Weyl orbits
\begin{equation}
\text{wt}(V_{\lambda} )=O(\lambda_{1}) \cup \cdots \cup O(\lambda_{n}),\qquad \text{with}\;\;\lambda_{1} =\lambda,
\end{equation}
where $O(\lambda_{i})$ is indexed by its unique anti-dominant representative $ \lambda_{i} $. For each orbit $O(\lambda_{i})$, all weights appear in $V_{\lambda}$ with the same multiplicity $  \mu(\lambda_{i})$, so \cite{Humphreys}
\begin{equation}
\dim V_{\lambda} =\sum^{n}_{i=1} \mu(\lambda_{i})|O(\lambda_{i}) |,
\end{equation}
where 
\begin{equation}
|O(\lambda_{i}) |=\frac{|W_{\mathcal{G}}|}{|W_{Z_{\mathcal{G}}(\lambda_{i})}|}
\end{equation}
is the orbit size. Here $ Z_{\mathcal{G}}(\lambda_{i}) $ is the centralizer of $\lambda_{i}$ in $\mathcal{G}$, $W_{\mathcal{G}}$ and $W_{Z_{\mathcal{G}}(\lambda_{i})}$ are Weyl groups of $\mathcal{G}$ and $Z_{\mathcal{G}}(\lambda_{i})$, respectively. Since $ Z_{\mathcal{G}}(\lambda_{i}) $ and $  \mathcal{G} $ have the same rank, it follows from \cite{HopfSamelson} that 
\begin{equation}
\chi [\mathcal{G}/Z_{\mathcal{G}}(\lambda_{i}) ]=\frac{|W_{\mathcal{G}}|}{|W_{Z_{\mathcal{G}}(\lambda_{i})}|}.
\end{equation} 
Therefore, 
\begin{equation}\label{semi3}
\dim V_{\lambda} =\sum^{n}_{i=1} \mu(\lambda_{i})\;\chi [\mathcal{G}/Z_{\mathcal{G}}(\lambda_{i}) ]. 
\end{equation}
\begin{itemize}
  \item
If $ \lambda$ is minuscule in $\mathcal{G}$, i.e. 
\begin{equation}
\langle \lambda,\boldsymbol{\alpha}\rangle\in\{0,-1\}\qquad   \forall\,\boldsymbol{\alpha} \in \Phi^{+}(\mathcal{G}),
\end{equation}
where $\Phi^{+}(\mathcal{G})$ is the set of positive roots, then $ n=1 $, $ \mu(\lambda)=1 $ and
\begin{equation}\label{semi}
\dim V_{\lambda} =\chi [\mathcal{G}/Z_{\mathcal{G}}(\lambda) ]. 
\end{equation}
 \item
If $ V_{\lambda}  $ is multiplicity-free with $ \mu(\lambda_{i})=1 $ for $1 \leq i \leq n$, then 
 \begin{equation}
\dim V_{\lambda} =\sum^{n}_{i=1} \chi [\mathcal{G}/Z_{\mathcal{G}}(\lambda_{i}) ]. 
\end{equation}

\end{itemize}
  
  \medskip

\subsubsection*{Ground state counting on the relative moduli space}

In $\mathcal{N}=4$ SYM, let $G$ be a compact, connected simple group with simple roots $\{\boldsymbol{\alpha}_{b}\}$. Suppose $G$ is broken to $H$ by a Higgs VEV $\Phi_{0}$. When $ \pi_{2}(G/H)=\mathbb{Z} $, $ H=H^{s} \times U(1)$ (up to a central quotient), where $H^{s}$, with simple roots $\{\boldsymbol{\alpha}_{b}  \}_{b\neq a} $, is the semisimple factor obtained by deleting the node $\boldsymbol{\alpha}_{a}$ from the Dynkin diagram of $G$. The Higgs VEV is aligned with the fundamental coweight $ \boldsymbol{\omega}_{a}^{\vee} $:
\begin{equation}
\Phi_{0} =v \boldsymbol{\omega}_{a}^{\vee} \cdot  \mathbf{T}.
\end{equation}
For a fixed topological charge $m$, the allowed magnetic charges form a finite union of $H^{s}$-orbits
\begin{equation}
\mathcal{K}=\bigcup_{i=1}^{n}C(k_{i}) ,
\end{equation}
where each representative can be written as
\begin{equation}
\mathbf{k}_{i}  =m \boldsymbol{\alpha}_{a}^{\vee}+\sum_{b \neq a} q_{b} \boldsymbol{\alpha}_{b}^{\vee}, \qquad m \in \mathbb{Z}_{>0}, \qquad q_{b} \in \mathbb{Z}_{\geq 0}
\end{equation}
with coefficients constrained by (\ref{asa}).

The projection of $\mathbf{k}_{i} $ to $ H^{s} $ gives
\begin{equation}\label{bk}
\overline{\mathbf{k}}_{i} =\sum_{c \neq a}(\mathbf{k}_{i} \cdot \boldsymbol{\alpha}_{c})\,\boldsymbol{\omega}_{c}^{\vee (H^{s})} =\sum_{c \neq a}(m A_{ac}+\sum_{b \neq a} q_{b} A_{bc})\,\boldsymbol{\omega}_{c}^{\vee (H^{s})}  , \qquad  1 \leq i \leq n,
\end{equation}
which is an anti-dominant weight of $(H^{\vee})^{s}$. Here $\{ \boldsymbol{\omega}_{c}^{\vee (H^{s})} \equiv \boldsymbol{\omega}_{c}^{\vee }|_{H^{s}}\}_{c \neq a}$ are restrictions of fundamental coweights to $H^{s}$, and hence comprise the fundamental weights of $(H^{\vee})^{s}$. The anti-dominance of $\overline{\mathbf{k}}_{i} $ in $(H^{\vee})^{s}$ is guaranteed by (\ref{asa}). If deleting $ \boldsymbol{\alpha}_{a} $ yields $ d $ connected components, then 
\begin{equation}
H^{s} =H^{s}_{(1)} \times \cdots \times H^{s}_{(d)}, \qquad   \overline{\mathbf{k}}_{i} =\overline{\mathbf{k}}_{i(1)} +\cdots+\overline{\mathbf{k}}_{i(d)}  ,
\end{equation}
with $\overline{\mathbf{k}}_{i(l)} $ an anti-dominant weight of $(H^{\vee}_{(l)})^{s}$.

For the principal charge $ \mathbf{k}_{1}  =m \boldsymbol{\alpha}_{a}^{\vee}$,
\begin{equation}\label{4281}
\overline{\mathbf{k}}_{1} =\sum_{c \neq a} m A_{ac}\,\boldsymbol{\omega}_{c}^{\vee (H^{s})}.
\end{equation}
Let $V_{\overline{\mathbf{k}}_{1}} $ denote the irreducible $(H^{\vee})^{s}$-representation of lowest weight $\overline{\mathbf{k}}_{1}$. By construction, the weight set of $V_{\overline{\mathbf{k}}_{1}} $ decomposes as 
\begin{equation}
\text{wt}(V_{\overline{\mathbf{k}}_{1}} )=O(\overline{\mathbf{k}}_{1}) \cup \cdots \cup O(\overline{\mathbf{k}}_{n}),
\end{equation}
with $\overline{\mathbf{k}}_{i}$ given by (\ref{bk}). If $ O(\overline{\mathbf{k}}_{i}) $ occurs with multiplicity $ \mu(\overline{\mathbf{k}}_{i})$, from (\ref{semi3}),  
\begin{equation}\label{wt1}
\dim V_{\overline{\mathbf{k}}_{1}} =\sum^{n}_{i=1} \mu(\overline{\mathbf{k}}_{i})\;\chi [H^{s}/Z_{H^{s}}(\overline{\mathbf{k}}_{i}) ]. 
\end{equation}

\begin{itemize}
  \item  For $m=1$, we have $\mathbf{k}_{1}= \boldsymbol{\alpha}_{a}^{\vee}  $. 
  
  \begin{itemize}[label=$\diamond$] 
      \item If $\overline{\mathbf{k}}_{1}$ is minuscule in $H^{s}$, which occurs when all simple roots have equal length ($ G=A,D,E $) or when $\boldsymbol{\alpha}_{a}$ is a long root, then (\ref{asa}) admits a unique solution $\mathbf{k}_{1}$, and 
\begin{equation}
 \mathcal{M}_{\mathrm{rel}} (1,\Phi_{0})=C(k_{1}) = H^{s}  / Z_{H^{s}}(k_{1}) .
\end{equation}
From (\ref{semi}),
\begin{equation}
\dim V_{\overline{\mathbf{k}}_{1}} = \chi [ H^{s}  / Z_{H^{s}}(\overline{\mathbf{k}}_{1})    ]    = \chi [ H^{s}  / Z_{H^{s}}(k_{1})    ]  = \chi [ \mathcal{M}_{\mathrm{rel}} (1,\Phi_{0})   ].
\end{equation}

     \item If $\overline{\mathbf{k}}_{1}$ is not minuscule, which happens when $\boldsymbol{\alpha}_{a}$ is a short root, then (\ref{asa}) has multiple solutions, and 
\begin{equation}
\mathcal{M}_{\mathrm{rel}} (1,\Phi_{0})=C(k_{1}) \cup  \mathcal{M}_{\mathrm{rel},2}(1,\Phi_{0})     \cup \cdots   \cup   \mathcal{M}_{\mathrm{rel},n}(1,\Phi_{0}).
\end{equation}
$ \mathcal{M}_{\mathrm{rel}} (1,\Phi_{0}) $ remains compact. Each component $ \mathcal{M}_{\mathrm{rel},i}(1,\Phi_{0})   $ is a fiber bundle over $C(k_i)$ with fiber $\mathcal M_{\mathrm{rel}}(1,\Phi_0,k_i)$. Since the fiber is contractible, the projection induces an isomorphism in cohomology \cite{BottTu}. Hence
\begin{equation}
H^{\ast}[ \mathcal{M}_{\mathrm{rel},i}(1,\Phi_{0}) ; \mathbb{R} ]=H^{\ast}[C(k_i); \mathbb{R}],
\end{equation}
and 
\begin{equation}
 \sum_{k}\dim H^{k}[ \mathcal{M}_{\mathrm{rel},i}(1,\Phi_{0}) ; \mathbb{R} ]=\chi[ \mathcal{M}_{\mathrm{rel},i}(1,\Phi_{0})  ]=\chi[C(k_i)].
\end{equation}
Therefore, by additivity of Euler characteristic over a finite stratification \cite{Hatcher},
\begin{equation}\label{wt2}
\chi[\mathcal{M}_{\mathrm{rel}} (1,\Phi_{0})]=\sum^{n}_{i=1} \chi [C(k_i)]=\sum^{n}_{i=1}  \chi [ H^{s}  / Z_{H^{s}}(\overline{\mathbf{k}}_{i})    ]   .
\end{equation}
On the other hand, when $m=1$, $V_{\overline{\mathbf{k}}_{1}}   $ is multiplicity-free with $  \mu(\overline{\mathbf{k}}_{i})=1 $ in (\ref{wt1}), so 
\begin{equation}\label{de}
\dim V_{\overline{\mathbf{k}}_{1}} = \chi [ \mathcal{M}_{\mathrm{rel}} (1,\Phi_{0})   ]
\end{equation}
still holds.

\end{itemize}

   \item When $m>1$, $\mathbf{k}_{1}=m \boldsymbol{\alpha}_{a}^{\vee}  $ is non-minuscule, and (\ref{asa}) admits multiple solutions $\{\mathbf{k}_{1},\ldots,\mathbf{k}_{n} \}$. Accordingly,
\begin{equation}
\mathcal{M}_{\mathrm{rel}} (m,\Phi_{0})=\mathcal{M}_{\mathrm{rel},1}(m,\Phi_{0})     \cup  \mathcal{M}_{\mathrm{rel},2}(m,\Phi_{0})     \cup \cdots   \cup   \mathcal{M}_{\mathrm{rel},n}(m,\Phi_{0}),
\end{equation}
where $ \mathcal{M}_{\mathrm{rel},i}(m,\Phi_{0})   $ is a fiber bundle over $C(k_i)$ with fiber $\mathcal M_{\mathrm{rel}}(m,\Phi_0,k_i)$.

In the full moduli space $ \mathcal{M} (m,\Phi_{0}) $, when $m>1$, each fiber $  \mathcal M (m,\Phi_0,k_i)$ contains $m$ identical massive monopoles (with some additional massless monopoles when $i \neq 1$) and thus has the structure \cite{Sen:1994, 1}
   \begin{equation}
\mathcal M (m,\Phi_0,k_i)=\mathbb{R}^{3} \times \frac{\mathbb{S}^{1}\times    \mathcal M^{0}_{\mathrm{rel}}(m,\Phi_0,k_i) }{\mathbb{Z}_{m}},
   \end{equation}
where $\mathbb{Z}_{m}$ is generated by an element $ g $ acting on both $\mathbb{S}^{1}$ and $\mathcal M^{0}_{\mathrm{rel}}(m,\Phi_0,k_i) $. A state of electric charge $ p $ has wavefunction $e^{ip\chi}$ on $\mathbb{S}^{1}$. Let $\omega$ be the wavefunction on $\mathcal M^{0}_{\mathrm{rel}}(m,\Phi_0,k_i) $. Then under the action of $g$, 
\begin{equation}
e^{ip\chi} \rightarrow   e^{ 2 \pi i p/m} e^{ip\chi},\qquad \omega \rightarrow e^{- 2 \pi i p/m} \omega. 
\end{equation}
For the moment we focus on the neutral sector $p=0$, for which, $\omega$ is $\mathbb{Z}_{m}$-invariant and $ \mathcal M_{\mathrm{rel}}(m,\Phi_0,k_i) =   \mathcal M^{0}_{\mathrm{rel}}(m,\Phi_0,k_i) / \mathbb{Z}_{m}$.

    $ \mathcal{M}_{\mathrm{rel}} (m,\Phi_{0}) $ is noncompact. The compact-support Euler characteristic is
\begin{equation}\label{eu1}
\chi_{c}[\mathcal{M}_{\mathrm{rel}} (m,\Phi_{0})]=\sum^{n}_{i=1} \chi_{c} [ \mathcal{M}_{\mathrm{rel},i}(m,\Phi_{0})   ]=\sum^{n}_{i=1} \chi_{c} [\mathcal M_{\mathrm{rel}}(m,\Phi_0,k_i)  ]  \chi [C(k_i)]
\end{equation}
by additivity over disjoint unions and multiplicativity for locally trivial bundles over compact bases \cite{BottTu, Hatcher}.

In fact, when $m>1$ with $ \mathcal{M}_{\mathrm{rel}} (m,\Phi_{0}) $ noncompact, bound states are normalizable harmonic forms, which are typically not captured by the ordinary de Rham cohomology. In our situation, the metric scale along $  C(k_i)$ is infinite, so the usual normalizability criterion must be adapted to the fiber-base decomposition. Let $ N_{\mathcal{M}} $ denote the number of bound states on $\mathcal{M}$. Instead of (\ref{eu1}), a suitable counting is 
\begin{equation}\label{N}
      N_{\mathcal{M}_{\mathrm{rel}} (m,\Phi_{0})}=\sum^{n}_{i=1}N_{ \mathcal M_{\mathrm{rel}}(m,\Phi_0,k_i)}\;\chi[C(k_i)]. 
\end{equation}      
Each $\mathcal M_{\mathrm{rel}}(m,\Phi_0,k_i)$ is a well-defined hyper-K\"ahler manifold obtained as a smooth limit of the maximally broken case \cite{1}. Concretely, let $\mathcal M_{\mathrm{rel}}(\tilde{\Phi}_0,k_i)$ be the relative moduli space for a VEV $\tilde{\Phi}_0$ that breaks $G$ to $ U(1)^{r} $ so that all components of $\mathbf{k}_{i}$ (as in (\ref{ki})) are topological. As $  \tilde{\Phi}_0 \rightarrow \Phi_0$, 
\begin{equation}
 \mathcal M_{\mathrm{rel}}(\tilde{\Phi}_0,k_i) \rightarrow  \mathcal M_{\mathrm{rel}}(m,\Phi_0,k_i),
\end{equation}
and all topological charges become holomorphic except for $m$.\footnote{In the $SU(3)$ example \ref{su31} for $m=2$, take
$\tilde{\Phi}_0=\diag(2v,-v+\varepsilon,-v-\varepsilon)$ with $0<\varepsilon<3v$.
Both $\mathcal M_{\mathrm{rel}}(\tilde{\Phi}_0,k_1)$ and $\mathcal M_{\mathrm{rel}}(2,\Phi_0,k_1)$ are the relative moduli space of two massive monopoles of identical charge $\diag(1,-1,0)$, whereas $\mathcal M_{\mathrm{rel}}(\tilde{\Phi}_0,k_2)$ contains an additional massive monopole of charge $\diag(0,1,-1)$.
As $\varepsilon\to 0$, the $\diag(0,1,-1)$ sector becomes massless and form the non-Abelian cloud in $\mathcal M_{\mathrm{rel}}(2,\Phi_0,k_2)$ \cite{1}.
Here no $\mathbf{k}_i$ is a root, so $N_{\mathcal M_{\mathrm{rel}}(2,\Phi_0)}=0$.} We may expect\footnote{When $  \tilde{\Phi}_0 \rightarrow \Phi_0$, harmonic forms on $\mathcal M_{\mathrm{rel}}(\tilde{\Phi}_0,k_i)$ descend to $\mathcal M_{\mathrm{rel}}(m,\Phi_0,k_i)$. This limiting procedure has subtleties, see \cite{1}. We will return to these issues in Example \ref{a2a}.} 
 \begin{equation}
     N_{\mathcal M_{\mathrm{rel}}(\tilde{\Phi}_0,k_i)}=N_{ \mathcal M_{\mathrm{rel}}(m,\Phi_0,k_i)}.
\end{equation}     
In the maximally broken regime, matching the monopole spectrum to the dual W-bosons requires 
\begin{equation}\label{max}
 N_{\mathcal M_{\mathrm{rel}}(\tilde{\Phi}_0,k_i)}
=
\begin{cases}
1, & \text{if }\mathbf{k}_{i} \text{ is a root of } G^{\vee},\\[2pt]
0, & \text{otherwise,}
\end{cases}
\end{equation}
which has been verified when the root contains no repeated simple roots (i.e. all coefficients are $0$ or $1$) \cite{Sen:1994, GauntlettLowe1996, LWY:SU3, LWY:ManyBPS, Gibbons:1996}. When some simple root appears with coefficient $ \geq 2 $, the corresponding harmonic form has not yet been constructed \cite{Yi:1996}. Assuming (\ref{max}), if 
\begin{equation}\label{lab1}
\{\mathbf{k}_{1},\ldots,\mathbf{k}_{n}\} \,\cap\, \Phi^{+}(G^{\vee})=\{\mathbf{k}_{i_{1}},\ldots,\mathbf{k}_{i_{N}}\},
\end{equation}
then
\begin{equation}\label{gn}
N_{\mathcal{M}_{\mathrm{rel}} (m,\Phi_{0})}
=\sum^{N}_{p=1}\chi [C(k_{i_{p}})]=\sum^{N}_{p=1}\chi [ H^{s}  / Z_{H^{s}}(\overline{\mathbf{k}}_{i_{p}})  ].
\end{equation}

\end{itemize}

When $m=1$, every solution $\mathbf{k}_{i}$ is a positive root of $G^{\vee}$, so (\ref{gn}) agrees with (\ref{wt2}). We may take (\ref{gn}) as the expression for the number of ground states on $\mathcal{M}_{\mathrm{rel}} (m,\Phi_{0})$ for all $m \in \mathbb{Z}_{>0}$.

\medskip

\subsubsection*{Dimension of the dual W-boson representation}

In the dual theory, the symmetry breaks from $G^{\vee}$ to $H^{\vee}=(H^{\vee})^{s} \times U(1)$ (up to a finite central quotient) by removing the node $\boldsymbol{\alpha}^{\vee}_{a} $. The semisimple factor $ (H^{\vee})^{s}  $ has simple roots $\{\boldsymbol{\alpha}^{\vee}_{b}\}_{b\neq a}$. The $ U(1) $ generator may be chosen as 
\begin{equation}
Y=\boldsymbol{\omega}_{a},
\end{equation}
so that for every root $\boldsymbol{\beta}$, the charge is integral, i.e. $q(\boldsymbol{\beta})=\langle  \boldsymbol{\beta},\boldsymbol{\omega}_{a}\rangle \in \mathbb{Z}$. $\Phi_{0} =v \boldsymbol{\omega}_{a} \cdot  \mathbf{T}  $. Under $(H^{\vee})^{s} \times U(1)   $, the adjoint decomposes as
\begin{equation}\label{365}
\text{Adj}_{G^{\vee}}=(\text{Adj}_{(H^{\vee})^{s}})_{q=0}\;\;\oplus\;\; (1)_{q=0}\;\;\bigoplus^{d_{a}}_{q=1}\;\; 
\left[ \left( V_{\Lambda_{q}}\right)_{q} \oplus  \left( \overline{V}_{\Lambda_{q}}\right)_{-q}\right] , 
\end{equation}
where $q$ is the $U(1)$ charge and $ V_{\Lambda_{q}}$ is the $(H^{\vee})^{s}$-representation of lowest weight $ \Lambda_{q} $. For $G^{\vee}$ with highest root $  \theta_{G^{\vee}} =\sum d_{b}\boldsymbol{\alpha}^{\vee}_{b} $, the maximal $U(1)$ charge is $d_{a}$ (the mark/Kac label of $ \boldsymbol{\alpha}^{\vee}_{a} $) \cite{Humphreys}. The W-boson mass scales as $M\propto |q| v$, and S-duality identifies the charge-$q$ W-bosons with monopoles of topological charge $m=q$.

\begin{itemize}
  \item
For $q=1$, the lowest weight $ \Lambda_{1} $ is the projection of $\boldsymbol{\alpha}^{\vee}_{a} $ to the $(H^{\vee})^{s}$ weight space:
\begin{equation}
\Lambda_{1} =\sum_{b \neq a} A_{ab}\boldsymbol{\omega}^{\vee (H^{s})}_{b}=\sum_{b \neq a} A_{ab}\boldsymbol{\omega}^{ (H^{\vee})^{s}}_{b},
\end{equation}
which is just $ \overline{\mathbf{k}}_{1} $ in (\ref{4281}) with $m=1$. From (\ref{de}), 
\begin{equation}
\dim V_{\Lambda_{1}} =\dim V_{\overline{\mathbf{k}}_{1}} = \chi [ \mathcal{M}_{\mathrm{rel}} (1,\Phi_{0})   ].
\end{equation}

  \item Generically, $\forall \; q \in \mathbb{Z}_{>0}$, define
\begin{equation}
\Phi^{+}_{q}=\{ \boldsymbol{\beta}=q\boldsymbol{\alpha}_{a}^{\vee} + \sum_{b\neq a} n_{b} \boldsymbol{\alpha}_{b}^{\vee} \,|\,\boldsymbol{\beta} \in  \Phi^{+}(G^{\vee})  \} ,
\end{equation}
the set of positive roots of $G^{\vee}$ with $U(1)$ charge $q$. Then 
\begin{equation}
\dim V_{\Lambda_{q}}=|\Phi^{+}_{q}|.
\end{equation}
Let $ \overline{\boldsymbol{\beta}} $ denote the projection of $\boldsymbol{\beta}$ to $H^{s}$. The weight set of $V_{\Lambda_{q}}$ is
\begin{equation}
\text{wt}(V_{\Lambda_{q}})=\bar{\Phi}^{+}_{q}=\{ \overline{\boldsymbol{\beta}} \,|\, \boldsymbol{\beta} \in \Phi^{+}_{q}\}=O(\overline{\boldsymbol{\beta}}_{1}) \cup \cdots \cup O(\overline{\boldsymbol{\beta}}_{N}),
\end{equation}
which splits into Weyl orbits $  O(\overline{\boldsymbol{\beta}}_{p})$. Here $ \overline{\boldsymbol{\beta}}_{p} $ is the anti-dominant representative selected by the condition 
\begin{equation}
\overline{\boldsymbol{\beta}}_{p} \cdot \boldsymbol{\alpha}_{c}=\boldsymbol{\beta}_{p}\cdot \boldsymbol{\alpha}_{c} \leq 0 ,  \qquad c \neq a.
\end{equation}
Among $\{\overline{\boldsymbol{\beta}}_{1}, \ldots, \overline{\boldsymbol{\beta}}_{N}\}$, the unique minimum in the $(H^{\vee})^{s}$ dominance order is the lowest weight $\Lambda_{q}$. For fixed $q$, the map $\boldsymbol{\beta}\rightarrow \overline{\boldsymbol{\beta}}$ is injective, so $V_{\Lambda_{q}}$ is multiplicity-free with 
\begin{equation}\label{gn1}
\dim V_{\Lambda_{q}}=\sum^{N}_{p=1}   \chi [H^{s}/Z_{H^{s}}(\overline{\boldsymbol{\beta}}_{p}) ].
\end{equation}

Comparing with (\ref{lab1}), when $m=q$, we have 
\begin{equation}
\{\boldsymbol{\beta}_{1}, \ldots, \boldsymbol{\beta}_{N}\}=\{\mathbf{k}_{i_{1}},\ldots,\mathbf{k}_{i_{N}}\}. 
\end{equation}
Therefore, by (\ref{gn}) and (\ref{gn1}), 
\begin{equation}
\dim V_{\Lambda_{q}}=N_{\mathcal{M}_{\mathrm{rel}} (q,\Phi_{0})}, \qquad \forall \; q \in \mathbb{Z}_{>0}.
\end{equation}

 \end{itemize}

\medskip

\subsubsection*{Generalization to $\pi_{2}(G/H)\cong \mathbb{Z}^{t}$}

The generalization to the $\pi_{2} (G/H) \cong \mathbb{Z}^{t}$ ($1 \leq  t \leq r$) situation is straightforward. Let $ \{\boldsymbol{\alpha}_{a}\}^{r}_{a=1} $ be the simple roots of $G$. Removing $\{\boldsymbol{\alpha}_{a_{1}},\ldots,\boldsymbol{\alpha}_{a_{t}}\}$ breaks the symmetry to $ H^{s} \times U(1)^{t}$, where the semisimple factor $ H^{s} $ has simple roots 
\begin{equation}
\{\boldsymbol{\alpha}_{b}\mid b=1,\ldots,r,\; b\notin\{a_{1},\ldots,a_{t}\}\}.
\end{equation}
Choose $ \Phi_{0}=\mathbf{h}  \cdot \mathbf{T}  $ with
\begin{equation}
\mathbf{h} =\sum^{t}_{l=1}v_{l} \boldsymbol{\omega}_{a_{l}}^{\vee} ,\qquad v_{l}  >0.
\end{equation}
The magnetic charge is $\mathbf{k} =\sum^{r}_{a=1}n_{a} \boldsymbol{\alpha}_{a}^{\vee}   $ with $n_{a}\in \mathbb{Z}_{\geq 0}$, in which 
\begin{equation}
m=(n_{a_{1}},\ldots ,n_{a_{t}})
\end{equation}
are topological charges.

For fixed $m$, if the anti-dominant condition 
\begin{equation}\label{295}
\mathbf{k}\cdot \boldsymbol{\alpha}_{c} \leq 0, \qquad \forall \;c\notin\{a_{1},\ldots,a_{t}\}
\end{equation}
admits $n$ solutions $\{\mathbf{k}_{1},\ldots,\mathbf{k}_{n}\}$, and among which, $\{\mathbf{k}_{i_{1}},\ldots,\mathbf{k}_{i_{N}}\}$ are roots of $G^{\vee}$, then 
 \begin{equation}\label{2891}
N_{\mathcal{M}_{\mathrm{rel}} (m,\Phi_{0})} =\sum^{N}_{p=1}\chi [ C(k_{i_{p}})].
 \end{equation}

In the dual theory, where $G^{\vee}$ breaks to $(H^{\vee})^{s} \times U(1)^{t}$, the $U(1)^{t}$ generator is taken as 
\begin{equation}
 Y=(\boldsymbol{\omega}_{a_{1}},\ldots ,\boldsymbol{\omega}_{a_{t}}),
\end{equation}
and a root $\boldsymbol{\beta} \in\Phi^{+}(G^{\vee})$ carries the $U(1)^{t}$ charge vector 
\begin{equation}
q(\boldsymbol{\beta})=\big(\langle\boldsymbol{\beta},\boldsymbol{\omega}_{a_{1}}\rangle,\ldots,
\langle\boldsymbol{\beta},\boldsymbol{\omega}_{a_{t}}\rangle\big).
\end{equation}
The W-boson sector $V_{\Lambda_{m}}$ of $U(1)^{t}$ charge $m$ arises from roots $ \boldsymbol{\beta} \in \Phi^{+}(G^{\vee})$ satisfying $\langle\boldsymbol{\beta},\boldsymbol{\omega}_{a_{l}}\rangle=n_{a_{l}}$ for $l=1,\ldots,t$, and therefore has
\begin{equation}\label{2892}
\dim V_{\Lambda_{m}} = \sum^{N}_{p=1}\chi [ C(k_{i_{p}})]
\end{equation}
as well.

\medskip

\subsubsection*{Extension to dyons}\label{dyon}

When the symmetry breaks from $G$ to $H^{s} \times U(1)^{t}$, the full moduli space $ \mathcal{M} (m,\Phi_{0})$ with $m=(n_{a_{1}},\ldots ,n_{a_{t}})$ has a $U(1)^{t}$ isometry, so wavefunctions on $  \mathcal{M} (m,\Phi_{0}) $ can also carry electric charges $e=(n'_{a_{1}},\ldots ,n'_{a_{t}})$. Let $N^{e}_{\mathcal{M}}$ denote the number of ground states on $\mathcal{M}$ with electric charge $e$. (\ref{N}) generalizes to
\begin{equation}\label{gnl}
      N^{e}_{\mathcal{M} (m,\Phi_{0})}=\sum^{n}_{i=1}N^{e}_{ \mathcal M(m,\Phi_0,k_i)}\;\chi[C(k_i)].
\end{equation}

In the maximally broken phase where $\tilde{\Phi}_0$ breaks $G$ to $U(1)^{r}$, consider a wavefunction on $ \mathcal M (\tilde{\Phi}_0,k_i) $ carrying electric charge $ \mathbf{k}'_{i} $. For a given topological charge
\begin{equation}
\mathbf{k}_{i} =\sum^{r}_{a=1}n_{a} \boldsymbol{\alpha}_{a}^{\vee} =m_{0}\sum^{r}_{a=1}l_{a} \boldsymbol{\alpha}^{\vee}_{a}, \qquad \gcd (l_{1},\ldots,l_{r})=1,
\end{equation}
the $\tfrac12$ BPS condition requires the electric charge to be parallel \cite{WittenOlive:1978}:
\begin{equation}
\mathbf{k}'_{i} =\sum^{r}_{a=1}n'_{a} \boldsymbol{\alpha}_{a}=e_{0}\sum^{r}_{a=1}l_{a} \boldsymbol{\alpha}_{a}. 
\end{equation}
Moreover, if $\gcd(e_{0},m_{0})=1$ and $\sum^{r}_{a=1}l_{a} \boldsymbol{\alpha}^{\vee}_{a}$ is a root of $G^{\vee}$, then $ N^{\mathbf{k}'_{i} }_{\mathcal M (\tilde{\Phi}_0,k_i)} =1$; otherwise, $ N^{\mathbf{k}'_{i} }_{\mathcal M (\tilde{\Phi}_0,k_i)}  =0$ \cite{Sen:1994, GauntlettLowe1996, LWY:SU3, LWY:ManyBPS}.

As $\tilde{\Phi}_0\to\Phi_{0}$, the symmetry is enhanced to $H^{s}\times U(1)^{t}$, and $\mathcal M(\tilde{\Phi}_0,k_i)$ degenerates to $\mathcal M(m,\Phi_{0},k_i)$. When $ N^{\mathbf{k}'_{i} }_{\mathcal M (\tilde{\Phi}_0,k_i)} =1$, the surviving topological and electric charges are
\begin{equation}\label{me}
m=m_{0}(l_{a_{1}},\ldots ,l_{a_{t}}),\qquad e=e_{0}(l_{a_{1}},\ldots ,l_{a_{t}}),\qquad\gcd(e_{0},m_{0})=1.  
\end{equation}
Thus, on $ \mathcal M (m, \Phi_{0},k_i) $ with electric charge $e$, if $(e,m)$ satisfies (\ref{me}) and $\mathbf{k}_{i}/m_{0}$ is a root of $G^{\vee}$, then $N^{e}_{\mathcal M (m, \Phi_{0},k_i) }=1$; otherwise, $N^{e}_{\mathcal M (m, \Phi_{0},k_i) } =0$. Since $C(k_i)=C(k_i/m_{0})$, combining (\ref{2891}), (\ref{2892}) and (\ref{gnl}) yields
\begin{equation}
 N^{e}_{\mathcal{M} (m,\Phi_{0})}=  N_{\mathcal{M}_{\mathrm{rel}} (l,\Phi_{0})}=\dim V_{\Lambda_{l}},
\end{equation}
where $l=(l_{a_{1}},\ldots ,l_{a_{t}})$ is determined by (\ref{me}), and $ V_{\Lambda_{l}}  $ is the W-boson sector of $U(1)^{t}$ charge $l$.

\medskip

\medskip

\medskip

\paragraph{Example: $G=G_{2}$}
\leavevmode\par\vspace{0.8em}

\begin{center}
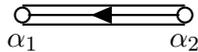

\tikzset{
  big arrow/.style={
    postaction={
      decorate,
      decoration={
        markings,
        mark=at position 0.55 with {\arrow[scale=1.0]{triangle 45 reversed}}
      }
    }
  }
}

\begin{tikzpicture}[scale=1.2, thick, >=stealth]

  \node[circle,draw,fill=white,inner sep=2pt,label=below:{\(\alpha_1\)}] (a1) at (0,0) {};
  \node[circle,draw,fill=white,inner sep=2pt,label=below:{\(\alpha_2\)}] (a2) at (1.8,0) {};

  \coordinate (top1) at ([yshift=3pt]a1);
  \coordinate (top2) at ([yshift=3pt]a2);
  \coordinate (mid1) at (a1);
  \coordinate (mid2) at (a2);
  \coordinate (bot1) at ([yshift=-3pt]a1);
  \coordinate (bot2) at ([yshift=-3pt]a2);

  \draw (top1) -- (top2);
  \draw[big arrow, shorten >=0.75mm, shorten <=0.75mm] (mid1) -- (mid2); 
  \draw (bot1) -- (bot2);

\end{tikzpicture}
\captionof{figure}{Dynkin diagram of \(G_{2}\) with simple roots \(\alpha_1,\alpha_2\).}
\label{fig9}
\end{center}

\medskip

Simple roots of $G_{2}$ are 
\begin{equation}
\boldsymbol{\alpha}_{1}=\boldsymbol{e}_{1}-\boldsymbol{e}_{2},\qquad \boldsymbol{\alpha}_{2}=-2\boldsymbol{e}_{1}+\boldsymbol{e}_{2}+\boldsymbol{e}_{3}. 
\end{equation}
Removing the short root $ \boldsymbol{\alpha}_{1} $ breaks $G_{2}$ to $H^{s}\times U(1)=SU(2) \times U(1)$.

For $m=1$, there are two solutions to (\ref{asa}):
\begin{equation}
\mathbf{k}_{1} = \boldsymbol{\alpha}^{\vee}_{1},\qquad   \mathbf{k}_{2} = \boldsymbol{\alpha}^{\vee}_{1}+\boldsymbol{\alpha}^{\vee}_{2},
\end{equation}
with $C(k_{1})=C(k_{2})  \cong \mathbb{CP}^{1}$. Hence
\begin{equation}
 \mathcal{M}_{\mathrm{rel}} (1,\Phi_{0})=\mathcal{M}_{\mathrm{rel},1}(1,\Phi_{0})\cup  \mathcal{M}_{\mathrm{rel},2}(1,\Phi_{0}) 
\end{equation}
 with
\begin{equation}
 \chi [ \mathcal{M}_{\mathrm{rel}} (1,\Phi_{0})]=\chi [ C(k_{1})]+\chi [C(k_{2})]=4 .
\end{equation}
On the other hand, 
\begin{equation}
\overline{\mathbf{k}}_{1} = A_{12}\boldsymbol{\omega}_{2}^{\vee (H^{s})} = -3 \boldsymbol{\omega}_{2}^{\vee (H^{s})}
\end{equation}
is the lowest weight of the $ 4$-representation of $SU(2) $. $ \dim V_{\overline{\mathbf{k}}_{1} }=  \chi [ \mathcal{M}_{\mathrm{rel}} (1,\Phi_{0})]$.

For $m=2$, there are $4$ solutions to (\ref{asa}):
\begin{equation}
\mathbf{k}_{i} =2 \boldsymbol{\alpha}^{\vee}_{1}+(i-1)\boldsymbol{\alpha}^{\vee}_{2} ,\qquad    i=1,2,3,4,
\end{equation}
among which $ \mathbf{k}_{4} $ is a root of $(G_{2})^{\vee}=G_{2}$. Therefore, 
\begin{equation}
 \mathcal{M}_{\mathrm{rel}} (2,\Phi_{0})=\mathcal{M}_{\mathrm{rel},1}(2,\Phi_{0})\cup \cdots \cup \mathcal{M}_{\mathrm{rel},4}(2,\Phi_{0}) 
\end{equation}
with  
\begin{equation}
N_{ \mathcal{M}_{\mathrm{rel}} (2,\Phi_{0})} =\chi [ C(k_{4})]=1 .
\end{equation}
The projection of $\mathbf{k}_{4} $ to $SU(2)$ is 
\begin{equation}
\overline{\mathbf{k}}_{4} =0,
\end{equation}
which is the lowest weight of the $ 1$-representation of $SU(2) $. $ \dim V_{0 }=  N_{ \mathcal{M}_{\mathrm{rel}} (2,\Phi_{0})} $.

In the dual theory, with $ \boldsymbol{\alpha}^{\vee}_{1} $ removed, $G_{2}$ breaks to $SU(2) \times U(1)$. W-bosons transform as 
\begin{equation}
(\mathbf{4})_{+1}\ \oplus\ (\mathbf{4})_{-1}\ \oplus\ (\mathbf{1})_{+2}\ \oplus\ (\mathbf{1})_{-2}.
\end{equation}

\medskip

\medskip

\medskip

\paragraph{Example: $SU(N+2)\rightarrow   SU(N)  \times U(1)^{2} $ }\label{a2a}
\leavevmode\par\vspace{0.8em}

Simple roots of $SU(N+2)$ are 
\begin{equation}
\boldsymbol{\alpha}_{1}= \boldsymbol{e}_{1}-\boldsymbol{e}_{2}, \qquad \boldsymbol{\alpha}_{2}=\boldsymbol{e}_{2}-\boldsymbol{e}_{3},\qquad \ldots \qquad \boldsymbol{\alpha}_{N+1}=\boldsymbol{e}_{N+1}-\boldsymbol{e}_{N+2}.
\end{equation}
Removing $\boldsymbol{\alpha}_{1}$ and $\boldsymbol{\alpha}_{N+1}$ breaks $SU(N+2)$ to $  SU(N)  \times U(1)^{2} $. The topological charge is $m=(n_{1},n_{N+1})$. When $m=(1,1)$, solutions to (\ref{295}) are 
\begin{equation}
 \mathbf{k}_{1}  = \boldsymbol{\alpha}^{\vee}_{1}+ \boldsymbol{\alpha}^{\vee}_{N+1} ,\qquad \mathbf{k}_{2}  = \boldsymbol{\alpha}^{\vee}_{1}+\cdots +  \boldsymbol{\alpha}^{\vee}_{N+1},
\end{equation}
with 
\begin{equation}
C(k_{1})\cong \frac{SU(N)}{S[U(1)\times U(N{-}2)\times U(1)]},
\qquad
C(k_{2})\ \text{a point}.
\end{equation}
Hence,

\begin{equation}
\mathcal{M}_{\mathrm{rel}}\big((1,1),\Phi_{0}\big)
=\mathcal{M}_{\mathrm{rel},1}\big((1,1),\Phi_{0}\big)\ \cup\ 
\mathcal{M}_{\mathrm{rel},2}\big((1,1),\Phi_{0}\big),
\end{equation}
with
\begin{equation}
\dim \mathcal{M}_{\mathrm{rel},1}\big((1,1),\Phi_{0}\big)= 4N-2,
\qquad 
\dim \mathcal{M}_{\mathrm{rel},2}\big((1,1),\Phi_{0}\big)=4N.
\end{equation}
Only $\mathbf{k}_{2} $ is a coroot of $SU(N+2)$, so 
 \begin{equation}
N_{\mathcal{M}_{\mathrm{rel}} ((1,1),\Phi_{0})} =\chi [ C(k_{2})]=1 .
 \end{equation}
In the dual theory, under $SU(N)  \times U(1)^{2} $, the massive gauge bosons transform as
\begin{equation}
(\mathbf{N})_{(-1,0)}\ \oplus\ (\overline{\mathbf{N}})_{(1,0)}  \ \oplus\  (\mathbf{N})_{(0,1)}\ \oplus\ (\overline{\mathbf{N}})_{(0,-1)}  \ \oplus\  (\mathbf{1})_{(1,1)}\ \oplus\ (\mathbf{1})_{(-1,-1)} .
\end{equation}

The bound state on $\mathcal{M}_{\mathrm{rel}} ((1,1),\Phi_{0})$ comes from the stratum 
\begin{equation}
\mathcal{M}_{\mathrm{rel},2}((1,1),\Phi_{0})= \mathcal M_{\mathrm{rel}}((1,1),\Phi_0,k_2)  ,
\end{equation}
which could be realized as a smooth limit of $   \mathcal M_{\mathrm{rel}}(\tilde{\Phi}_0,k_2)   $ with maximal breaking. As $\tilde{\Phi}_0 \rightarrow \Phi_0$, the unique normalizable harmonic form $ \omega(\tilde{\Phi}_0,k_2)$ on $\mathcal M_{\mathrm{rel}}(\tilde{\Phi}_0,k_2)  $ descends to a harmonic form $  \omega ((1,1),\tilde{\Phi}_0 \to \Phi_0,k_2) $ on $  \mathcal M_{\mathrm{rel}}((1,1),\Phi_0,k_2)  $. In \cite{1} this descent was analyzed and it was shown that $  \omega ((1,1),\tilde{\Phi}_0 \to\Phi_0,k_2) $ is self-dual but not normalizable and not $SU(N)$-invariant. Each $  \tilde{\Phi}_0$ selects a Cartan subgroup $T \subset SU(N)$, and the descended form is only $T$-invariant. There are infinitely many self-dual harmonic forms on $\mathcal{M}_{\mathrm{rel},2}$ related by the unbroken $SU(N)$.

With the stratification in mind, we may impose one further constraint. As the cloud size $a \rightarrow \infty$, the asymptotic ``angular'' slice of $\mathcal{M}_{\mathrm{rel},2}$ has dimension $4N-1$. To match the $4N-2$ dimensions of $\mathcal{M}_{\mathrm{rel},1}$, the angular slice must be modded out by the angle-dependent circle
\begin{equation}
U(1)_{\hat\Omega}=g(\hat\Omega)\,U(1)\,g(\hat\Omega)^{-1}\subset SU(N).
\end{equation}
Accordingly, harmonic forms are required to be $U(1)_{\hat\Omega}$-invariant on the asymptotic slice, which actually enforces the full $SU(N)$-invariance. An $SU(N)$-invariant representative is obtained by superposing the descended forms $ \omega ((1,1),\tilde{\Phi}_0 \to\Phi_0,k_2)$. It is unique and self-dual, but may still be non-normalizable.

\section{Quantum mechanics on $ \mathcal{M} (m,\Phi_{0}) $}\label{qm}

In the maximally broken case, the low-energy dynamics of BPS monopoles in $\mathcal N=4$ SYM is governed by an $\mathcal N=8$ supersymmetric quantum mechanics on a hyper-K\"ahler moduli space \cite{Gauntlett, Blum, GauntlettKimLeeYi2001}. In the non-maximally broken phase, we have shown that when the relevant moduli space is $\mathcal{M}(m,\Phi_{0})$ introduced in Section~\ref{str}, harmonic forms on the relative moduli space match the dual W-boson spectrum. For completeness, in this section we derive the effective monopole action on $\mathcal{M}(m,\Phi_{0})$ via a collective-coordinate expansion. The resulting Lagrangian has the same structure as in the maximally broken case, but the supersymmetry is reduced and some metric components diverge. Nevertheless, in the relative sector, the Hamiltonian remains one half of the Hodge Laplacian, and the ground state wave functions are still harmonic forms. When counting harmonic forms, the divergent part of the metric, which lies entirely along the orbits $C(k_{i})$, causes no difficulty, since only the topology of $C(k_{i})$ enters. We also construct operators on the monopole Hilbert space that serve as building blocks for the magnetic gauge transformation generators in Section~\ref{st}.

The $\mathcal{N}=4$ SYM, when viewed as the $4d$ reduction of $10d$ $\mathcal{N}=1$ SYM, contains a gauge field $A_M$ ($M=0,\dots,9$) and a Majorana-Weyl gaugino $\Psi$, with SUSY variations
\begin{equation}
\delta A_M = -i\,\bar{\varepsilon}\,\gamma_M\,\Psi,
\qquad
\delta \Psi = \tfrac12\,F_{MN}\,\gamma^{MN}\,\varepsilon,
\end{equation}
where $ \varepsilon $ is a constant Majorana-Weyl spinor.

Static BPS monopoles satisfy the Bogomolny equations
\begin{equation}\label{B}
F_{0i} = 0,\qquad
B_i = \tfrac12\,\varepsilon_{ijk} F_{jk} = D_i \Phi,
\quad i=1,2,3,
\end{equation}
where the scalar $ \Phi $ is identified as $\Phi \equiv A_4$ and the rest scalars $  \Phi_{I} \equiv A_{I}$ $( I=5,\ldots,9 )$ are set to zero. Such configurations preserve $8$ of $16$ supercharges obeying 
\begin{equation}
\gamma^{1234} \varepsilon = \varepsilon.
\end{equation}

Solutions of (\ref{B}) are specified by $\Phi_{0}$ in the boundary condition (\ref{frf}) and the topological charge $m=(m_{1},\ldots,m_{t})$. For the fixed $\Phi_{0}$ and $m$, under the framing (\ref{fr}), all solutions (modulo local gauge transformations) form a connected space $\mathcal{M}(m, \Phi_{0})$. As reviewed in Section \ref{str}, 
\begin{equation}
\mathcal{M} (m,\Phi_{0}) =\bigcup_{G_{0} \in \mathcal{K}}  \mathcal{M} (m,\Phi_{0}, G_{0}).
\end{equation}
Each component $\mathcal{M} (m,\Phi_{0}, G_{0})$ is hyper-K\"ahler, whereas the union $ \mathcal{M} (m,\Phi_{0}) $ need not be. Only in the maximal symmetry breaking case is $ G_{0} $ unique, so that $\mathcal{M} (m,\Phi_{0})= \mathcal{M} (m,\Phi_{0}, G_{0}) $.

The monopole dynamics is governed by an effective action obtained from a collective-coordinate expansion of the $\mathcal{N}=4$ SYM Lagrangian (\ref{b1}). For simplicity, we first consider the bosonic Lagrangian, where $\Phi^{I}$ can also be consistently truncated, 
\begin{equation}\label{510}
\mathcal{L}_{\rm bos}
= \tfrac{1}{2}\,\tr \left( E_i E_i    +D_0\Phi\,D_0\Phi \right) 
 -\tfrac{1}{2}\,\tr  \left( B_i B_i+D_i\Phi\,D_i\Phi 
  \right).
\end{equation}
Let 
\begin{equation}\label{5.6}
A_i(x,t) = A^{\rm BPS}_i\big(x^{i}; z^{m}(t)\big), \qquad
\Phi(x,t) = \Phi^{\rm BPS}\big(x^{i}; z^{m}(t)\big),
\end{equation}
where $ A_{i}^{\rm BPS}$ and $  \Phi^{\rm BPS}$ are BPS solutions labeled by the moduli $z^{m}$. The full field evolution is thereby reduced to a trajectory $z^{m}(t)$ on the monopole moduli space.

A conceptual point is what should be regarded as the moduli space. If we take a fixed hyper-K\"ahler component $\mathcal{M} (m,\Phi_{0}, G_{0})$ as the moduli space, all constructions are well-defined, but the resulting effective action only captures the dynamics with the fixed $ G_{0}$. For instance, under $ SU(N+1)\rightarrow SU(N)\times U(1) $, one has $\mathcal{M} (1,\Phi_{0}, G_{0})=  \mathbb{R}^{3} \times \mathbb{S}^{1} $, irrespective of the rank $N$. This is classically acceptable, since $ G_{0}$ may be conserved along the classical motion, but at the quantum level, superpositions of $G_{0}$ are allowed, and the ground states can mix different components. From a symmetry perspective, the moduli space should be invariant under the unbroken $H$. However, $ \mathcal{M} (m,\Phi_{0}, G_{0}) $ is invariant under $H$ only when $ G_{0} $ lies in the centralizer of $H$. Since configurations with distinct $ G_{0} $ can be continuously deformed into each other, we will allow the trajectory $z^{m}(t)$ to pass between components $\mathcal{M} (m,\Phi_{0}, G_{0})  $ and take the full $\mathcal{M} (m,\Phi_{0})   $ as the moduli space. The price to pay is that certain subtleties will arise.

From (\ref{5.6}), 
\begin{equation}\label{513}
E_i = F_{0i} =  \dot z^{m}\,\partial_m A_i - D_i A_0 , \qquad D_0\Phi =  \dot z^{m}\,\partial_m \Phi -i [A_0,\Phi] .
\end{equation}
Substituting (\ref{513}) into (\ref{510}) and integrating over space, 
\begin{equation}\label{513q}
\int d^3x\, \mathcal{L}_{\rm bos}
=\tfrac{1}{2}\,g^{\rm raw}_{mn}\,\dot z^{m}\dot z^{n}
 + \dot z^{m}\,\int d^3x\,\tr \big(A_0\,\mathcal{G}_m\big)
 - \tfrac{1}{2}\,\int d^3x\,\tr \big(A_0\,K\,A_0\big)
 - 2 \pi \tr  ( \Phi_{0}G_{0}
).
\end{equation}
Here
\begin{align}
g^{\rm raw}_{mn}
&= \int d^3x\,\tr\!\big(\partial_m A_i\,\partial_n A_i + \partial_m \Phi\,\partial_n \Phi\big),\\
\mathcal{G}_m &= D_i\,\partial_m A_i - i[\Phi,\partial_m \Phi], \qquad
K = D_i D_i - [\Phi,[\Phi,\,\cdot\,]],
\end{align}
and the term $ - 2 \pi\tr  ( \Phi_{0}G_{0})  $ is a constant in $\mathcal{M} (m,\Phi_{0})  $. Integrating out $A_0$ is equivalent to solving 
\begin{equation}\label{515}
K\,A_0 - \dot z^{m}\,\mathcal{G}_m = 0
\;\;\Rightarrow\;\;
A_0 = \,\dot z^{m}\,\alpha_m,\qquad   \alpha_m =K^{-1} \mathcal{G}_m.
\end{equation}
Inserting (\ref{515}) back into (\ref{513}) yields 
\begin{equation}\label{516}
E_i = \dot z^{m} \delta_m A_i,\qquad D_0\Phi 
=\dot z^{m} \delta_m \Phi,
\end{equation}
with
\begin{equation}\label{ba1}
\delta_m A_i = \partial_m A_i - D_i \alpha_m,\qquad
\delta_m \Phi = \partial_m \Phi -i [\alpha_m,\Phi]
\end{equation}
satisfying the background gauge condition 
\begin{equation}
D_i \delta_m A_i -i [\Phi,\delta_m \Phi]
= \mathcal{G}_m - K\alpha_m
= 0.
\end{equation}
Using (\ref{516}) in (\ref{510}), we finally obtain
\begin{equation}
\int d^3x\, \mathcal{L}_{\rm bos}
=\tfrac{1}{2}\,g_{mn}\,\dot z^{m}\dot z^{n}
 - 2 \pi \tr  ( \Phi_{0}G_{0}
),
\end{equation}
where the moduli-space metric is
\begin{equation}\label{ba11}
g_{mn}(z)  
    = \int  d^3x\;
    \tr\!\left( \delta_m A_i\,\delta_n A_i
          + \delta_m \Phi\,\delta_n \Phi \right)= \int  d^3x\;
    \tr\!\left( \delta_m A_{\hat{\mu}}\,\delta_n A_{\hat{\mu}}
   \right),\qquad \hat{\mu}=1,2,3,4.
\end{equation}

For an arbitrary BPS trajectory $\big(  A^{\rm BPS}_i (x^{i}; z^{m}(t)) ,\Phi^{\rm BPS}(x^{i}; z^{m}(t))\big) $, integrating out $A_{0}$ automatically projects the moduli derivatives onto $(\delta_m A_i , \delta_m \Phi)$ in background gauge. When $(\partial_m A_i , \partial_m \Phi)$ corresponds to a variation of $G_{0}$ along its adjoint orbit, the same procedure yields the background gauge representative $(\delta_m A_i , \delta_m \Phi)$. For such variations, $(\partial_m A_i , \partial_m \Phi) $ does not decay sufficiently fast at spatial infinity, so the raw metric components $g^{\rm raw}_{mn}$ diverge. As shown in \cite{Abouelsaood1983a}, no local gauge transformation can bring these modes into the background gauge and imposing background gauge projects them out.

In the eight-parameter family of $SO(5)\rightarrow U(2)$ solutions reviewed in Appendix~\ref{reg}, the set of magnetic charges is $\mathcal{K}=C(k_{1})\cup\{k_{2}\}$. For finite $a$, $G_{0}=k_{2}$ and all fluctuation modes are well-defined. As $a\to\infty$, $G_{0}$ jumps to $C(k_{1})$, and the solution becomes gauge equivalent to the standard $SU(2)$-embedding with $G_{0}\in C(k_{1})$, for which the non-Abelian directions are ill-defined. We may take $a$ as a regulator for these directions, with the physical result obtained by taking $a\to\infty$. The non-Abelian orientation modes in background gauge are given by (\ref{2.233}), with $(\delta' A_i,\delta'\Phi)\to 0$ as $a\to\infty$, which is consistent with \cite{Abouelsaood1983a}. Meanwhile, the associated metric components $g_{mn}\to\infty$, so $(\delta' A_i,\delta'\Phi)$ are not identically zero. Intuitively, since non-normalizable modes are delocalized, their pointwise amplitudes may tend to zero while the spatial norm diverges. In light of the examples in Appendix~\ref{reg}, we may expect that for such non-normalizable modes, $(\delta_m A_i,\delta_m\Phi)$ tends to zero locally while the corresponding metric components diverge.

In \cite{Weinberg2, Weinberg1}, the monopole moduli (fluctuations satisfying both the linearised BPS equations and the background gauge condition) were counted via index theory. In non-maximal symmetry breaking case, the index receives a contribution from the continuum, and the final result matches $ \dim  \mathcal{M} (m,\Phi_{0}) $, with the continuum piece accounting for the non-normalizable modes.

From the bosonic zero modes $\delta_m A_{\hat{\mu}}$, the fermionic zero modes are constructed as $ \chi_m
  = \gamma^{\hat{\mu}}\,\delta_m A_{\hat{\mu}}
   \,\varepsilon_0 $. In a static monopole background, the Dirac equation reduces to
\begin{equation}
\slashed{\mathcal{D}}\,\Psi
  = \gamma^i D_i \Psi -i \gamma^4[\Phi,\Psi]  = \gamma^{\hat{\mu}} D_{\hat{\mu}} \Psi= 0.
\end{equation}
$ \slashed{\mathcal{D}}\,\chi_m=0 $ if and only if 
\begin{equation}\label{1234}
\gamma^{1234} \varepsilon_0 =- \varepsilon_0.
\end{equation}
$ \chi_m $ is a c-number Majorana-Weyl spinor of the same chirality as $ \Psi $, so $\varepsilon_0 $ must be a Majorana-Weyl of opposite chirality. (\ref{1234}) has $8$ independent solutions $ \varepsilon^{\mathcal{A}}_0 $ ($ \mathcal{A}=1,\ldots,8 $) normalized by
\begin{equation}\label{419a}
( \varepsilon^{\mathcal{A}}_0 )^{\dagger} \varepsilon^{\mathcal{B}}_0 =\delta^{\mathcal{A}\mathcal{B}}. 
\end{equation}
Accordingly, for each $m$, there are $ 8 $ fermionic zero modes:
\begin{equation}\label{ch}
\chi^{\mathcal{A}}_m
  = \gamma^{\hat{\mu}}\,\delta_m A_{\hat{\mu}}
   \,\varepsilon^{\mathcal{A}}_0.
\end{equation}
Let 
\begin{equation}
I^{(1)}=
\begin{pmatrix}
0&0&0&1\\
0&0&1&0\\
0&-1&0&0\\
-1&0&0&0
\end{pmatrix},\qquad
 I^{(2)}=
\begin{pmatrix}
0&0&1&0\\
0&0&0&-1\\
-1&0&0&0\\
0&1&0&0
\end{pmatrix},\qquad
 I^{(3)}=
\begin{pmatrix}
0&1&0&0\\
-1&0&0&0\\
0&0&0&1\\
0&0&-1&0
\end{pmatrix}
\end{equation}
be the self-dual 't Hooft matrices satisfying 
\begin{equation}
I^{(i)}\, I^{(j)} \;=\; -\,\delta_{ij}\,\mathbf 1 \;+\; \varepsilon^{ijk}\, I^{(k)}.
\end{equation} 
Then 
\begin{equation}\label{chia}
\frac14\,(I^{(i)})_{\hat\mu\hat\nu}\,\gamma^{\hat\mu\hat\nu}\,\varepsilon^{\mathcal{A}}_{0}
= -(J^{(i)})^{\mathcal{A}}{}_{\mathcal{B}}\,\varepsilon^{\mathcal{B}}_{0},
\qquad \gamma^{\hat\mu\hat\nu}\,\varepsilon^{\mathcal{A}}_{0}
= -(I^{(i)})^{\hat\mu\hat\nu}(J^{(i)})^{\mathcal{A}}{}_{\mathcal{B}}\,\varepsilon^{\mathcal{B}}_{0},
\end{equation}
where $J^{(i)}$ generate an $SU(2)$ action on the real 8-space spanned by $\{\varepsilon^{\mathcal{A}}_{0}  \}$ with
\begin{equation}
 J^{(i)}\, J^{(j)} \;=\; -\,\delta_{ij}\,\mathbf 1 \;+\; \varepsilon^{ijk}\, J^{(k)}.
\end{equation}
We may choose a basis in which $ J^{(i)} $ is block diagonal,
\begin{equation}\label{ji}
J^{(i)}=
\left(\begin{matrix} I^{(i)}&0_{4}\\ 0_{4}&I^{(i)}\end{matrix}\right). 
\end{equation}

Up to this point, no hyper-K\"ahler input has been used. If the moduli space is hyper-K\"ahler, the endomorphisms \cite{HarveyStrominger1993}
\begin{equation}\label{k1}
(K^{(i)})_{m}{}^{n}=g^{ln}\int d^{3}x \;(I^{(i)})^{\hat{\mu}\hat{\nu}}\tr(\delta_{m}A_{\hat{\mu}}\,\delta_{l}A_{\hat{\nu}}  )
\end{equation}
form the quaternionic triplet of complex structures,
\begin{equation}
K^{(i)}\, K^{(j)} \;=\; -\,\delta_{ij}\,\mathbf 1 \;+\; \varepsilon^{ijk}\, K_{(k)}.
\end{equation}
The action of $ K^{(i)} $ on bosonic zero modes is
\begin{equation}
(K^{(i)})_{m}{}^{n}\delta_{n}A_{\hat{\mu}}=-(I^{(i)})_{\hat{\mu}}{}^{\hat \nu}\delta_{m}A_{\hat{\nu}}.
\end{equation}
Using (\ref{ch}) and (\ref{chia}), the action of $K^{(i)}$ on fermionic zero modes is
\begin{equation}\label{mo}
(K^{(i)})_{m}{}^{n}\chi^{\mathcal{A}}_{n}=-(J^{(i)})^{\mathcal{A}}{}_{\mathcal{B}}\chi^{\mathcal{B}}_{m}.
\end{equation}
So $\chi^{\mathcal{A}}_{m}$ are not linear independent. For example,
\begin{align}
\chi^{2}_{m} &= -(K^{(3)})_{m}{}^{n}\chi^{1}_{n}, &
\chi^{3}_{m} &= -(K^{(2)})_{m}{}^{n}\chi^{1}_{n}, &
\chi^{4}_{m} &= -(K^{(1)})_{m}{}^{n}\chi^{1}_{n}, \nonumber\\
\chi^{6}_{m} &= -(K^{(3)})_{m}{}^{n}\chi^{5}_{n}, &
\chi^{7}_{m} &= -(K^{(2)})_{m}{}^{n}\chi^{5}_{n}, &
\chi^{8}_{m} &= -(K^{(1)})_{m}{}^{n}\chi^{5}_{n}.
\end{align}
Choose two arbitrary unit spinors $\{\varepsilon^{\alpha}_0\,|\, \alpha=1,2\}$ from the first and second blocks (e.g. $\varepsilon^{1}_{0}$ and $\varepsilon^{5}_{0}$), respectively. An independent basis of fermionic zero modes is then
\begin{equation}\label{431}
\{\chi^{\alpha}_{m}= \gamma^{\hat{\mu}}\,\delta_m A_{\hat{\mu}}
   \,\varepsilon^{\alpha}_0\,|\, \alpha=1,2\,;\,m=1,\ldots,\dim  \mathcal{M} (m,\Phi_{0}) \}.
\end{equation}

In our setting, $\mathcal{M}(m,\Phi_{0})$ is not necessarily hyper-K\"ahler, so the above argument does not apply. Nevertheless, the same index computation in \cite{Weinberg2, Weinberg1} that yields $\dim \mathcal{M}(m,\Phi_{0})$ for the bosonic moduli also fixes the number of fermionic zero modes \cite{Index, Callias1978} (including the non-normalizable modes). In $\mathcal{N}=4$ SYM, this number is always $2\,\dim \mathcal{M}(m,\Phi_{0})$, even when $\dim \mathcal{M}(m,\Phi_{0})$ is not divisible by four. We will therefore continue to use (\ref{431}) as our fermionic zero-mode basis.

From (\ref{chia}), 
\begin{equation}\label{435}
(\varepsilon^{\alpha}_{0})^{\dagger}\gamma^{\hat\mu\hat\nu}\,\varepsilon^{\beta}_{0}=0,\qquad \alpha,\beta=1,2,
\end{equation}
so 
\begin{equation}\label{436}
\int d^{3}x\; (\chi^{\alpha}_{m})^{\dagger} \chi^{\beta}_{n}
=\delta^{\alpha\beta}  g_{mn}.
\end{equation}
On a BPS background, the fermion fields are expanded as
\begin{equation}\label{ps}
\Psi(x,t)=\chi^{\alpha}_{m}(x;z(t))\,\xi_{\alpha}^{m}(t),
\qquad
\bar\Psi(x,t) =\bar\chi^{\alpha}_m(x;z(t))\,  \xi_{\alpha}^{m}(t)  ,
\end{equation}
where $\xi_{\alpha}^{m}(t)$ are real two-component Grassmann variables. Substituting (\ref{5.6}) and (\ref{ps}) into the Lagrangian (\ref{b1}) and integrating out $A_{0}$ and $\Phi^{I}$, we obtain the monopole effective Lagrangian
\begin{equation}\label{L}
\mathcal{L}_{\rm eff}
= \tfrac{1}{2}\,g_{mn} \dot z^{m}\dot z^{n} +\tfrac{i}{2} \,g_{mn} \,\xi^{m}_{\alpha}\nabla_{t}\xi_{\alpha}^{n}
 +\tfrac{1}{8}\, R_{mnpq}  \, (\xi^{m}_{\alpha}\xi^{n}_{\alpha})(\xi^{p}_{\beta}\xi^{q}_{\beta})  - 2 \pi \tr  \left(  \Phi_{0}G_{0}\right) ,
\end{equation}
where $\nabla_{t}\xi_{\alpha}^{n}=\dot\xi_{\alpha}^{n}
+  \Gamma^{n}{}_{pl} \,  \dot z^{p} \xi_{\alpha}^{l}$. (\ref{L}) is the standard Lagrangian of a one-dimensional supersymmetric sigma model with at least two real supercharges \cite{Bagger:1984ge}. For completeness, we derive (\ref{L}) via a collective coordinate expansion in Appendix \ref{EFF}, without assuming the hyper-K\"ahler structure\footnote{As in $\mathcal{N}=2$ gauge theories \cite{Gauntlett}, integrating out the additional five scalars can contribute to the effective action even when they do not acquire vacuum expectation values. A collective-coordinate expansion in $\mathcal{N}=4$ SYM that includes these scalars was carried out in \cite{WeinbergYi2007}; however, that derivation uses the hyper-K\"ahler structure of the moduli space, which is not available in our case.}.

Introducing complex Grassmann variables
\begin{equation}
\lambda^{m}=\tfrac{1}{\sqrt{2}}(\xi_{1}^{m}+i\xi_{2}^{m}),\qquad \lambda^{\dagger           
  m} =\tfrac{1}{\sqrt{2}}(\xi_{1}^{m}-i\xi_{2}^{m}),
\end{equation}
(\ref{L}) becomes 
\begin{equation}
\mathcal{L}_{\rm eff}
= \tfrac{1}{2}\,g_{mn} \dot z^{m}\dot z^{n} +i \,g_{mn} \,\lambda^{\dagger m} \nabla_{t} \lambda^{n}
 -\tfrac{1}{4}\, R_{mnpq}  \, \lambda^{m}\lambda^{n} \lambda^{\dagger   p}\lambda^{\dagger  q} - 2 \pi \tr  \left(  \Phi_{0}G_{0}\right) .
\end{equation}
The full monopole moduli space factorizes as 
\begin{equation}
 \mathcal{M} (m,\Phi_{0}) =  \mathcal{M}_{\mathrm{rel}} (m,\Phi_{0}) \times  \mathbb{R}^{3}\times \mathbb{S}^{1} .
\end{equation}
Let $ z^{\mu} $ $(1 \leq \mu \leq D=\dim \mathcal{M}_{\mathrm{rel}}  (m,\Phi_{0})  )$ and $ z^{A} $ $(D+1 \leq A \leq D+4=\dim  \mathcal{M} (m,\Phi_{0})  )$ be coordinates on $ \mathcal{M}_{\mathrm{rel}} (m,\Phi_{0})$ and $\mathbb{R}^{3}\times \mathbb{S}^{1}$, respectively, with $\lambda^{\mu}$ and $\lambda^{A}$ the associated fermionic collective coordinates. Then $ \mathcal{L}_{\rm eff}=\mathcal{L}_{0}+\mathcal{L}_{\rm int} $, where 
\begin{equation}
\mathcal{L}_{0}
= \tfrac{1}{2}\dot z^{A}\dot z^{A} +i \lambda^{\dagger A} \dot \lambda^{A}
- 2 \pi \tr  \left(  \Phi_{0}G_{0}\right),\qquad \mathcal{L}_{\rm int}
= \tfrac{1}{2}g_{\mu\nu} \dot z^{\mu}\dot z^{\nu} +ig_{\mu\nu}\lambda^{\dagger \mu} \nabla_{t} \lambda^{\nu}
 -\tfrac{1}{4} R_{\mu\nu\rho\sigma}   \lambda^{\mu}\lambda^{\nu} \lambda^{\dagger   \rho}\lambda^{\dagger  \sigma} .
\end{equation}

Upon quantization, the canonical (anti)commutation relations are
\begin{equation}
[\hat z^{A},\hat p_{B}]=i\delta^{A}_{B},\qquad \{\hat \lambda^{A}, \hat \lambda^{\dagger B}\}=\delta^{AB},\qquad [\hat z^{\mu},\hat p_{\nu}]=i\delta^{\mu}_{\nu},\qquad \{\hat \lambda^{\mu}, \hat \lambda^{\dagger \nu}\}=g^{\mu\nu}.
\end{equation}
The free Hamiltonian is 
\begin{equation}
 \hat  H_{0}=\tfrac{1}{2}\hat p_{A}\hat p_{A},
\end{equation}
where $\hat p_{A}$ is the momentum on $\mathbb{R}^{3}\times \mathbb{S}^{1}$. The action of $\hat\lambda^{A}, \hat\lambda^{\dagger A}$ generates a $16$-fold degeneracy. The interacting Hamiltonian is
\begin{equation}
\hat   H_{\rm int}=\tfrac{1}{2}\{ \hat   Q , \hat  Q^{\dagger}  \} ,
\end{equation}
with supercharges
\begin{equation}
\hat Q=\hat \lambda^{\dagger m}\hat \Pi_{m},\qquad \hat Q^{\dagger}= \hat \lambda^{ m}\hat \Pi_{m},
\end{equation}
where $\hat  \Pi_{m} =\hat p_{m}-i \Gamma_{mnl} \hat \lambda^{\dagger n} \hat  \lambda^{l}$ is the covariant momentum. The Hilbert space may be realized as the space of differential forms on $\mathcal{M}_{\mathrm{rel}}(m,\Phi_{0})$ with \cite{Witten:1982}
\begin{equation}
\hat \lambda^{\dagger m} \longleftrightarrow dz^{m} \wedge,\qquad  \hat \lambda_{m}=g_{mn}\hat \lambda^{n} \longleftrightarrow \iota_{\partial_{m}},\qquad   \hat Q \longleftrightarrow -i \,d  ,\qquad   \hat Q^{\dagger} \longleftrightarrow i \,d^{\dagger},
\end{equation}
and 
\begin{equation}
\hat   H_{\rm int} \longleftrightarrow \tfrac{1}{2} (dd^{\dagger}+ d^{\dagger}d) ,
\end{equation}
so the ground states are harmonic forms.

When $ \mathcal{M}_{\mathrm{rel}} (m,\Phi_{0}) $ is K\"ahler, it carries a complex structure $ K^{(3)} $ and admits an additional pair of supercharges
\begin{equation}
\hat Q^{(3)}=\hat \lambda^{\dagger m} (K^{(3)})_{m}{}^{n}  \hat \Pi_{n},\qquad \hat Q^{(3)\dagger}= \hat \lambda^{ m}(K^{(3)})_{m}{}^{n} \hat \Pi_{n}. 
\end{equation}
Acting on scalars,
\begin{equation}
[\hat Q , \hat z^{m}]=-i \lambda^{\dagger m},\qquad [\hat Q^{(3)} , \hat z^{m}]=-i \hat \lambda^{\dagger n}(K^{(3)})_{n}{}^{m}.
\end{equation}
The supersymmetry algebra is given by
\begin{equation}
\{\hat Q,\hat Q^{\dagger}\}= \{\hat Q^{(3)},\hat Q^{(3)\dagger}\}=2 \hat H ,\qquad\{\hat Q,\hat Q\}=\{\hat Q^{\dagger},\hat Q^{\dagger}\}=\{\hat Q^{(3)},\hat Q^{(3)}\}=\{\hat Q^{(3)\dagger},\hat Q^{(3)\dagger}\}=0.
\end{equation}

In the following, we construct states and operators on the relative moduli space \(\mathcal{M}_{\mathrm{rel}} (1,\Phi_{0})\equiv M\) that will be used in Section \ref{st}. From operators \(\hat z^{\mu}\) and \(\hat \lambda_{\mu}\), the position eigenstate \(\vert z \rangle\) and the fermionic vacuum \(\vert \Omega \rangle\) are defined by
\begin{equation}
\hat z^{\mu}\vert z \rangle=z^{\mu} \vert z \rangle ,\qquad \hat \lambda_{\mu} \vert \Omega \rangle =0. 
\end{equation}
For a \(p\)-form 
\begin{equation}
f=\frac{1}{p!}f_{\mu_{1}\cdots \mu_{p}}\,dz^{\mu_{1}}\wedge \cdots \wedge dz^{\mu_{p}},
\end{equation}
the associated state is
\begin{equation}
\vert f\rangle=\frac{1}{p!}\int_{M} d^{D}z \sqrt{g}\;f_{\mu_{1}\cdots \mu_{p}}(z)\, \vert z\rangle \,\hat \lambda^{\dagger \mu_{1}} \cdots\hat \lambda^{\dagger \mu_{p}}\vert \Omega\rangle.
\end{equation}
In particular, for the \(0\)-form \( f=1 \),
\begin{equation}\label{1s}
\vert 1\rangle=\int_{M} d^{D}z \sqrt{g}\; \vert z\rangle \vert \Omega\rangle.
\end{equation}
The operator realization of \(f\) is 
\begin{equation}
O[f]=\frac{1}{p!}f_{\mu_{1}\cdots \mu_{p}}(\hat z)\,\hat \lambda^{\dagger \mu_{1}} \cdots\hat \lambda^{\dagger \mu_{p}},\qquad 
O^{\dagger}[f]=\frac{1}{p!}f^{\mu_{1}\cdots \mu_{p}}(\hat z)\,\hat\lambda_{\mu_{p}} \cdots\hat\lambda_{\mu_{1}} ,
\end{equation}
with
\begin{equation}
O[f]O[f']=O[ f \wedge f']
\end{equation}
and \(  \vert f \rangle =O[f] \vert 1\rangle\). When acting on states,
\begin{equation}
O[f]\ket{f'}=\ket{f\wedge f'} ,\qquad 
O^{\dagger}[f]\ket{f'}=
\begin{cases}
0, & p>p',\\[4pt]
\ket{\,f\;\lrcorner\; f'\,}, & p\le p',
\end{cases}
\end{equation}
where \(p'=\deg f'\) and 
\begin{equation}
(f \mathbin{\lrcorner} f')_{\mu_1\cdots \mu_{p'-p}}
 =\frac{1}{p!}f^{\nu_1\cdots \nu_p}
f'_{\nu_1\cdots \nu_p\,\mu_1\cdots \mu_{p'-p}}.
\end{equation}
So \( O[f] \) acts by wedging with \( f \), while \(O^{\dagger}[f]\) acts by contraction with \(f\).

Ground states are harmonic forms. Let \( \{f_k\}_{k=1}^K \) be an orthogonal basis of harmonic forms on $\mathcal{M}_{\mathrm{rel}}(1,\Phi_{0})$, all of which have even degree. For two basis elements \( f_{k} \) and \(f_{k'}\) of the same degree, 
\begin{equation}
 f_{k}\wedge (\ast f_{k'} )=(f_{k}, f_{k'}) \,\varepsilon ,
\end{equation}
where $( \cdot ,\cdot)$ is the pointwise inner product on $p$-forms and 
\begin{equation}
\varepsilon=\frac{1}{D!}\varepsilon_{\mu_{1}\cdots \mu_{D}}dz^{\mu_{1}}\wedge \cdots \wedge dz^{\mu_{D}}
\end{equation}
is the Riemannian volume form. If all harmonic forms are parallel with $\nabla f_k=0$,\footnote{In the single-stratum case, $\mathcal M_{\rm rel}(1,\Phi_0)=C(k_1)$ is a compact symmetric space, where all harmonic forms are parallel. But harmonic forms on $\mathcal M_{\rm rel}(m,\Phi_0)$ may not be parallel in general.} $ (f_{k}, f_{k'})$ is a constant and orthogonality yields 
\begin{equation}\label{ft1}
(f_{k}, f_{k'})=\delta_{k,k'} \, c^{2}_{k},
\end{equation}
where \(  c_{k} \) is the constant pointwise norm of \(f_{k}\). For each $f_{k}$, a normalized state is defined as $ \vert k \rangle \equiv \tfrac{1}{c_{k}}\vert f_{k} \rangle $.

In this representation, the matrix units can be constructed as 
\begin{equation}\label{e}
E_{m,n}=\frac{1}{c_{m}c_{n}}O[f_{m}] O^{\dagger}[\varepsilon]O[\varepsilon]O^{\dagger}[f_{n}]
=\frac{1}{c_{m}c_{n}}O[f_{m}] O^{\dagger}[\varepsilon]O[\ast f_{n}],\qquad m,n=1,2,\ldots,K,
\end{equation}
satisfying \( E^{\dagger}_{m,n}=E_{n,m} \) and
\begin{align}
E_{m,n}E_{k,l} 
&= \frac{1}{c_{m}c_{n}c_{k}c_{l}}O[f_{m}] O^{\dagger}[\varepsilon]O[(\ast f_{n}) \wedge f_{k}] O^{\dagger}[\varepsilon]O[\ast f_{l}]\nonumber  \\
&=  \frac{\delta_{n,k}}{c_{m}c_{l}}O[f_{m}] O^{\dagger}[\varepsilon]O[\varepsilon] O^{\dagger}[\varepsilon]O[\ast f_{l}]\nonumber \\
&= \frac{\delta_{n,k}}{c_{m}c_{l}}O[f_{m}] O^{\dagger}[\varepsilon]O[\ast f_{l}]
=\delta_{n,k} E_{m,l},
\end{align}
where we used (\ref{ft1}). Acting on the basis \( \vert k \rangle  \),
\begin{align}\label{e1}
E_{m,n}\vert k \rangle 
&= \frac{1}{c_{m}c_{n}c_{k}}O[f_{m}] O^{\dagger}[\varepsilon]O[\ast f_{n}]O[f_{k}] \vert 1\rangle
=\frac{1}{c_{m}c_{n}c_{k}}O[f_{m}] O^{\dagger}[\varepsilon]O[(\ast f_{n}) \wedge f_{k}] \vert 1\rangle\nonumber \\
&= \frac{\delta_{n,k}}{c_{m}}O[f_{m}] O^{\dagger}[\varepsilon]O[\varepsilon] \vert 1\rangle
=\frac{\delta_{n,k}}{c_{m}}O[f_{m}]  \vert 1\rangle
=\delta_{n,k} \vert m\rangle .
\end{align}
In the next section, we will use $E_{m,n}$ to construct the generators of the magnetic gauge transformations.

\section{Generators of magnetic gauge transformations and the $H^{s}\times (H^{\vee})^{s}$ symmetry}\label{st}

As shown in Subsection~\ref{sur}, for $\mathcal{N}=4$ SYM with simple gauge group $G$, when $ G $ is broken to $H$ with semisimple factor $H^{s}$, the degeneracy of ground states on $ \mathcal{M}_{\mathrm{rel}} (q,\Phi_{0})   $ equals the dimension of the charge-$q$ W-boson multiplet in the dual theory. To promote this numerical match into a one-to-one correspondence, and moreover, to construct operators generating the $(H^{\vee})^{s}$-action, we still need to assign an $(H^{\vee})^{s}$-weight to each harmonic form.

Let us recall some standard facts about equal-rank homogeneous spaces \cite{BottTu, Humphreys, Helgason:1978, FultonHarris:1991}. Let $\mathcal{G}$ be a compact, connected Lie group and $ \mathcal{H}\subset \mathcal{G} $ a connected, closed subgroup of the same rank.  Then $H^{\ast}(\mathcal{G}/\mathcal{H}; \mathbb{R})$ is concentrated in even degrees and 
\begin{equation}
\sum_{k}\dim H^{k}(\mathcal{G}/\mathcal{H}; \mathbb{R})= \chi[\mathcal{G}/\mathcal{H}]=|W_{\mathcal{G}}/W_{\mathcal{H}}|= |W_\mathcal{G}| / |W_\mathcal{H}| ,
\end{equation}
where $W_{\mathcal{G}}$ and $W_{\mathcal{H}}$ are the Weyl groups of $\mathcal{G}$ and $\mathcal{H}$. If moreover $\mathcal H=Z_{\mathcal G}(S)$ is the centralizer of a torus $S\subset \mathcal G$, then $\mathcal G/\mathcal H$ is a generalized flag manifold. In this case the Schubert classes form a basis of $H^\ast(\mathcal G/\mathcal H; \mathbb{R})$ indexed by the minimal-length representatives $w$ of $ W_{\mathcal G}/W_{\mathcal H}$. Let $\omega_w$ denote the $\mathcal G$-invariant harmonic representative of the Schubert class indexed by $w$. Then
\begin{equation}\label{co}
w\in \ \{\text{minimal rep of $W_{\mathcal{G}}/W_{\mathcal{H}}$}\}
\;\;\longleftrightarrow\;\;
\omega_{w}\in \{\text{Schubert harmonic basis on } \mathcal{G}/\mathcal{H}\}.
\end{equation}
The degree of $\omega_{w}$ is 
\begin{equation}
\deg (\omega_{w})= 2\ell (w),
\end{equation}
with $\ell (w) $ the Coxeter length. In particular, the identity coset has minimal representative $e$, corresponding to the $0$-form $\omega_{e}=1$.

If $ \mathbf{k} $ is a coweight of $\mathcal{G}$ and $\mathcal{H}=Z_{\mathcal{G}}(\mathbf{k})$ is its centralizer, then $W_{\mathcal{H}}=  W_{Z_{\mathcal{G}}(\mathbf{k})}$ is the stabilizer of $ \mathbf{k} $ in $ W_{\mathcal{G}} $. There is a natural bijection
\begin{equation}
w\in \ \{\text{minimal rep of $W_{\mathcal{G}}/W_{\mathcal{H}}$}\}\;\;\longleftrightarrow \;\; \text{wt}_{w} =w( \mathbf{k}) \in O( \mathbf{k} ),
\end{equation}
where $O( \mathbf{k} )$ is the Weyl orbit of $ \mathbf{k}$ with $|O( \mathbf{k} )|=|W_\mathcal{G}| / |W_{Z_{\mathcal{G}}(\mathbf{k})}| $. Correspondingly, the harmonic form $\omega_{w}$ is assigned the $ \mathcal{G}^{\vee} $-weight $w( \mathbf{k}) $. The $0$-form $\omega_{e} =1 $ carries weight $ \mathbf{k}$. For convenience, we take $\mathbf{k}$ to be anti-dominant, so that $\omega_{e}=1$ is the lowest-weight state. If we instead choose $\mathbf{k}'=u(\mathbf{k})$ for some $u\in W_{\mathcal G}$, then the weight assigned to $\omega_{w}$ is
\begin{equation}
w(\mathbf{k}')=w\!\big(u(\mathbf{k})\big)=(wu)(\mathbf{k}),
\end{equation}
i.e. the entire weight assignment is shifted by the Weyl transformation $u$.

In our setting, $ \mathcal{M}_{\mathrm{rel}} (q,\Phi_{0}) $ is a stratified manifold with 
\begin{equation}
\mathcal{M}_{\mathrm{rel}} (q,\Phi_{0})=\mathcal{M}_{\mathrm{rel},1}(q,\Phi_{0})     \cup  \mathcal{M}_{\mathrm{rel},2}(q,\Phi_{0})     \cup \cdots   \cup   \mathcal{M}_{\mathrm{rel},n}(q,\Phi_{0}),
\end{equation}
where each $ \mathcal{M}_{\mathrm{rel},i}(q,\Phi_{0})   $ is a fiber bundle over $H^{s}/Z_{H^{s}}(\overline{\mathbf{k}}_{i})$ with fiber $\mathcal M_{\mathrm{rel}}(q,\Phi_0,k_i)$. Among $\{\mathbf{k}_{1},\ldots,\mathbf{k}_{n}\}$, if 
\begin{equation}
\{\mathbf{k}_{i_{1}},\ldots,\mathbf{k}_{i_{N}}\}
\end{equation}
are roots of $G^{\vee}$, then each $\mathcal{M}_{\mathrm{rel},i_{p}}(q,\Phi_{0})   $ $(1 \leq p \leq N)$ contributes $ \chi [H^{s}/Z_{H^{s}}(\overline{\mathbf{k}}_{i_{p}})] $ harmonic forms indexed by $W_{H^{s}}/W_{Z_{H^{s}}(\overline{\mathbf{k}}_{i_{p}})}$. Altogether, a basis of harmonic forms on $\mathcal{M}_{\mathrm{rel}} (q,\Phi_{0})$ may be written as 
\begin{equation}
\Bigl\{
\;(\Omega_{p}\wedge\omega_{w_{p}})_{\text{harm}}
\;\Big|\;
w_{p}\in W_{H^{s}}/W_{Z_{H^{s}}(\overline{\mathbf{k}}_{i_{p}})},
\; p=1,\ldots,N
\Bigr\},
\end{equation}
realizing the $(H^{\vee})^{s}$-representation $V_{\Lambda_{q}}$. Here $ (\Omega_{p}\wedge\omega_{w_{p}})_{\text{harm}} $ denotes the harmonic representative of the cohomology class $[\Omega_{p}\wedge\omega_{w_{p}}]$.

Weights of $V_{\Lambda_{q}}$ split as 
\begin{equation}
\text{wt}(V_{\Lambda_{q}})=O(\overline{\mathbf{k}}_{i_{1}}) \cup \cdots \cup O(\overline{\mathbf{k}}_{i_{N}}), 
\end{equation}
which are assigned to the basis elements by
\begin{equation}
\text{wt}((\Omega_{p}\wedge\omega_{w_{p}})_{\text{harm}})=\text{wt}(\Omega_{p}\wedge\omega_{w_{p}})=\text{wt}(\omega_{w_{p}})=w_{p}(\overline{\mathbf{k}}_{i_{p}} )   \in O(\overline{\mathbf{k}}_{i_{p}}) .
\end{equation}
The degree is $ \deg ( \Omega_{p}\wedge\omega_{w_{p}})=\deg ( \Omega_{p})+2\ell (w_{p}) $ with
\begin{equation}\label{4100}
\deg(\Omega_{p}) =
\begin{cases}
\dim \mathcal{M}_{\mathrm{rel}}(q,\Phi_{0},k_{i}), & \text{if $ q $ is a unite vector},\\[6pt]
\frac{1}{2}\,\dim \mathcal{M}_{\mathrm{rel}}(q,\Phi_{0},k_{i}), & \text{otherwise}.
\end{cases}
\end{equation}
If $q$ is a unit vector, $\mathcal{M}_{\mathrm{rel}}(q,\Phi_{0})$ is compact, and the fiber contributes its top-degree cohomology class (the Thom class). Otherwise, $\mathcal{M}_{\mathrm{rel}}(q,\Phi_{0})$ is noncompact and the bound state on $\mathcal{M}_{\mathrm{rel}}(q,\Phi_{0},k_{i})$ is represented by the (anti-)self-dual harmonic form in middle degree \cite{Sen:1994}.

The simplest case is when $\overline{\mathbf{k}}_{1}$ is minuscule in $H^{s}$. Then $n=1$ and 
\begin{equation}
\mathcal{M}_{\mathrm{rel}} (1,\Phi_{0})=C(k_{1})=H^{s}  / Z_{H^{s}}(\overline{\mathbf{k}}_{1})   .
\end{equation}
Harmonic forms are $ \{\omega_{w}\,|\, w \in W_{H^{s}}/W_{Z_{H^{s}}(\overline{\mathbf{k}}_{1})}\} $ with $\deg (\omega_{w})= 2\ell (w)  $, carrying weights $ \text{wt}_{w} =w(\overline{\mathbf{k}}_{1})$ in the $(H^{\vee})^{s}$-representation $ V_{\overline{\mathbf{k}}_{1}} $ with lowest weight $\overline{\mathbf{k}}_{1}$.

In what follows, for examples in Subsections \ref{si} and \ref{mut}, we explicitly compute the weights carried by the harmonic forms in the associated $(H^{\vee})^{s}$-representation. In particular, in Subsection \ref{si}, where $\mathcal{M}_{\mathrm{rel}}(1,\Phi_{0})=C(k_{1})$ is a compact Hermitian symmetric space equipped with its standard invariant K\"ahler metric, all harmonic forms are parallel. Moreover, the Schubert harmonic representatives $\{\omega_{w}\}$ are orthogonal \cite{KuennemannTamvakis:2002}. Using this orthogonal parallel harmonic basis, we construct the magnetic generators of the $(H^{\vee})^{s}$-action on the monopole Hilbert space via the matrix units $E_{m,n}$ in \eqref{e}. When $\mathcal{M}_{\mathrm{rel}}(1,\Phi_{0})$ is multi-stratified, we do not have a general proof for the parallelness of harmonic forms, so \eqref{e} may not apply.

\subsection{$(H^{\vee})^{s}$-weight decomposition of harmonic forms}

\medskip
\begin{enumerate}[label=\textbf{(\arabic*)}, leftmargin=2em, itemsep=1.2em]

\item \textit{$SU(N+1)\rightarrow U(N) \Longleftrightarrow  SU(N+1)/\mathbb{Z}_{N+1}\rightarrow U(N)$}

\medskip

In Figure \ref{fig1}, removing $\boldsymbol{\alpha}_{N}$ breaks $SU(N+1)$ to $U(N)$. $ H^{s} =SU(N) $. $ \mathbf{k}_{1}  =   \boldsymbol{\alpha}_{N}^{\vee} = \boldsymbol{e}_{N}-\boldsymbol{e}_{N+1}   $. The $H^{s}$-projection is
\begin{equation}
 \overline{\mathbf{k}}_{1} =\boldsymbol{e}_{N}  -\frac{1}{N}\sum^{N}_{a=1} \boldsymbol{e}_{a} ,
\end{equation}
which is minuscule in $SU(N)$ and is the lowest weight of $\mathbf{N}$-representation of $SU(N)$. The centralizer is $ Z_{H^{s}}(\overline{\mathbf{k}}_{1})=S[U(N-1)\times U(1) ]$.

The relative moduli space is 
\begin{equation}
\mathcal{M}_{\mathrm{rel}} (1,\Phi_{0})\cong SU(N)/S[U(N-1)\times U(1) ]\cong \mathbb{CP}^{N-1},\qquad \dim_{\mathbb{R}} \mathcal{M}_{\mathrm{rel}} (1,\Phi_{0})=2N-2
\end{equation}
with Euler characteristic
\begin{equation}
 \chi[ \mathcal{M}_{\mathrm{rel}} (1,\Phi_{0})]=\frac{|W_{SU(N)}|}{|W_{S[U(N-1)\times U(1)]}|}=N .
\end{equation}

Let $\omega$ be the K\"ahler form on $\mathbb{CP}^{N-1}$. A Schubert basis of harmonic forms is
\begin{equation}
f_{k}=\omega^{k-1},\qquad
\deg(f_{k})=2k-2,\qquad
1\le k\le N .
\end{equation}
The corresponding minimal-length representatives in $W_{H^{s}}/W_{Z_{H^{s}}(\overline{\mathbf{k}}_{1})}$ may be chosen as
\begin{equation}
w_{k}=s_{N-k+1}s_{N-k+2}\cdots s_{N-1},\qquad\ell(w_{k})= k-1  ,\qquad 1\leq k\leq N,
\end{equation}
where $s_{k}$ denotes the simple reflection associated with the simple root $\boldsymbol{\alpha}_{k}$. The resulting $SU(N)$-weights are
\begin{equation}
\text{wt}_{k}=w_{k}( \overline{\mathbf{k}}_{1})= \boldsymbol{e}_{N+1-k} -\frac{1}{N}\sum^{N}_{a=1} \boldsymbol{e}_{a}.
\end{equation}

The dual group $SU(N)^{\vee}\cong SU(N)/\mathbb{Z}_{N}$ is generated by operators $ E_{m,n} $ constructed in (\ref{e}). In particular, the simple root generators are
\begin{equation}
E_{\boldsymbol{\alpha}_{i}}=E_{N+1-i,N-i},\qquad E_{-\boldsymbol{\alpha}_{i}}=E^{\dagger}_{\boldsymbol{\alpha}_{i}}=E_{N-i,N+1-i},\qquad i=1,\ldots,N-1.
\end{equation}

\item \textit{$USp(2N+2) \rightarrow U(N+1)    \Longleftrightarrow  SO(2N+3)\rightarrow U(N+1)$}

\medskip

In Figure \ref{fig4}, removing $\boldsymbol{\alpha}_{N+1}$ breaks $USp(2N+2)$ to $  U(N+1)$ with the semisimple part $ H^{s} =SU(N+1) $. $ \mathbf{k}_{1}  = \boldsymbol{\alpha}_{N+1}^{\vee} = \boldsymbol{e}_{N+1} $. 
\begin{equation}
\overline{\mathbf{k}}_{1} =\boldsymbol{e}_{N+1} -\frac{1}{N+1}\sum^{N+1}_{a=1} \boldsymbol{e}_{a}
\end{equation}
is minuscule in $SU(N+1)$ and is the lowest weight of $(\mathbf{N{+}1})$-representation of $SU(N+1)$. 

The remaining discussion parallels the $SU(N{+}1)\to U(N)$ case.

\item \textit{$SO(2N+3)\rightarrow SO(2N+1) \times U(1)\Longleftrightarrow USp(2N+2)\rightarrow USp(2N) \times U(1)$}

\medskip

In Figure \ref{fig3}, removing $\boldsymbol{\alpha}_{1}$ breaks $SO(2N+3)$ to $SO(2N+1) \times U(1)$. $ H^{s} =SO(2N+1) $. $  \mathbf{k}_{1}  = \boldsymbol{\alpha}_{1}^{\vee} = \boldsymbol{e}_{1}-\boldsymbol{e}_{2}   $. The $H^{s}$-projection is
\begin{equation}
 \overline{\mathbf{k}}_{1} =-\boldsymbol{e}_{2}  ,
\end{equation}
which is a minuscule coweight of $SO(2N+1)$, and is the lowest weight of the $\mathbf{2N}$-representation in $(H^{\vee})^{s} =USp(2N)$.

The relative moduli space is 
\begin{equation}
\mathcal{M}_{\mathrm{rel}} (1,\Phi_{0})\cong  SO(2N+1)/[SO(2N-1) \times U(1)],\qquad \dim_{\mathbb{R}} \mathcal{M}_{\mathrm{rel}} (1,\Phi_{0})=4N -2,
\end{equation}
with the Euler characteristic
\begin{equation}
 \chi[ \mathcal{M}_{\mathrm{rel}} (1,\Phi_{0})]=\frac{|W_{ SO(2N+1)}|}{|W_{SO(2N-1) \times U(1)}|}=2N .
\end{equation}
After the relabeling $i \rightarrow i-1$, simple roots of $USp(2N)$ are 
\begin{equation}
\boldsymbol{\alpha}_{1}= \boldsymbol{e}_{1}-\boldsymbol{e}_{2}  ,\qquad \boldsymbol{\alpha}_{2}= \boldsymbol{e}_{2}-\boldsymbol{e}_{3}  ,\qquad \ldots \qquad \boldsymbol{\alpha}_{N-1}= \boldsymbol{e}_{N-1}-\boldsymbol{e}_{N}  ,\qquad 
\boldsymbol{\alpha}_{N}=2 \boldsymbol{e}_{N},
\end{equation}
and the lowest weight of the $\mathbf{2N}$-representation is
\begin{equation}
 \overline{\mathbf{k}}_{1}=- \boldsymbol{e}_{1} .
\end{equation}

A Schubert basis of harmonic forms is
\begin{equation}
\{1,\omega,\ldots,\omega^{2N-1}\},
\end{equation}
with $ \omega $ the K\"ahler form. The corresponding minimal-length coset representatives in $ W_{H^{s}}/W_{Z_{H^{s}}(\overline{\mathbf{k}}_{1})}  $ are  
\begin{equation}
\{w_{-1},  w_{-2} ,\ldots , w_{-N},  w_{+N}   ,  \ldots  ,  w_{+2}   ,  w_{+1}   \},
\end{equation}
with
\begin{align}
w_{-k}  &=s_{k-1}s_{k-2}\cdots s_{1},   &\;  &\;\;\;\;\ell(w_{-k})= k-1, \\
w_{+k} &=  s_{k}s_{k+1}\cdots s_{N-1} s_{N}s_{N-1}\cdots s_{2}   s_{1}  ,   &  \; &\;\;\;\;\ell(w_{+k})= 2N-k.
\end{align}
Since
\begin{equation}
w_{-k}(\overline{\mathbf{k}}_{1})=- \boldsymbol{e}_{k}, \qquad w_{+k}(\overline{\mathbf{k}}_{1})= \boldsymbol{e}_{k},
\end{equation}
the weights carried by harmonic forms in the $ USp(2N) $ fundamental representation are read off as 
\begin{equation}
\{-\boldsymbol{e}_{1} ,-\boldsymbol{e}_{2} ,\ldots,-\boldsymbol{e}_{N} ,\boldsymbol{e}_{N} ,\boldsymbol{e}_{N-1} ,\ldots,\boldsymbol{e}_{1}\}.
\end{equation}

Let $f_{m}=\omega^{m-1}$, $ 1 \leq m \leq 2N $. The simple root generators of $USp(2N)$ are 
\begin{equation}
E_{\boldsymbol{\alpha}_{i}}=E_{i+1,i}-E_{2N+1-i,2N-i},\qquad i=1,\ldots,N-1,\qquad E_{\boldsymbol{\alpha}_{N}}=E_{N+1,N}
\end{equation}
with $E_{m,n}$ given by (\ref{e}).

\item \textit{$SO(2N+2)\rightarrow U(N+1) \Longleftrightarrow  SO(2N+2)\rightarrow U(N+1) $}

\medskip

In Figure \ref{fig2}, removing $\boldsymbol{\alpha}_{N+1}$ breaks $SO(2N+2)$ to $U(N+1)$. $ H^{s} =SU(N+1) $. $  \mathbf{k}_{1}  = \boldsymbol{\alpha}_{N+1}^{\vee} = \boldsymbol{e}_{N}+\boldsymbol{e}_{N+1} $. The $H^{s}$-projection is
\begin{equation}
\overline{\mathbf{k}}_{1}  =\boldsymbol{e}_{N}+\boldsymbol{e}_{N+1} -\frac{2}{N+1}\sum^{N+1}_{a=1} \boldsymbol{e}_{a} ,
\end{equation}
which is minuscule in $SU(N+1)$, and is the lowest weight of the $2$-index antisymmetric representation $ \Lambda^{2}\,\mathbf{N{+}1} $.

Simple roots of \( SU(N+1) \) are
\begin{equation}
\boldsymbol{\alpha}_{i} = \boldsymbol{e}_{i} - \boldsymbol{e}_{i+1}, \qquad i = 1, \ldots, N. 
\end{equation}
The relative moduli space is 
\begin{equation}
\mathcal{M}_{\mathrm{rel}} (1,\Phi_{0}) \cong SU(N+1)/S[U(N-1) \times U(2)  ]\cong \mathrm{Gr}(2, N+1),\;\; \dim_{\mathbb{R}} \mathcal{M}_{\mathrm{rel}} (1,\Phi_{0}) = 4N - 4,
\end{equation}
with Euler characteristic
\begin{equation}
\chi\left[ \mathcal{M}_{\mathrm{rel}} (1,\Phi_{0}) \right] = \frac{|W_{SU(N+1)}|}{|W_{S[U(N-1) \times U(2)  ]}|} =\tfrac{1}{2} N(N+1).
\end{equation}

The cohomology ring of $M=\mathrm{Gr}(2,N{+}1)$ is generated by the special Schubert classes \(\sigma_{1}\in H^{2}(M)\) and \(\sigma_{2}\in H^{4}(M)\). Let \(\omega\) and \(\Omega\) be the unique harmonic representatives of \(\sigma_{1}\) and \(\sigma_{2}\), respectively.\footnote{\(\omega\) is the K\"ahler form. When \(N=2\), \(\Omega\propto\omega^{2}\). When \(N\ge 3\), \(\Omega=a\,\omega^{2}+\psi\) for some constant $a$. Here \(\psi\) is the primitive harmonic \(4\)-form, unique up to scale. } Then the Schubert harmonic basis is $\{f_{ij} \; |\; 1\le i<j\le N+1\}$ with \cite{Kostant:1963, Fulton:YoungTableaux}
\begin{equation}\label{fij}
 f_{ij}
=
\sum_{m=0}^{[\frac{j-i-1}{2}]}
(-1)^m\binom{j-i-1-m}{m}\;
\omega^{\,j-i-1-2m}\wedge \Omega^{\,i-1+m}.
\end{equation}

Minimal-length representatives of \( W_{H^{s}}/W_{Z_{H^{s}}(\overline{\mathbf{k}}_{1})}  \) can be chosen as
\begin{equation} 
w_{i,j} = (s_{N+2-i} \cdots s_{N-1}s_{N})(s_{N+2-j} \cdots s_{N-2}s_{N-1}), \qquad \ell(w_{i,j}) = i + j - 3 = \tfrac{1}{2} \deg(f_{ij}).
\end{equation}
Under the correspondence \( f_{ij} \longleftrightarrow w_{i,j} \), the weights carried by \( f_{ij} \) in $  \Lambda^{2}\,\mathbf{N{+}1}  $ are
\begin{equation}
\text{wt}_{ij} =w_{i,j} (\overline{\mathbf{k}}_{1})=\boldsymbol{e}_{N+2-i} + \boldsymbol{e}_{N+2-j}-\frac{2}{N+1} \sum^{N+1}_{k=1} \boldsymbol{e}_{k}.
\end{equation}

The simple root generators of \( SU(N+1) \) acting on the basis \( \{f_{ij}\} \) are realized as
\begin{equation}
E_{\boldsymbol{\alpha}_{k}} = \sum^{N-k}_{i=1} E_{i\;(N+2-k)\,,\, i\;(N+1-k)} + \sum^{N+1}_{j=N+3-k} E_{(N+2-k)\;j\,,\, (N+1-k)\;j}, 
\end{equation}
where \( E_{ij,kl} \) denotes the matrix unit \(E_{m,n}\) defined in~(\ref{e}), with the single-index basis \( f_{m} \) replaced by the double-index basis \( f_{ij} \).

\item \textit{$SO(2N+2)\rightarrow SO(2N) \times U(1) \Longleftrightarrow  SO(2N+2)\rightarrow SO(2N) \times U(1)$}

\medskip

In Figure \ref{fig2}, removing $\boldsymbol{\alpha}_{1}$ breaks $SO(2N+2)$ to $SO(2N) \times U(1)$. $ H^{s} =SO(2N)$. $  \mathbf{k}_{1}  = \boldsymbol{\alpha}_{1}^{\vee} = \boldsymbol{e}_{1}-\boldsymbol{e}_{2} $. 
\begin{equation}
 \overline{\mathbf{k}}_{1}  =-\boldsymbol{e}_{2} 
\end{equation}
is minuscule in $SO(2N)$ and is the lowest weight of the $\mathbf{2N}$-representation of $SO(2N)$.

With a relabeling $i \rightarrow i-1$, simple roots of $SO(2N)$ are 
\begin{equation}
\boldsymbol{\alpha}_{1}= \boldsymbol{e}_{1}-\boldsymbol{e}_{2}  ,\qquad \boldsymbol{\alpha}_{2}= \boldsymbol{e}_{2}-\boldsymbol{e}_{3}  ,\qquad \ldots \qquad \boldsymbol{\alpha}_{N-1}= \boldsymbol{e}_{N-1}-\boldsymbol{e}_{N}  ,\qquad \boldsymbol{\alpha}_{N}=\boldsymbol{e}_{N-1}+\boldsymbol{e}_{N},
\end{equation}
and the lowest weight of the $\mathbf{2N}$-representation becomes
\begin{equation}
 \overline{\mathbf{k}}_{1}= -\boldsymbol{e}_{1} .
\end{equation}
The relative moduli space is 
\begin{equation}
\mathcal{M}_{\mathrm{rel}} (1,\Phi_{0})\cong SO(2N)/[SO(2N-2) \times U(1) ], \qquad\dim_{\mathbb{R}} \mathcal{M}_{\mathrm{rel}} (1,\Phi_{0})=4N-4,
\end{equation}
with Euler characteristic 
\begin{equation}
 \chi[ \mathcal{M}_{\mathrm{rel}} (1,\Phi_{0})]=\frac{|W_{ SO(2N)}|}{|W_{SO(2N-2) \times U(1)}|}=2N .
\end{equation}

The Schubert harmonic basis may be chosen as
\begin{equation}
\{1,\omega,\ldots,\omega^{N-2},\omega^{N-1}+\psi,\omega^{N-1}-\psi,\omega^{N},\ldots,\omega^{2N-2}\}
=\{f_m\}_{m=1}^{2N},
\end{equation}
where $\omega$ is the K\"ahler form and $\psi$ is the unique primitive middle-degree harmonic form satisfying $\omega\wedge\psi=0$, normalized so that $\omega^{N-1}\pm\psi$ are two middle-degree Schubert harmonic forms. The associated minimal-length representatives in $W_{H^{s}}/W_{Z_{H^{s}}(\overline{\mathbf{k}}_{1})}  $ are
\begin{equation} 
\{w_{-1},  w_{-2} , \ldots,  w_{-N},  w_{+N},  w_{+(N-1)} , \ldots,  w_{+1} \},
\end{equation}
with
\begin{align}
w_{-k}  &=s_{k-1}s_{k-2}\cdots s_{1},   &\;  &\;\;\;\;\ell(w_{-k})= k-1, \\
w_{+k} &=  s_{k}s_{k+1}\cdots s_{N-1} s_{N}s_{N-2}\cdots  s_{2}  s_{1},   &  \; &\;\;\;\;\ell(w_{+k})= 2N-k-1.
\end{align}
Since
\begin{equation}
w_{-k}(\overline{\mathbf{k}}_{1})=- \boldsymbol{e}_{k}, \qquad w_{+k}(\overline{\mathbf{k}}_{1})= \boldsymbol{e}_{k},
\end{equation}
the weights carried by harmonic forms in the $SO(2N)$ vector representation are given by
\begin{equation}
\{-\boldsymbol{e}_{1} ,-\boldsymbol{e}_{2} ,\ldots,-\boldsymbol{e}_{N} ,\boldsymbol{e}_{N} ,\boldsymbol{e}_{N-1} ,\ldots,\boldsymbol{e}_{1}\}.
\end{equation}

The simple root generators of $SO(2N)$ in this basis can be constructed as 
\begin{equation}
E_{\boldsymbol{\alpha}_{i}}=E_{i+1,i}+E_{2N-i+1,2N-i}, \qquad i=1,\ldots,N-1, \qquad E_{\boldsymbol{\alpha}_{N}}=E_{N+1,N-1}+E_{N+2,N},
\end{equation}
with $E_{m,n}$ in (\ref{e}).

\item \textit{ $USp(2N+2)\rightarrow USp(2N)\times U(1)  \Longleftrightarrow   SO(2N+3)\rightarrow SO(2N+1) \times U(1)  $}

\medskip

In Figure \ref{fig4}, removing $\boldsymbol{\alpha}_{1}$ reduces $USp(2N+2)$ to $USp(2N) \times U(1)$. $ H^{s} =USp(2N)$. $  \mathbf{k}_{1}  = \boldsymbol{\alpha}_{1}^{\vee} = \boldsymbol{e}_{1}-\boldsymbol{e}_{2} $. Its projection
\begin{equation}
 \overline{\mathbf{k}}_{1}  =-\boldsymbol{e}_{2} 
\end{equation}
is the lowest weight of $(\mathbf{2N{+}1})$-representation of $ (H^{\vee})^{s} =SO(2N+1)$, but is not a minuscule coweight of $USp(2N)$. There is a second solution
\begin{equation}
\mathbf{k}_{2}  =\boldsymbol{e}_{1}, \qquad  \text{with} \qquad  \overline{\mathbf{k}}_{2}  =0.
\end{equation}

As shown in \ref{mut1}, $ C(k_{1})\cong USp(2N) /[USp(2N-2) \times U(1)] \cong \mathbb{CP}^{2N-1}$, $ C(k_{2})=\{k_{2}\} $,
\begin{equation}
 \mathcal{M}_{\mathrm{rel}} (1,\Phi_{0})\cong\mathbb{CP}^{2N-1} \cup  \mathbb{C}^{2N}\approx \mathbb{CP}^{2N},\qquad \dim_{\mathbb{R}} \mathcal{M}_{\mathrm{rel}} (1,\Phi_{0})=4N .
\end{equation}
The Euler characteristic is
\begin{equation}
 \chi[ \mathcal{M}_{\mathrm{rel}} (1,\Phi_{0})]=\chi[C(k_{1})]+\chi[C(k_{2})]=2N+1= \chi[ \mathbb{CP}^{2N}].
\end{equation}
Since $   \mathcal{M}_{\mathrm{rel}} (1,\Phi_{0})$ is homeomorphic to $\mathbb{CP}^{2N}$, they have the isomorphic cohomology rings. The basis of harmonic forms on $  \mathcal{M}_{\mathrm{rel}} (1,\Phi_{0})  $ can be taken as\footnote{Here we take $\mathbb{CP}^{2N-1}\cup \mathbb{C}^{2N}$ as $\mathbb{C}^{2N}$ with the sphere $\mathbb{S}^{4N-1}_\infty$ quotiented by the Hopf $\mathbb{S}^{1}$-action, so that $\mathbb{CP}^{2N-1}$ arises at infinity. A form on $\mathbb{C}^{2N}$ descends, upon restriction to $\mathbb{S}^{4N-1}_\infty$, to a well-defined form on $\mathbb{CP}^{2N-1}$ if and only if its restriction is basic for the Hopf fibration, i.e.\ $\mathbb{S}^{1}$-invariant and horizontal. Moreover, a harmonic form on $\mathbb{C}^{2N}$ whose boundary values on $\mathbb{S}^{4N-1}_\infty$ stay finite as $r\to\infty$ is necessarily parallel. These conditions single out \eqref{lal}, where $\tilde{\omega} = \frac{i}{2}\sum_{a=1}^{2N} dz^{a} \wedge d\bar z^{a}$ is the flat K\"ahler form on $\mathbb{C}^{2N}$.}
\begin{equation}\label{lal}
\{1,\tilde{\omega},\ldots,\tilde{\omega}^{2N}\}. 
\end{equation}

With the relabeling $i \rightarrow i-1$, 
\begin{equation}
 \overline{\mathbf{k}}_{1}  =-\boldsymbol{e}_{2}\;\;  \longrightarrow \;\; \overline{\mathbf{k}}_{1}  =-\boldsymbol{e}_{1} ,
\end{equation}
simple roots of $SO(2N+1)$ become
\begin{equation}\label{so1a}
\boldsymbol{\alpha}_{1}= \boldsymbol{e}_{1}-\boldsymbol{e}_{2}  ,\qquad \boldsymbol{\alpha}_{2}= \boldsymbol{e}_{2}-\boldsymbol{e}_{3}  ,\qquad \ldots  \qquad \boldsymbol{\alpha}_{N-1}= \boldsymbol{e}_{N-1}-\boldsymbol{e}_{N}  ,\qquad 
\boldsymbol{\alpha}_{N}= \boldsymbol{e}_{N}.
\end{equation}
The $ (\mathbf{2N{+}1}) $-representation $V_{-\boldsymbol{e}_{1} }$ of $ SO(2N+1) $ has weight decomposition
\begin{equation}\label{so11}
\text{wt}(V_{-\boldsymbol{e}_{1} } )=O(-\boldsymbol{e}_{1})\cup O(0)=  \{-\boldsymbol{e}_{1} ,-\boldsymbol{e}_{2} ,\ldots,-\boldsymbol{e}_{N} ,\boldsymbol{e}_{N} ,\boldsymbol{e}_{N-1} ,\ldots,\boldsymbol{e}_{1}\} \cup  \{0\} .
\end{equation}
The corresponding minimal-length representatives in $W_{H^{s}}/W_{Z_{H^{s}}(-\boldsymbol{e}_{1})}$ and $W_{H^{s}}/W_{Z_{H^{s}}(0)}$ are
\begin{equation}
\{w_{-1},  w_{-2} , \ldots,  w_{-N},  w_{+N},  w_{+(N-1)} , \ldots,  w_{+1} \} \qquad \text{and} \qquad  \{e\},
\end{equation}
where 
\begin{align}
w_{-k}  &=s_{k-1}s_{k-2}\cdots s_{1},   &\;  &\;\;\;\;\ell(w_{-k})= k-1, \\
w_{+k} &=  s_{k}s_{k+1}\cdots s_{N-1} s_{N}s_{N-1}\cdots  s_{2}  s_{1},   &  \; &\;\;\;\;\ell(w_{+k})= 2N-k. 
\end{align}
Consequently, harmonic forms split as 
\begin{equation}
\{1,\tilde{\omega},\ldots,\tilde{\omega}^{2N}\}=\{1,\tilde{\omega},\ldots,\tilde{\omega}^{2N-1}\} \cup \{\tilde{\omega}^{2N}\},
\end{equation}
with weights assigned according to (\ref{so11}). $ \deg ( \tilde{\omega}^{2N} )=\dim \mathcal{M}_{\mathrm{rel}}(1,\Phi_{0},k_{2})$, which is consistent with (\ref{4100}).

Let $f_{k}=\tilde{\omega}^{k-1}$, which are parallel and mutually orthogonal. The simple-root generators of $SO(2N+1)$ in this basis are
\begin{equation}
E_{\boldsymbol{\alpha}_{i}}=E_{i+1,i}+E_{2N+1-i,2N-i},\qquad i=1,\ldots,N-1,\qquad E_{\boldsymbol{\alpha}_{N}}=E_{2N+1,N}+2E_{N+1,2N+1},
\end{equation}
with $E_{m,n}$ in (\ref{e}).

\item \textit{ $SO(2N+3)\rightarrow U(N+1) \Longleftrightarrow  USp(2N+2)\rightarrow U(N+1)  $}

\medskip

In Figure \ref{fig3}, removing $\boldsymbol{\alpha}_{N+1}$ breaks $SO(2N+3)$ to $U(N+1)$ with semisimple factor $ H^{s} =SU(N+1)$. $  \mathbf{k}_{1}  = \boldsymbol{\alpha}_{N+1}^{\vee}=2\boldsymbol{e}_{N+1}$.
\begin{equation}
 \overline{\mathbf{k}}_{1}  =2\boldsymbol{e}_{N+1}-\frac{2}{N+1}\sum^{N+1}_{a=1} \boldsymbol{e}_{a}
\end{equation}
is the lowest weight of $  \mathrm{Sym}^2 \, \mathbf{N{+}1} $-representation of $SU(N+1)$, but is not minuscule in $SU(N+1)$. The second solution is
\begin{equation}
\mathbf{k}_{2}=\boldsymbol{e}_{N}+\boldsymbol{e}_{N+1}, \qquad  \text{with} \qquad  \overline{\mathbf{k}}_{2}  =\boldsymbol{e}_{N}+\boldsymbol{e}_{N+1}-\frac{2}{N+1}\sum^{N+1}_{a=1} \boldsymbol{e}_{a}.
\end{equation}

As shown in \ref{mut2}, $C(k_{1})\cong SU(N+1)/S[U(N)\times U(1)]\cong\mathbb{CP}^{N}$, $   C(k_{2})\cong SU(N+1)/S[U(N-1) \times U(2) ]\cong \mathrm{Gr}(2,N+1)$, 
\begin{equation}
 \mathcal{M}_{\mathrm{rel}} (1,\Phi_{0}) \cong\mathbb{CP}^{N} \;\cup\; \mathrm{Tot}(S^\vee) \;\approx\; \mathrm{Gr}(2,N+2)\cong SU(N+2)/S[U(N) \times U(2) ],
\end{equation}
where $ \mathrm{Tot}(S^\vee)$ is locally isomorphic to $\mathbb{C}^2 \times \mathrm{Gr}(2,N+1)$. $\dim_{\mathbb{R}} \mathcal{M}_{\mathrm{rel}} (1,\Phi_{0})=4N $. 
\begin{equation}
  \chi[ \mathcal{M}_{\mathrm{rel}} (1,\Phi_{0})]=  \chi[ C(k_{1})]+  \chi[C(k_{2})] =\tfrac{1}{2}(N+1)(N+2)=\chi[ \mathrm{Gr}(2,N+2)]. 
\end{equation}

$\mathcal{M}_{\mathrm{rel}}(1,\Phi_{0})$ and $\mathrm{Gr}(2,N{+}2)$ have isomorphic cohomology rings. Let $[\tilde{\omega}]$ and $[\tilde{\Omega}]$ be the images of the degree-$2$ and degree-$4$ special Schubert generators under this isomorphism. Then the cohomology classes on $\mathcal{M}_{\mathrm{rel}}(1,\Phi_{0})$ are $[f'_{ij}]$ for $1\le i<j\le N+2$, where
\begin{equation}
f'_{ij}
=
\sum_{m=0}^{[\frac{j-i-1}{2}]}
(-1)^m\binom{j-i-1-m}{m}\;
\tilde{\omega}^{\,j-i-1-2m} \wedge \tilde{\Omega}^{\,i-1+m},
\end{equation}
as in \eqref{fij}. Let $(f'_{ij})_{\mathrm{harm}}$ be the harmonic representative of $[f'_{ij}]$. Setting $f_{ij}= f'_{i\,j+1}$, the harmonic basis on $\mathcal{M}_{\mathrm{rel}}(1,\Phi_{0})$ is
\begin{equation}
\{(f_{ij})_{\mathrm{harm}}\mid 1\le i \le j\le N+1\}=\{(f'_{ij})_{\mathrm{harm}}\mid 1\le i<j\le N+2\},
\end{equation}
which furnishes the $ \mathrm{Sym}^2 \, \mathbf{N{+}1}$ representation.

Simple roots of $SU(N+1)$ are taken to be 
\begin{equation}
\boldsymbol{\alpha}_{i}= \boldsymbol{e}_{i}-\boldsymbol{e}_{i+1} ,\qquad i=1,\ldots,N.
\end{equation}
The weights of $ \mathrm{Sym}^2 \, \mathbf{N{+}1} $ can be written as
\begin{equation}
\text{wt}_{ij}=\boldsymbol{e}_{N+2-i}  +\boldsymbol{e}_{N+2-j} -\frac{2}{N+1}\sum^{N+1}_{a=1} \boldsymbol{e}_{a},
\end{equation}
which decompose into two Weyl-orbits:
\begin{eqnarray}
 \nonumber && \{\text{wt}_{ij}\,|\, 1 \leq i \leq  j \leq N+1 \}= O(\overline{\mathbf{k}}_{1}  )\cup O(\overline{\mathbf{k}}_{2}  )\\&=&  \{\text{wt}_{jj}\,|\, 1 \leq  j \leq N+1 \}\,\cup\, \{\text{wt}_{ij}\,|\, 1 \leq i <  j \leq N+1 \} .
\end{eqnarray} 
Correspondingly, the harmonic forms split as 
\begin{eqnarray}
 \nonumber && \left\lbrace (f_{ij})_{\mathrm{harm}}\mid 1\le i \le j\le N+1\right\rbrace\\
&=& \left\lbrace (f_{1j})_{\mathrm{harm}}\mid 1 \le j\le N+1\right\rbrace\cup
     \left\lbrace (\tilde{\Omega} \wedge f'_{ij})_{\mathrm{harm}}\mid 1\le i < j\le N+1\right\rbrace,
\end{eqnarray}
where the restrictions $\{[f_{1j}]|_{C(k_1)}\}$ form a basis of $H^\ast(C(k_1))$, while
$\{[\tilde{\Omega}  \wedge f'_{ij}]\}$ is the Thom-shifted image of the Schubert basis of $  H^\ast(C(k_2))$. Here $ \deg( \tilde{\Omega} )=4 = \dim\mathcal{M}_{\mathrm{rel}}(1,\Phi_{0},k_{2}) $ as in (\ref{4100}). The weight assignment is  
\begin{equation}
\text{wt}((f_{ij})_{\text{harm}}) =
\begin{cases}
 \text{wt}_{jj},
& i = 1,\; 1 \leq j \leq N+1,\\[4pt]
 \text{wt}_{i-1\;j},
&  2 \leq  i   \leq j \leq N+1.
\end{cases}
\end{equation}

\medskip

\end{enumerate}

\subsection{Non-abelian electric and magnetic gauge transformations}

 In $\mathcal N=4$ SYM with the symmetry breaking $G \rightarrow  H^{s} \times U(1)^{t}$, $H^{s}$ acts isometrically on the relative moduli space $\mathcal{M}_{\mathrm{rel}}(m,\Phi_{0})$, with generators given by Killing vector fields. The Hilbert space of states on $\mathcal{M}_{\mathrm{rel}}(m,\Phi_{0})$ also admits an action of the dual group $(H^{\vee})^{s}$. While $H^{s}$ acts geometrically by metric-preserving diffeomorphisms, $(H^{\vee})^{s}$ acts algebraically by shifting weights within the representation space. On differential forms, this algebraic action is realized by wedge ($ O[f]$) and contraction ($ O^{\dagger}[f]$) operations.

Such geometric/algebraic duality is analogous to a familiar pair in quantum mechanics: the translation operator $e^{iPa}$ acts on position space as
\begin{equation}\label{xp}
 e^{iPa} \vert x \rangle = \vert x+a\rangle,
\end{equation}
while the dual $ e^{iXa} $ acts on momentum spectrum by
\begin{equation}\label{xp1}
  e^{iXa} \vert p \rangle =  \vert p+a \rangle.
\end{equation}

When $m=1$, $\mathcal{M}_{\mathrm{rel}}(1,\Phi_{0})$ is compact and carries an isometric action of the connected group $H^{s}$, so every harmonic form is $H^{s}$-invariant. When $m>1$, $\mathcal{M}_{\mathrm{rel}}(m,\Phi_{0})$ remains $H^{s}$-invariant but is necessarily non-compact and stratified. Since all strata are even-dimensional, one must impose an extension (gluing) condition across the boundary to obtain well-defined differential forms on $\mathcal{M}_{\mathrm{rel}}(m,\Phi_{0})$. As in Example~\ref{a2a}, this extension condition projects out the $H^{s}$-invariant harmonic representatives.

Since all admissible harmonic forms are $H^{s}$-invariant, the operators implementing the $(H^{\vee})^{s}$-action can be chosen to commute with both $H^{s}$ and the Laplacian $\Delta$. Let $h$ and $\tilde h$ denote the induced actions for generators of $H^{s}$ and $(H^{\vee})^{s}$ on differential forms. Then
\begin{equation}
[h,\tilde{h}]=0,\qquad
[h,\Delta]=0,\qquad
[\tilde{h},\Delta]=0,
\qquad
\forall\, h\in \mathrm{Lie}(H^{s}),\ \forall\, \tilde{h}\in \mathrm{Lie}((H^{\vee})^{s}) .
\end{equation}
Consequently, each $\Delta$-eigenspace $ \mathcal H_{\alpha} $ is stable under $H^{s}\times (H^{\vee})^{s}$ and hence carries an $H^{s} \times (H^{\vee})^{s}$-representation. In particular, the harmonic subspace $\mathcal H_{0}$ is an $H^{s}$-singlet but may carry a nontrivial $(H^{\vee})^{s}$-representation, whereas for $\alpha>0$ the $H^{s}$-action on $\mathcal H_{\alpha}$ need not be trivial.

In what follows, we construct the explicit $H^{s}\times (H^{\vee})^{s}$-representation on the Hilbert space of $\mathcal{M}_{\mathrm{rel}}(1,\Phi_{0})=C(k_{1})$. In this case all harmonic forms are parallel, so a primitive (Lefschetz) decomposition is available.

Let $\{g_1,\dots,g_s\}$ be parallel forms generating the harmonic ring of $\mathcal{M}_{\mathrm{rel}}(1,\Phi_{0})$, and set
\begin{equation}
L_m =g_m\wedge (\,\cdot\,),\qquad \Lambda_m= g_m \mathbin{\lrcorner} (\,\cdot\,)=L_m^\dagger ,\qquad m=1,2,\ldots,s.
\end{equation}
Then $[\Delta,L_m]=[\Delta,\Lambda_m]=0$. For a fixed Hodge Laplacian eigenvalue $\alpha$, define the joint-primitive subspace  
\begin{equation}
P_\alpha=\Big(\bigcap_{m=1}^s \ker\Lambda_m\Big)\cap \mathcal{H}_{\alpha},
\end{equation}
and choose an orthonormal basis $\{\beta_\alpha^{(a)}\}_{a=1}^{r_\alpha}  $ of $P_\alpha$. Let
\begin{equation}
W=\operatorname{span}\Big\{\,L_1^{i_1}\cdots L_s^{i_s}\,\beta_\alpha^{(a)}
\ \Big|\ i_m\in\mathbb Z_{\ge 0},\ 1\le a\le r_\alpha\Big\}\subseteq \mathcal H_\alpha .
\end{equation}
By the standard Lefschetz/primitive decomposition argument \cite{GriffithsHarris}, we have $W=\mathcal H_\alpha$.\footnote{If $W\neq\mathcal H_\alpha$, choose $\psi\in W^\perp\cap \mathcal{H}_{\alpha}$ of minimal form degree. Since $W$ is $L_m$-stable, $\forall \;w\in W$, we have $\langle \Lambda_m\psi,w\rangle=\langle \psi,L_m w\rangle=0$, hence $\Lambda_m\psi\in W^\perp$. Besides, $[\Delta,\Lambda_m]=0$ implies $\Lambda_m\psi\in\mathcal H_\alpha$. If $\Lambda_m\psi\neq 0$ for some $m$, then $\deg(\Lambda_m\psi)<\deg\psi$, contradicting the minimality of $\deg\psi$. Thus $\Lambda_m\psi=0$ for all $m$, so $\psi\in P_\alpha\subset W$, contradicting $\psi\in W^\perp$. } In particular, the harmonic subspace is
\begin{equation}
\mathcal H_0=\operatorname{span}\Big\{\,L_1^{i_1}\cdots L_s^{i_s}\,1
\ \Big|\ i_m\in\mathbb Z_{\ge 0}  \Big\}.
\end{equation}

Both $L_m$ and $\Lambda_m$ are $H^{s}$-invariant, so $P_\alpha$ is $H^{s}$-stable. We may decompose $P_\alpha$ into irreducible $H^{s}$-multiplets and choose the basis
\begin{equation}
\{\beta_\alpha^{(a)}\}_{a=1}^{r_\alpha}  =\{\beta_\alpha^{(1_{k})}\}_{k=1}^{r^{(1)}_\alpha} \cup \{\beta_\alpha^{(2_{k})}\}_{k=1}^{r^{(2)}_\alpha} \cup \cdots  \cup \{\beta_\alpha^{(n_{k})}\}_{k=1}^{r^{(n)}_\alpha} ,
\end{equation}
where the forms in each $\{\beta_\alpha^{(j_k)}\}_{k=1}^{r_\alpha^{(j)}}$ have the same degree. Then $\mathcal H_\alpha$ is spanned by
\begin{equation}
\bigcup^{n}_{j=1}\Big\{\,L_1^{i_1}\cdots L_s^{i_s}\,\beta_\alpha^{(j_{k})}
\;\Big|\;  i_{m} \in \mathbb{Z}_{\geq 0}, \, 1 \leq k \leq r^{(j)}_{\alpha} \,\Big\}.
\end{equation}
Since $L_{m}$ is $H^{s}$-invariant, in each 
\begin{equation}
\Big\{\,L_1^{i_1}\cdots L_s^{i_s}\,\beta_\alpha^{(j_{k})}
\;\Big|\;  i_{m} \in \mathbb{Z}_{\geq 0}, \, 1 \leq k \leq r^{(j)}_{\alpha} \,\Big\},
\end{equation}
the $H^{s}$-representation is entirely carried by $ \beta_\alpha^{(j_{k})} $. For fixed $ \beta_\alpha^{(j_{k})} $, the action of $L_{m}$ generates an $(H^{\vee})^{s}$-multiplet.

By the degree bound
\begin{equation}
\deg\!\big(L_1^{i_1}\cdots L_s^{i_s}\beta_\alpha^{(a)}\big)\le \dim M,
\end{equation}
if $\deg(\beta_\alpha^{(j_k)})>0$, then the admissible wedge tower is strictly shorter than the harmonic case. The $(H^{\vee})^{s}$-multiplet generated by $L_m$ is a truncation of the harmonic one and, in general, need not match any nontrivial $(H^{\vee})^{s}$-representation. By contrast, when $\deg(\beta_\alpha^{(j_k)})=0$, the resulting $(H^{\vee})^{s}$-multiplet is isomorphic to the harmonic case. Therefore, only the $0$-form multiplets $\{\beta_\alpha^{(j_k)}\}$ can support a nontrivial $(H^{\vee})^{s}$-representation, and in that case the representation matches the harmonic sector.

The operators generating such $(H^{\vee})^{s}$-transformations are precisely $E_{m,n}$ constructed in (\ref{e}). Acting on a generic $p$-form $\vert f \rangle=O[f]\vert 1 \rangle$, we have
\begin{eqnarray}
E_{m,n}\vert f \rangle \nonumber &=& \frac{1}{c_{m}c_{n}}O[f_{m}] O^{\dagger}[\varepsilon]O[\ast f_{n} \wedge f]\vert 1 \rangle\\ \nonumber &=& \frac{\delta_{\deg (f_{n}),p}}{c_{m}c_{n}}O[f_{m}] O^{\dagger}[\varepsilon]O[\varepsilon]O[\langle f_{n},f\rangle]\vert 1 \rangle\\ &=&  \frac{\delta_{\deg (f_{n}),p}}{c_{m}c_{n}}\vert \langle f_{n},f\rangle  f_{m} \rangle,
\end{eqnarray} 
where $\langle f_{n},f\rangle$ is the pointwise inner product (for $\deg(f_n)=p$),
\begin{equation}
\langle f_{n},f\rangle (z)=\frac{1}{p!}f^{\mu_{1}\cdots \mu_{p}}_{n}(z)f_{\mu_{1}\cdots \mu_{p}}(z).
\end{equation}

Using the pointwise adjointness $\langle \Lambda_{m}w,v\rangle (z)=\langle w,L_{m}v \rangle (z)$ and the fact that
$f_n\in\mathcal H_0=\operatorname{span}\{L_1^{j_1}\cdots L_s^{j_s}1\}$, for
$f=L_1^{i_1}\cdots L_s^{i_s}\beta_\alpha^{(a)}$ with $\deg(\beta_\alpha^{(a)})>0$, we have
\begin{eqnarray}
\langle f_{n},f\rangle
&=&\big\langle f_{n},\,L_1^{i_1}\cdots L_s^{i_s}\beta_\alpha^{(a)} \big\rangle
=\big\langle \Lambda_s^{i_s}\cdots \Lambda_1^{i_1} f_{n},\,\beta_\alpha^{(a)} \big\rangle \nonumber\\
&=&\sum_{\vec j} c_{\vec j}\,\big\langle L_1^{j_1}\cdots L_s^{j_s}1,\,\beta_\alpha^{(a)}\big\rangle
=\sum_{\vec j} c_{\vec j}\,\big\langle 1,\,\Lambda_s^{j_s}\cdots \Lambda_1^{j_1}\beta_\alpha^{(a)}\big\rangle
=0, \nonumber
\end{eqnarray}
where we used $\Lambda_m\beta_\alpha^{(a)}=0$. Hence $E_{m,n}\vert f \rangle=0$. When $ \deg (\beta_\alpha^{(a)})=0 $, $ f=\frac{1}{c_{k}}f_{k} \wedge \beta_\alpha^{(a)}$ for some label $k$, and 
\begin{equation}
E_{m,n}\vert \frac{1}{c_{k}}f_{k} \wedge \beta_\alpha^{(a)}\rangle  =\delta_{n,k}  \vert  \frac{1}{c_{m}} f_{m} \wedge \beta_\alpha^{(a)} \rangle,
\end{equation}
in agreement with (\ref{e1}).

\section{Discussion}\label{cda}

We now discuss some features of the magnetic gauge symmetry from the perspective of moduli-space description.

In $\mathcal N=4$ SYM with symmetry breaking $G\to H^{s}\times U(1)^{t}$ induced by $\Phi_0$, finite-energy configurations split into disconnected sectors labeled by the $t$-component topological charge $m$. In the moduli-space approximation, the low-energy dynamics in a fixed charge sector is described by supersymmetric quantum mechanics on $\mathcal{M} (m,\Phi_{0}) $.

In a unit-charge sector, where $m$ has a single nonzero component equal to $1$, the dynamics reduces to a free particle, with no interactions. For maximal symmetry breaking $G\to U(1)^{r}$, $\mathcal M(m,\Phi_0)=\mathbb R^{3}\times \mathbb S^{1}$, so semiclassical quantization yields a particle with no internal structure. When $t<r$, the relative moduli space $\mathcal M_{\mathrm{rel}}(m,\Phi_0)$ can be nontrivial, and the magnetic particle may carry internal degrees of freedom, on which the magnetic gauge symmetry acts.

The simplest case is $SU(N+1)\to U(N)$, where the relative moduli space is the magnetic charge orbit $\mathcal M_{\mathrm{rel}}(1,\Phi_0)=C(k_{1})\cong \mathbb{CP}^{N-1}$, with position eigenstates $\vert G_{0} \rangle$. Each $\vert G_{0} \rangle$ represents a field configuration with magnetic charge $G_{0}$. For BPS particles, the internal states are $N$ harmonic wavefunctions on $C(k_{1})$ denoted by $\{\ket{k}\}_{k=1}^{N}$, which form an $SU(N)$ isospin multiplet. The electric $SU(N)$ action moves $\ket{G_{0}}$ along the orbit, leaving $\{\ket{k}\}_{k=1}^{N}$ fixed. By contrast, the magnetic $SU(N)$ action rotates the BPS isospin multiplet and leaves $G_{0}$ fixed: $\vert G_{0}\rangle \vert \Omega \rangle \rightarrow \vert G_{0}\rangle \exp \{i \theta (G_{0},\hat \lambda^{\dagger }  ,\hat \lambda )\}\vert \Omega \rangle$.

In general, the BPS monopole states are $H^{s}$-invariant, so S-duality implies that in the dual theory with symmetry breaking $G^{\vee}\to (H^{\vee})^{s}\times U(1)^{t}$, the corresponding W-bosons are invariant under the magnetic symmetry group $H^{s}$. Equivalently, in the original theory with $G\to H^{s}\times U(1)^{t}$, W-bosons are $(H^{\vee})^{s}$-invariant. Adding a $U(1)^{t}$ electric charge turns monopoles into dyons, which are still $H^{s}$-invariant, since $U(1)^{t}$ commutes with $H^{s}$. To summarize, for $\mathcal N=4$ SYM with $G\to H^{s}\times U(1)^{t}$, among BPS vector multiplets, W-bosons furnish a representation of $H^{s}$ but are $(H^{\vee})^{s}$-singlets, whereas monopoles and dyons are $H^{s}$-singlets but transform in the same $(H^{\vee})^{s}$-representation. Chromodyonic states carrying nontrivial $H^{s}\times (H^{\vee})^{s}$ representations can also arise. They belong to the positive-energy spectrum of the relative Hamiltonian and thus fall into long multiplets, with the additional multiplicity generated by the relative supercharges. These states are non-BPS, so their masses\footnote{In the large-size limit of $C(k)$, the excitation gap can be arbitrarily small.} and any accidental degeneracies are unprotected.

\section*{Acknowledgements}
The work is supported in part by NSFC under the Grant No. 11605049.

\vspace{6mm}

\begin{appendix}

\section{$SO(5)\to U(2)$}\label{reg}

Simple roots of $SO(5)$ can be taken as
\begin{equation}
\boldsymbol{\gamma}= \boldsymbol{e}_{1}-\boldsymbol{e}_{2},\qquad \boldsymbol{\mu} = \boldsymbol{e}_{2},
\end{equation}
with remaining positive roots
\begin{equation}
\boldsymbol{\alpha}= \boldsymbol{e}_{1}+\boldsymbol{e}_{2},\qquad\boldsymbol{\beta}= \boldsymbol{e}_{1}.
\end{equation}
Removing $\boldsymbol{\mu}$ breaks $SO(5)$ to $U(2)$. $\mathbf{h}=v \boldsymbol{\alpha}$. $ \pi_{2}(SO(5)/U(2)) \cong \mathbb{Z}$, $t=1$. For fundamental monopoles with $m=1$, there are two solutions to (\ref{ad}):
\begin{equation}
 \mathbf{k}_{1} =\boldsymbol{\mu}^{\vee} =2\boldsymbol{e}_{2},\qquad\mathbf{k}_{2} =\boldsymbol{\mu}^{\vee}+\boldsymbol{\gamma}^{\vee}=\boldsymbol{e}_{1}+\boldsymbol{e}_{2},
\end{equation}
with $ C(k_{1})\cong SU(2)/U(1) $ and $C(k_{2})=\{k_{2}\}$. $\mathcal{K}= C(k_{1})\, \cup \,\{k_{2} \}$. 
\begin{equation}
 \mathcal{M} (1,\Phi_{0})=\mathcal{M}_{1}(1,\Phi_{0})\cup \mathcal{M}_{2}(2,\Phi_{0}),
\end{equation}
where 
\begin{equation}
\dim \mathcal{M}_{1}(1,\Phi_{0}) = 6,\qquad \dim \mathcal{M}_{2}(1,\Phi_{0}) = 8.
\end{equation}

Any root $\boldsymbol{\nu}  $ determines an $ SU(2) $ subgroup generated by 
\begin{equation}\label{Tk}
t^1(\boldsymbol{\nu}) = \frac{1 }{ \sqrt{2
{\boldsymbol{\nu}}^2}} (E_{\boldsymbol{\nu}} +
E_{-\boldsymbol{\nu}})       ,   \qquad 
t^2(\boldsymbol{\nu}) = -\frac{i }{ \sqrt{2{
\boldsymbol{\nu}}^2}} (E_{\boldsymbol{\nu}} -
E_{-{\boldsymbol{\nu}}})         ,     \qquad 
t^3(\boldsymbol{\nu}) = \frac{1}{2} {\boldsymbol{\nu}}^{\vee}
\cdot  \mathbf{T}  .
\end{equation}
For the $SO(5)$ roots, we have 
\begin{equation}\label{eq1}
[t^{i}(\boldsymbol{\alpha}),t^{j}(\boldsymbol{\gamma})]=0,\qquad [t^{i}(\boldsymbol{\beta}),t^{3}(\boldsymbol{\mu})]=0,\qquad 
[t^{i}(\boldsymbol{\mu}),t^{3}(\boldsymbol{\beta})]=0, \qquad i,j=1,2,3.
\end{equation}

The eight-parameter family of solutions for $\mathcal{M}_{2}(1,\Phi_{0})$ is \cite{Weinberg1982}
\begin{align}\label{8}
A_{i}&= \epsilon_{ijk}\,\hat r^{j}\,[A(r)t^{k}(\boldsymbol{\alpha})
      +  A(r) L(a, r)t^{k}(\boldsymbol{\gamma})] \nonumber\\
&\quad + \sqrt{2}F(r)L(a, r)^{1/2}\big(
      -\delta_{i1} t^{2}(\boldsymbol{\mu})
      + \delta_{i2} t^{1}(\boldsymbol{\mu})
      - \delta_{i3} t^{2}(\boldsymbol{\beta})\big), \nonumber\\[4pt]
\Phi &= \hat r_i \big[ H(r)t^{i}(\boldsymbol{\alpha})
               +  A(r) L(a, r)t^{i}(\boldsymbol{\gamma}) \big]
   + \sqrt{2} F(r)L(a, r)^{1/2} t^{1}(\boldsymbol{\beta}),
\end{align}
with
\begin{equation}
A(r) = {v \over \sinh vr} - {1\over r}, \qquad 
H(r) = v \coth vr  - {1\over r}, \qquad 
F(r) = { v \over \sqrt{8} \cosh (vr/2) } ,
\end{equation}
and
\begin{equation}
   L(a, r) = \left[ 1 +  (r/a) \coth(vr/2) \right]^{-1} ,\qquad a \in [0, +\infty),
\end{equation}
where $a$ is the modulus controlling the size of the non-Abelian cloud. In addition to translations and variations of $a$, the configurations admit an action of the residual $U(2)$ that leaves $k_{2}$ invariant.

For large $r$,
\begin{align}
A_{i} =
\begin{cases}
- \dfrac{1}{r}\epsilon_{ijk}\,\hat r^{j}\,t^{k}(\boldsymbol{\alpha}) +\mathcal{O}\!\left(\dfrac{1}{r^{2}}\right), & a \ \text{finite}, \\[6pt]
- \dfrac{1}{r} \epsilon_{ijk}\,\hat r^{j}\left[t^{k}(\boldsymbol{\alpha}) + t^{k}(\boldsymbol{\gamma})\right] +\mathcal{O}\!\left(\dfrac{1}{r^{2}}\right), & a = \infty,
\end{cases}
\end{align}
\begin{align}\label{larg}
\Phi =
\begin{cases}
v \hat r_i t^{i}(\boldsymbol{\alpha})  - \dfrac{1}{r}\hat r_i t^{i}(\boldsymbol{\alpha}), & a \ \text{finite}, \\[6pt]
v \hat r_i t^{i}(\boldsymbol{\alpha})  - \dfrac{1}{r}\hat r_i\big[ t^{i}(\boldsymbol{\alpha})+t^{i}(\boldsymbol{\gamma})  \big], & a = \infty.
\end{cases}
\end{align}
When $ a $ is finite, 
\begin{equation}
G_{0}=2t^{3}( \boldsymbol{\alpha} )= {\boldsymbol{\alpha}}^{\vee}
\cdot  \mathbf{T} =\mathbf{k}_{2}\cdot  \mathbf{T} =k_{2};
\end{equation}
when $a= \infty$, 
\begin{equation}
G_{0}=2[  t^{3}(\boldsymbol{\alpha})+t^{3}(\boldsymbol{\gamma}) ]=( {\boldsymbol{\alpha}}^{\vee}+{\boldsymbol{\gamma}}^{\vee})\cdot  \mathbf{T}={\boldsymbol{\beta}}^{\vee}\cdot  \mathbf{T}=2  t^{3}(\boldsymbol{\beta}),
\end{equation}
which is Weyl-conjugate to $k_{1}=\mathbf{k}_{1} \cdot  \mathbf{T} ={\boldsymbol{\mu}}^{\vee}\cdot  \mathbf{T}=2  t^{3}(\boldsymbol{\mu})$. So as $a \rightarrow \infty$, $ G_{0} $ jumps from $k_{2}$ to $C(k_{1})$. The residue $U(2)$ fixes $G_{0}$ for finite $a$, but rotates it along $C(k_{1})$ when $a=\infty$.

At $a=0$, $  L(a, r)=0$,
\begin{equation}
A_{i}= \epsilon_{ijk}\,\hat r^{j}\, A(r)t^{k}(\boldsymbol{\alpha})\qquad \Phi = \hat r_i  H(r)t^{i}(\boldsymbol{\alpha})
\end{equation}
is the standard $\boldsymbol{\alpha}  $-embedded $SU(2)$ solution with magnetic charge $k_{2}$. As $a\rightarrow\infty$, $  L(a, r)\rightarrow 1$,
\begin{align}\label{ai1}
A_{i} &= \epsilon_{ijk}\,\hat r^{j}\, [t^{k}(\boldsymbol{\alpha})
      + t^{k}(\boldsymbol{\gamma})]A(r) + \sqrt{2}F(r)\big(
      -\delta_{i1} t^{2}(\boldsymbol{\mu})
      + \delta_{i2} t^{1}(\boldsymbol{\mu})
      - \delta_{i3} t^{2}(\boldsymbol{\beta})\big) \nonumber\\[4pt]
\Phi &= \hat r_i \big[ H(r)t^{i}(\boldsymbol{\alpha})
               + A(r)t^{i}(\boldsymbol{\gamma}) \big]
   + \sqrt{2} F(r) t^{1}(\boldsymbol{\beta})
\end{align}
is gauge equivalent to the $\boldsymbol{\beta}  $-embedded $SU(2)$ solution
\begin{align}\label{ai}
A_i &= \epsilon_{ijk}\, \hat{r}^{j}
 \, t^{k}(\boldsymbol{\beta})\left(   \dfrac{v}{2 \sinh \left(  v r/2\right)  }  - \frac{1}{r} \right), \nonumber \\[4pt]
\Phi &= \hat{r}_i \,  t^{i}(\boldsymbol{\beta})
\left( \dfrac{1}{2} v \coth \left( v r/2 \right) - \frac{1}{r} \right)
+  \frac{v}{2}\; t^{3} (\boldsymbol{\mu})  ,
\end{align}
and, by the Weyl reflection $\boldsymbol{\beta} \leftrightarrow \boldsymbol{\mu}  $, to the $\boldsymbol{\mu}  $-embedded $SU(2)$ solution with magnetic charge $k_{1} $. The two strata $\mathcal{M}_{1}(1,\Phi_{0})$ and $ \mathcal{M}_{2}(2,\Phi_{0})$ glue at $a=\infty$.

For the $\boldsymbol{\beta}  $-embedded solution (\ref{ai}), at spatial infinity, 
\begin{equation}
\Phi (\hat{\mathbf{r}},\infty)=  \frac{v}{2} [t^{3} (\boldsymbol{\mu})+ \hat{r}_i \,  t^{i}(\boldsymbol{\beta})]=u(\theta,\phi)[ v  t^{3}(\boldsymbol{\alpha})]u^{-1}(\theta,\phi) , 
\end{equation}
where
\begin{equation}\label{u}
u(\theta,\phi)=e^{-i \phi t^{3}(\boldsymbol{\beta})}e^{-i \theta t^{2}(\boldsymbol{\beta})}e^{i \phi t^{3}(\boldsymbol{\beta})}, \qquad \theta \in [0,\pi],\;\phi \in [0,2\pi ).
\end{equation}
Since $[t^{i}(\boldsymbol{\gamma}), t^{3}(\boldsymbol{\alpha})]=0  $, for \begin{equation}
 t^{i}(\theta,\phi)=u(\theta,\phi)t^{i}(\boldsymbol{\gamma})u^{-1}(\theta,\phi),
\end{equation}
we have $[  t^{i}(\theta,\phi) , \Phi (\hat{\mathbf{r}},\infty)]=0  $. It seems that $  t^{i}(\theta,\phi) $ generate a direction-dependent unbroken $SU(2)$ subgroup, while the electric $U(1) \subset U(2)$ is generated by $\Phi$.

However, there are two issues. First, $ t^{3}(\theta,\phi) $ generates the same $U(1)$ subgroup as $\Phi$. From (\ref{eq1}), $[ u(\theta,\phi), t^{3}(\boldsymbol{\mu})]=0  $, so
\begin{align}
t^{3}(\theta,\phi) &= u(\theta,\phi)\, t^{3}(\boldsymbol{\gamma})\, u^{-1}(\theta,\phi)
= u(\theta,\phi)\left(t^{3}(\boldsymbol{\alpha}) - t^{3}(\boldsymbol{\mu})\right) u^{-1}(\theta,\phi) \notag \\
&= u(\theta,\phi)\, t^{3}(\boldsymbol{\alpha})\, u^{-1}(\theta,\phi) - t^{3}(\boldsymbol{\mu})
= \frac{1}{v}\Phi (\hat{\mathbf{r}},\infty)   - t^{3}(\boldsymbol{\mu}).
\end{align}
Since $ t^{3}(\boldsymbol{\mu}) $ commutes with the entire solution (\ref{ai}), the action of $t^{3}(\theta,\phi)$ coincides with that of $\Phi(\hat{\mathbf{r}}, \infty)/v$. Second, along the negative $ \hat{r}_{3} $-axis, where $\theta= \pi$, 
\begin{equation}
u(\pi,\phi)=e^{-i \phi t^{3}(\boldsymbol{\beta})}e^{-i \pi t^{2}(\boldsymbol{\beta})}e^{i \phi t^{3}(\boldsymbol{\beta})}
\end{equation}
depends on $ \phi $ and thus is singular. As a result, $  t^{1}(\pi,\phi) $ and $ t^{2}(\pi,\phi) $, which do not commute with $ u(\pi,\phi) $, are also singular. This is the well-known problem that for non-Abelian monopoles, generators of the unbroken subgroup that do not commute with $G_{0}$ cannot be globally defined \cite{Balachandran1971, Balachandran1984a, Balachandran1984b, Abouelsaood1984b, NelsonManohar1983}.

On the other hand, for the solution (\ref{ai1}), which is gauge equivalent to (\ref{ai}), the second issue does not arise. Here, 
\begin{equation}
\Phi (\hat{\mathbf{r}},\infty)=v \,\hat r_i t^{i}(\boldsymbol{\alpha})  ,
\end{equation}
so the unbroken $SU(2)$ at infinity is generated by the constant $t^{i}(\boldsymbol{\gamma})$ on $\mathbb{S}^{2}_{\infty}$. Nevertheless, the magnetic charge is $ G_{0}=2  t^{3}(\boldsymbol{\beta}) $, which does not commute with $  t^{i}(\boldsymbol{\gamma})  $, so the gauge modes are still non-normalizable.

The solution (\ref{ai1}), obtained as the $a \rightarrow \infty$ limit of the well-defined non-Abelian monopole (\ref{8}), can be viewed as a regularization of (\ref{ai}). For the configuration (\ref{8}), the three global $SU(2)$ zero modes $(\delta' A_i, \delta' \Phi)$ and the cloud-size zero mode $ (\delta A_i = \partial_{a} A_i \delta a, \delta \Phi= \partial_{a} \Phi \delta a) $ are related by the hyper-K\"ahler $SU(2)$ rotation \cite{Abouelsaood1984a, 1}
\begin{equation}
\delta' A_i = \hat{n}_i\,\delta \Phi + \epsilon_{ijk}\,\hat{n}^j\,\delta A^k,
\qquad 
\delta' \Phi = -\hat{n}_i\,\delta A^i.
\end{equation}
The resulting gauge zero modes, already in background gauge $D_i \delta A_i + i[\Phi, \delta \Phi] =0$, are
\begin{equation}\label{2.233}
     \delta' A_i = D_i \Delta = \partial_i \Delta +i[A_i,\Delta],\qquad    \delta' \Phi =i[\Phi,\Delta] ,
\end{equation} 
with
\begin{equation}\label{2.23}
\Delta (a,r)=-  \hat{n}_{i} t^{i}(\boldsymbol{\gamma}) \frac{\partial_{a}L}{L} \delta a=\Lambda (a,r) \delta  \psi , \qquad  \delta \psi =\frac{\delta a}{a},
\end{equation} 
and 
\begin{equation}\label{ar}
\Lambda (a,r)= -  \hat{n}_{i} t^{i}(\boldsymbol{\gamma})\frac{r\,\coth\!\left(\frac{v r}{2}\right)}{a + r\,\coth\!\left(\frac{v r}{2}\right)}. 
\end{equation}

As $a\to\infty$, $\Lambda(a,r)\to 0$, the gauge orientation modes $(\delta' A_i , \delta' \Phi ) $ approaches $0$. As shown in \cite{1}, 
\begin{equation}
\mathcal{M}_{2}(1,\Phi_{0}) \cong \mathbb{R}^{3} \times\mathbb{S}^{1}  \times  \mathbb{R}^{4},
\end{equation}
where $\rho=2\sqrt{a}$ is the radial coordinate on the $\mathbb{R}^{4}$ factor. The slice at $a=\infty$ is $\mathbb{R}^{3}\times \mathbb{S}^{1}\times \mathbb{S}^{3}_{\infty}$. Quotienting $\mathbb{S}_{\infty}^{3}  \cong SU(2)$ by $U(1)$ along the Hopf fibers (identify $ U(1) \subset SU(2) $ with $U(1)$ in the $\mathbb{S}^{1}$ factor) gives
\begin{equation}
\mathbb{S}^{3}_{\infty}/U(1)\cong \mathbb{S}^{2}\cong \mathbb{CP}^{1},
\end{equation}
which matches the $C(k_{1})\cong \mathbb{CP}^{1}$ in $\mathcal{M}_{1}(1,\Phi_{0})$. $ \mathbb{CP}^{1}$ lies at infinity and thus inherits the divergent metric. So although $\delta' A_i \rightarrow 0$, $ \delta' \Phi \rightarrow 0 $ pointwise as $a\to\infty$, the $L^{2}$ norm of the corresponding gauge zero mode still diverges.

\section{Moduli space geometry}

This section derives expressions for the Christoffel connection $\Gamma^{m}{}_{nl}$ and the Riemann tensor $R_{mnpq}$ in terms of zero modes, using standard techniques from the literature (see e.g.~\cite{Gauntlett, HarveyStrominger1993, Cederwall1996}). The results are used in Appendix~\ref{EFF}.

Let $A_{\hat \mu}=(A_{i},\Phi)$ with $\hat{\mu}=1,2,3,4$, and $D_{\hat{\mu}}=\partial_{\hat{\mu}}-i[A_{\hat{\mu}},\,\cdot\,]$. Define the operators $ D^{+} $ and $D^{-}$ acting on adjoint scalars and adjoint 1-forms as 
\begin{equation}
(D^{+}\phi)_{\hat{\mu}}=D_{\hat{\mu}}\phi,\qquad D^{-}\eta = D_{\hat{\mu}}\eta_{\hat{\mu}}.
\end{equation}
With the inner products given by 
\begin{equation}
\langle \phi, \phi' \rangle = \int d^{3}x \; \tr(\phi\,\phi'),\qquad \langle\eta, \eta' \rangle = \int d^{3}x \; \tr(\eta_{\hat{\mu}}\,\eta'_{\hat{\mu}}),
\end{equation}
we have 
\begin{equation}
\langle \eta, D^{+}\phi \rangle = -\langle D^{-}\eta, \phi \rangle, 
\end{equation}
so that $  ( D^{+})^{\dagger}=-D^{-} $. Introduce 
\begin{eqnarray}\label{B4}
K &\equiv& D^{-}D^{+}= D_{\hat{\mu}}D_{\hat{\mu}}, \qquad
\mathcal{G}_m \equiv D^{-}\partial_{m} A=D_{\hat{\mu}}\,\partial_m A_{\hat{\mu}}, \nonumber\\
\alpha_m &\equiv& K^{-1} \mathcal{G}_m, \qquad   \qquad    \;\;\;\;\, 
J_{mn} \equiv [\,\delta_m A_{\hat{\mu}},\, \delta_n A_{\hat{\mu}}\,], 
\end{eqnarray}
and define the covariant derivative on moduli space 
\begin{equation}
s_{m}=\partial_{m}-i[\alpha_m,\,\cdot\,].
\end{equation}
The following identities hold:
\begin{eqnarray}\label{ID}
[ D_{\hat{\mu}}, s_{m}] &=& i\,\delta_{m}A_{\hat{\mu}}
\\
\phi_{mn}=[s_{m},s_{n}] &=& -2\,K^{-1}J_{mn}\label{ID}
\\
D_{\hat{\mu}}\phi_{mn} &=& 2i\,s_{[m}\delta_{n]}A_{\hat{\mu}}\,.
\end{eqnarray}

Consider $ s_{p}\delta_m A $, which can be decomposed as  
\begin{equation}
s_{p}\delta_m A=\Gamma^{q}{}_{pm}\delta_q A+ \eta_{pm}
\end{equation}
with
\begin{equation}
\langle  \delta_n A,\eta_{pm}\rangle=0.
\end{equation}
The metric on the moduli space is
\begin{equation}
g_{mn}
  = \int  d^3x\;
    \tr\!\left( \delta_m A_{\hat{\mu}}\,\delta_n A_{\hat{\mu}} \right)
    \equiv\langle \delta_m A ,\delta_n A \rangle,
\end{equation}
so
\begin{equation}
\langle \delta_{l}A, s_{p}\delta_m A \rangle =g_{lq}\Gamma^{q}{}_{pm} = \Gamma_{lpm}. 
\end{equation}
$  \Gamma^{q}{}_{pm}$ is the Christoffel connection, which is metric-compatible and torsionless:
\begin{eqnarray}
\partial_{p}g_{mn}
&=& \langle s_{p}\delta_m A ,\delta_n A \rangle +\langle \delta_m A ,s_{p}\delta_n A \rangle
= \Gamma_{npm}+\Gamma_{mpn}
\\
\Gamma_{l[pm]}
&=& \langle \delta_{l}A, s_{[p}\delta_{m]} A \rangle
=-\tfrac{i}{2}\langle \delta_{l}A_{\hat{\mu}}, D_{\hat{\mu}}\phi_{pm}\rangle
=\tfrac{i}{2}\langle D_{\hat{\mu}}\delta_{l}A_{\hat{\mu}}, \phi_{pm}\rangle
=0 .
\end{eqnarray}
For the fermionic zero modes 
\begin{equation}
\chi^{\alpha}_{m}=\gamma^{\hat\mu}\,\delta_{m}A_{\hat\mu}\,\varepsilon^{\alpha}_{0},\qquad
\bar\chi^{\alpha}_{m}=-\bar\varepsilon^{\alpha}_{0}\,\gamma^{\hat\mu}\,\delta_{m}A_{\hat\mu},
\end{equation}
we have
\begin{equation}
s_{p}\chi^{\alpha}_{m}=\partial_p\chi^{\alpha}_{m}-i[\alpha_p,\chi^{\alpha}_{m}]=\gamma^{\hat\mu}\,s_{p}\delta_{m}A_{\hat\mu}\,\varepsilon^{\alpha}_{0},
\end{equation}
and hence
\begin{equation}\label{b17}
\int d^{3}x\,\tr\Big(\bar\chi^{\alpha}_m\gamma^{0}s_p\chi^{\beta}_n \Big)=\bar\varepsilon^{\alpha}_{0}\,\gamma^{0}\gamma^{\hat\mu}\gamma^{\hat\nu}\varepsilon^{\beta}_{0}\int d^{3}x\,\tr\Big(\,\delta_{m}A_{\hat\mu}    \,s_{p}\delta_{n}A_{\hat\nu}\,\Big)=-\Gamma_{mpn}\delta^{\alpha\beta},
\end{equation}
which is an alternative expression for $ \Gamma_{mpn} $.

Next, consider
\begin{eqnarray}
\langle \delta_m A,  s_{l}s_{p}\delta_n A\rangle
&=& \left\langle \delta_m A,\,
\partial_{l}\Gamma^{r}{}_{pn}\,\delta_r A
+ \Gamma^{q}{}_{pn}\Gamma^{r}{}_{lq}\,\delta_r A
+ \Gamma^{q}{}_{pn}\,\eta_{lq}
+ s_{l}\eta_{pn}\right\rangle
\\
&=& g_{mr}\!\left(\partial_{l}\Gamma^{r}{}_{pn}+\Gamma^{r}{}_{lq}\Gamma^{q}{}_{pn}\right)
- \langle \eta_{lm}, \eta_{pn}\rangle.
\end{eqnarray}
The Riemann tensor is then
\begin{eqnarray}\label{r1}
R_{mnlp} &=& g_{mr}\!\left(\partial_{l}\Gamma^{r}{}_{np}+\Gamma^{r}{}_{lq}\Gamma^{q}{}_{np}
- \partial_{p}\Gamma^{r}{}_{nl}-\Gamma^{r}{}_{pq}\Gamma^{q}{}_{nl}\right) \nonumber\\
&=& \left\langle \delta_m A,\,[ s_{l},s_{p}]\,\delta_n A \right\rangle
+ \langle  \eta_{lm}, \eta_{pn}\rangle
- \langle  \eta_{pm}, \eta_{ln}\rangle  .
\end{eqnarray}
From (\ref{ID}), 
\begin{equation}\label{r11}
\left\langle \delta_m A,\,[ s_{l},s_{p}]\,\delta_n A \right\rangle=-2\left\langle \delta_m A,\,[K^{-1}J_{lp},\,\delta_n A] \right\rangle=2 \langle J_{mn},K^{-1}J_{lp}\rangle .
\end{equation}
Since 
\begin{equation}
D^{-}\delta_m A\equiv D_{\hat{\mu}}\delta_m A_{\hat{\mu}}=0,
\end{equation}
the action of $D^{-} $ on $\eta_{pm}  $ gives
\begin{equation}
D^{-} \eta_{pm}
=D^{-} s_{p}\delta_m A= D_{\hat{\mu}} s_{p}\delta_m A_{\hat{\mu}}
= [D_{\hat{\mu}} ,s_{p}]\delta_m A_{\hat{\mu}}
= i [\delta_p A_{\hat{\mu}},\delta_m A_{\hat{\mu}}]=i J_{pm}. 
\end{equation}
$ D^{+}(D^{-}D^{+})^{-1}D^{-} $ is the projection operator on the subspace orthogonal to the zero modes, so
\begin{equation}
\eta_{pm}=D^{+}(D^{-}D^{+})^{-1}D^{-} \eta_{pm}=iD^{+}K^{-1}J_{pm}.
\end{equation}
The inner products of $ \eta $ is then given by 
\begin{equation}\label{r111}
\langle \eta_{ln},\eta_{pm}\rangle
=\langle \eta_{ln}, iD^{+}K^{-1}J_{pm} \rangle= -i\langle D^{-}\eta_{ln}, K^{-1}J_{pm} \rangle
= \langle J_{ln}, K^{-1}J_{pm} \rangle. 
\end{equation}
Combing (\ref{r1}) (\ref{r11}) and (\ref{r111}) and using the symmetries of $\langle J_{mn},K^{-1}J_{lp}\rangle  $:
\begin{align}\label{b30}
\langle J_{mn},K^{-1}J_{lp}\rangle
&=-\langle J_{nm},K^{-1}J_{lp}\rangle
=-\langle J_{mn},K^{-1}J_{pl}\rangle
=\langle J_{lp},K^{-1}J_{mn}\rangle \nonumber\\
\langle J_{mn},K^{-1}J_{lp}\rangle
&+\langle J_{ml},K^{-1}J_{pn}\rangle
+\langle J_{mp},K^{-1}J_{nl}\rangle
=0, 
\end{align}
we arrive at 
\begin{equation}\label{b28}
R_{mnlp}=3 \langle J_{mn},K^{-1}J_{lp}\rangle .
\end{equation}

\section{Effective Lagrangian via the collective-coordinate expansion}\label{EFF}

Starting from the $\mathcal{N}=4$ SYM Lagrangian 
\begin{align}\label{b1}
\mathcal{L}
&= \tfrac{1}{2}\,\tr \left( E_i E_i    +D_0\Phi\,D_0\Phi \right) 
 -\tfrac{1}{2}\,\tr  \left( B_i B_i+D_i\Phi\,D_i\Phi 
  \right)\nonumber  \\[6pt]   
  &+ \tfrac{1}{2}\,\tr \left( D_0\Phi^I\,D_0\Phi^I\right) 
 -\tfrac{1}{2}\,\tr  \left(  D_i\Phi^I\,D_i\Phi^I -[\Phi,\Phi^I]^2-\tfrac{1}{2}\,[\Phi^I,\Phi^J]^2
  \right) \nonumber \\[6pt]   
&-\tfrac{i}{2}\,\tr \left(  \bar{\Psi}\,\gamma^{0}D_{0}\Psi\right) 
-\tfrac{i}{2}\,\tr \left(  \bar{\Psi}\,\gamma^{i}D_{i}\Psi  
-i \bar{\Psi}\,\gamma^{4}\,[\Phi,\Psi]-i \bar{\Psi}\,\gamma^{I}\,[\Phi^{I},\Psi]\right)  ,
\end{align}
with $ I=5,\ldots,9 $, let 
\begin{align}
A_i(x,t) &= A^{\rm BPS}_i\big(x; z(t)\big), &
\Phi(x,t) &= \Phi^{\rm BPS}\big(x; z(t)\big) ,\label{b3}\\
\Psi(x,t) &= \chi^{\alpha}_{m}(x;z(t))\,\xi^{m}_{\alpha}(t), &
\bar\Psi(x,t) &= \bar\chi^{\alpha}_{m}(x;z(t))\,\xi^{m}_{\alpha}(t).
\end{align}
Substituting (\ref{b3}) into (\ref{b1}), 
\begin{align}\label{oss1}
\mathcal{L}
&= \tfrac{1}{2}\,\tr \left( E_i E_i    +D_0\Phi\,D_0\Phi \right) 
 - 2 \pi \tr  \left(  \Phi_{0}G_{0}
\right)-\tfrac{i}{2}\,\tr \left(  \bar{\Psi}\,\gamma^{0}D_{0}\Psi\right) 
-\tfrac{1}{2}\,\tr \left(  \bar{\Psi}\,\gamma^{I}\,[\Phi^{I},\Psi]\right)  \nonumber   \\[6pt]   
  &+ \tfrac{1}{2}\,\tr \left( D_0\Phi^I\,D_0\Phi^I\right) 
 -\tfrac{1}{2}\,\tr  \left(  D_i\Phi^I\,D_i\Phi^I -[\Phi,\Phi^I]^2-\tfrac{1}{2}\,[\Phi^I,\Phi^J]^2
  \right) ,
\end{align}
where the unspecified $\Phi^{I}$ and $A_{0}$ are to be integrated out by solving the equations of motion
\begin{align}\label{c66}
&
D_i E_i - i\,[\Phi, D_0\Phi]- i\,[\Phi^I, D_0\Phi^I]+\tfrac{1}{2}\,[\,\bar\Psi\,\gamma^0,\Psi\,] =0,
\nonumber \\[6pt]
&
-D_0^2\Phi^I+D_i^2\Phi^I-[\,\Phi,\,[\Phi,\Phi^I]\,]+[\,\Phi^J,\,[\Phi^I,\Phi^J]\,]
+\tfrac{1}{2}\,[\,\bar\Psi\,\gamma^I,\Psi\,]=0.
\end{align}

The effective action can be expanded in the parameter $ n=n_{\partial}+\tfrac{1}{2}n_{f} $, where $n_{\partial}$ is the number of time derivatives and $n_{f}$ is the number of fermions \cite{HarveyStrominger1993}. At the order of $ n=1$, the solutions of (\ref{c66}) are
\begin{equation}\label{oss}
A_0=  \dot z^{m}\,\alpha_{m} +\tfrac{1}{2}\,K^{-1}[\,\bar\Psi\,\gamma^0,\Psi\,] ,\qquad  \Phi^I
=-\tfrac{1}{2}\,K^{-1}[\,\bar\Psi\,\gamma^I,\Psi\,],
\end{equation}
where $\alpha_{m}$ and $K$ are defined in (\ref{B4}).

With (\ref{oss}) inserted into (\ref{oss1}), the Lagrangian at the order of $n=2$ is
\begin{align}\label{b39}
\mathcal{L}_{2}
&= \tfrac{1}{2}\, \dot z^{m}\dot z^{n}\, \tr \left( \partial_m A_{\hat{\mu}}\partial_n A_{\hat{\mu}}+\alpha_{m}K\alpha_{n}\right) -\tfrac{i}{2}\,\tr \left [  \bar{\Psi}\,\gamma^{0}  ( \partial_{0}\Psi  -i[\, \dot z^{m}\,\alpha_{m} ,\Psi\,])\right] 
 - 2 \pi \tr  \left(  \Phi_{0}G_{0}
\right) \nonumber   \\[6pt]   
& 
 +\tfrac{1}{8}\,\tr \left( [\bar{\Psi}\,\gamma^{0},\Psi] K^{-1}[\bar{\Psi}\,\gamma^{0},\Psi]   \right) -\tfrac{1}{8}\,\tr \left( [\bar{\Psi}\,\gamma^{I},\Psi] K^{-1}[\bar{\Psi}\,\gamma^{I},\Psi] \right) . 
\end{align}
The next step is to perform an integration over space. 
\begin{equation}\label{b40}
 \int d^3x\,\tr \left( \partial_m A_{\hat{\mu}}\partial_n A_{\hat{\mu}}+\alpha_{m}K\alpha_{n}\right)=g_{mn}. 
\end{equation}
From (\ref{436}) and (\ref{b17}), 
\begin{eqnarray}\label{b41}
&&\int d^{3}x\,\tr \left[ \bar{\Psi}\gamma^{0} (\partial_{0}\Psi-i[ \dot z^{m}\alpha_{m} ,\Psi])\right] \nonumber \\
&=& \int d^{3}x\,\tr \left(  \bar\chi^{\beta}_{m}\gamma^{0}   \chi^{\alpha}_{n}\right) \xi^{m}_{\beta}\dot\xi_{\alpha}^{n}
+\int d^{3}x\, \tr \left(  \bar\chi^{\beta}_{m}\gamma^{0}   s_{p}\chi^{\alpha}_{l}\right)\dot z^{p} \xi^{m}_{\beta}\xi_{\alpha}^{l} \nonumber\\
&=& -g_{mn} \xi^{m}_{\alpha}\nabla_{t}\xi_{\alpha}^{n},
\end{eqnarray}
where $\nabla_{t}\xi_{\alpha}^{n}=\dot\xi_{\alpha}^{n}
+  \Gamma^{n}{}_{pl} \,  \dot z^{p} \xi_{\alpha}^{l}$. From (\ref{435}), 
\begin{equation}
 [\bar{\Psi}\,\gamma^{0},\Psi]=[\bar\chi^{\alpha}_{m} \,\gamma^{0},\chi^{\beta}_{n}]\,\xi^{m}_{\alpha}\xi^{n}_{\beta}=-[\delta_{m}A_{\hat\mu},\delta_{n}A_{\hat\mu}]\,\xi^{m}_{\alpha}\xi^{n}_{\alpha}=-J_{mn}\,\xi^{m}_{\alpha}\xi^{n}_{\alpha},
\end{equation}
so
\begin{equation}\label{b42}
\int d^{3}x\,  \tr \left( [\bar{\Psi}\,\gamma^{0},\Psi] K^{-1}[\bar{\Psi}\,\gamma^{0},\Psi]   \right)=\langle J_{mn},K^{-1}J_{pq}\rangle (\xi^{m}_{\alpha}\xi^{n}_{\alpha})(\xi^{p}_{\beta}\xi^{q}_{\beta}). 
\end{equation}

To evaluate the $[\bar{\Psi}\,\gamma^{I},\Psi]$ contribution, we return to the eight real spinors $\varepsilon^{\mathcal{A}}_{0}   $ $(\mathcal{A}=1,\ldots,8)$ in (\ref{419a}), and set $\mathcal{A}=1,5$ at the end to recover $\varepsilon^{\alpha}_{0}$. Let 
\begin{equation}
 (\varepsilon^{\mathcal{A}}_{0})^{\dagger}\gamma^{0}\gamma^{I}   \varepsilon^{\mathcal{B}}_{0} =( \Sigma^{I} )^{\mathcal{A}\mathcal{B}},    \qquad    I=5,\ldots,9,
\end{equation}
where $ \Sigma^{I} $ is a real $8 \times 8$ representation of $\gamma^{0 I}   $ obeying 
\begin{equation}
 \{\Sigma^{I},\Sigma^{J}\}=2\,\delta^{IJ}\,\mathbf{1}_{8},\qquad [J^{i},\Sigma^{I}]=0,    \qquad i=1,2,3,
\end{equation}
with $J^{i} $ given by (\ref{ji}). A convenient choice is 
\begin{equation}\label{c13}
\Sigma^{4+i}=
\left(\begin{matrix}0_{4}&-\bar{I}^{(i)}\\ \bar{I}^{(i)}&0_{4}\end{matrix}\right),
\quad
\Sigma^{8}=
\left(\begin{matrix}0_{4}&\mathbf 1_{4}\\ \mathbf 1_{4}&0_{4}\end{matrix}\right),
\quad
\Sigma^{9}=
\left(\begin{matrix}\mathbf 1_{4}&0_{4}\\ 0_{4}&-\mathbf 1_{4}\end{matrix}\right),
\end{equation}
where 
\begin{equation}
\bar{I}^{(1)}=\begin{pmatrix}
0&0&0&-1\\
0&0&1&0\\
0&-1&0&0\\
1&0&0&0
\end{pmatrix},\qquad
\bar{I}^{(2)}=\begin{pmatrix}
0&0&-1&0\\
0&0&0&-1\\
1&0&0&0\\
0&1&0&0
\end{pmatrix},\qquad
\bar{I}^{(3)}=\begin{pmatrix}
0&1&0&0\\
-1&0&0&0\\
0&0&0&-1\\
0&0&1&0
\end{pmatrix}
\end{equation}
are anti-self-dual 't Hooft matrices satisfying 
\begin{equation}
\bar{I}^{(i)}\bar{I}^{(j)}=-\delta^{ij}\,\mathbf 1_{4}+\varepsilon^{ijk}\,\bar{I}^{(k)},\qquad
[\bar{I}^{(i)},\,I^{(j)}]=0 .
\end{equation}
Moreover,
\begin{equation}\label{c17}
\Sigma^{4+i}J^{j}
=\begin{pmatrix}
\mathbf 0_{4} & -\,\bar{I}^{(i)}I^{(j)}\\[2pt]
\bar{I}^{(i)}I^{(j)} & \mathbf 0_{4}
\end{pmatrix},\qquad\Sigma^{8}J^{i}
=\begin{pmatrix}0_{4}&I^{(i)}\\ I^{(i)}&0_{4}\end{pmatrix},\qquad
\Sigma^{9}J^{i}
=\begin{pmatrix}I^{(i)}&0_{4}\\ 0_{4}&-I^{(i)}\end{pmatrix}.
\end{equation}

With the zero modes inserted, 
\begin{align}
 [\bar{\Psi}\,\gamma^{I},\Psi]
&=[\bar\chi^{\mathcal{A}}_{m}\,\gamma^{I},\chi^{\mathcal{B}}_{n}]\,\xi^{m}_{\mathcal{A}}\xi^{n}_{\mathcal{B}}
\nonumber\\
&=J_{mn}\,
\bar\varepsilon^{\mathcal{A}}_{0}\,\gamma^{I}   \,\varepsilon^{\mathcal{B}}_{0}\,\xi^{m}_{\mathcal{A}}\xi^{n}_{\mathcal{B}}
+
[\,\delta_{m}A_{\hat\mu},\,\delta_{n}A_{\hat\nu}\,]\;
\bar\varepsilon^{\mathcal{A}}_{0}\,\gamma^{I} \gamma^{\hat\mu \hat\nu}  \,\varepsilon^{\mathcal{B}}_{0}\,\xi^{m}_{\mathcal{A}}\xi^{n}_{\mathcal{B}} \nonumber\\
&=J_{mn}
(\Sigma^{I})^{ \mathcal{A}\mathcal{B}}\,\xi^{m}_{\mathcal{A}}\xi^{n}_{\mathcal{B}}
-[\delta_{m}A_{\hat\mu},\,\delta_{n}A_{\hat\nu}]  \;
(\Sigma^{I})^{ \mathcal{A}\mathcal{C}}(J^{(i)})^{\mathcal{B}}{}_{\mathcal{C}} (I^{(i)})^{\hat\mu\hat\nu}\;\xi^{m}_{\mathcal{A}}\xi^{n}_{\mathcal{B}}, 
\end{align}
where $(\Sigma^{I})^{ \mathcal{A}\mathcal{B}}$ and $(\Sigma^{I})^{ \mathcal{A}\mathcal{C}}(J^{(i)})^{\mathcal{B}}{}_{\mathcal{C}}$ are read off from (\ref{c13}) and (\ref{c17}). Now setting $ \mathcal{A},\mathcal{B} =1,5$, using (\ref{b30}), the Grassmann nature of $\xi$, and 
\begin{equation}
\sum_{i=1}^{3}
(I^{(i)})^{\hat\mu\hat\nu}(I^{(i)})^{\hat\rho\hat\sigma}
= 2\!\left(\delta^{\hat\mu\hat\rho}\delta^{\hat\nu\hat\sigma}
-\delta^{\hat\mu\hat\sigma}\delta^{\hat\nu\hat\rho}\right)
+\epsilon^{\hat\mu\hat\nu\hat\rho\hat\sigma},
\end{equation}
we obtain 
\begin{align}\label{b43}
\int d^{3}x\,  \tr \left( [\bar{\Psi}\,\gamma^{8},\Psi] K^{-1}[\bar{\Psi}\,\gamma^{8},\Psi] \right)
&=\langle J_{mn},K^{-1}J_{pq}\rangle\,(\xi^{m}_{1}\xi^{n}_{5}+\xi^{m}_{5} \xi^{n}_{1})(\xi^{p}_{1}\xi^{q}_{5}+\xi^{p}_{5} \xi^{q}_{1}) \nonumber\\
&= - \langle J_{pm},K^{-1}J_{nq}\rangle\, (\xi_{\alpha}^{p}\xi_{\alpha}^{m})(\xi_{\beta}^{n}\xi_{\beta}^{q}),
\end{align}
\begin{align}\label{b44}
\int d^{3}x\,  \tr \left( [\bar{\Psi}\,\gamma^{9},\Psi] K^{-1}[\bar{\Psi}\,\gamma^{9},\Psi] \right)
&=\langle J_{mn} ,K^{-1} J_{pq}\rangle\,(\xi^{m}_{1}\xi^{n}_{1}-\xi^{m}_{5} \xi^{n}_{5}) (\xi^{p}_{1}\xi^{q}_{1}-\xi^{p}_{5} \xi^{q}_{5}) \nonumber\\
&= -\langle J_{pm},K^{-1}J_{nq}\rangle\, (\xi_{\alpha}^{p}\xi_{\alpha}^{m})(\xi_{\beta}^{n}\xi_{\beta}^{q}) ,
\end{align}
\begin{eqnarray}\label{b45}
&&\sum^{3}_{i=1}\int d^{3}x\,  \tr \left( [\bar{\Psi}\,\gamma^{4+i},\Psi] K^{-1}[\bar{\Psi}\,\gamma^{4+i},\Psi] \right)\nonumber\\
&=&\sum^{3}_{i=1}
(I^{(i)})^{\hat\mu\hat\nu}(I^{(i)})^{\hat\rho\hat\sigma}\;
\langle [\delta_{m}A_{\hat\mu},\,\delta_{n}A_{\hat\nu}],K^{-1}[\delta_{p}A_{\hat\rho},\,\delta_{q}A_{\hat\sigma}] \rangle \, (\xi^{m}_{1}\xi^{n}_{5}-\xi^{m}_{5} \xi^{n}_{1})(\xi^{p}_{1}\xi^{q}_{5}-\xi^{p}_{5} \xi^{q}_{1})\nonumber\\
&=&
\Big( 
2\, \langle[\delta_{m}A_{\hat\mu},\,\delta_{n}A_{\hat\nu}],
K^{-1}\,
[\delta_{p}A_{\hat\mu},\,\delta_{q}A_{\hat\nu}]\rangle \nonumber\\
& + & 
\epsilon_{\hat\mu\hat\nu\hat\rho\hat\sigma}\;
\langle[\delta_{m}A_{\hat\mu},\,\delta_{n}A_{\hat\nu}],
K^{-1}\,
[\delta_{p}A_{\hat\rho},\,\delta_{q}A_{\hat\sigma}]\rangle
\Big) (\xi^{m}_{1}\xi^{n}_{5}-\xi^{m}_{5}\xi^{n}_{1})
(\xi^{p}_{1}\xi^{q}_{5}-\xi^{p}_{5}\xi^{q}_{1})=0.
\end{eqnarray}

Finally, combining (\ref{b40}), (\ref{b41}), (\ref{b42}), (\ref{b43}), (\ref{b44}), (\ref{b45}) and (\ref{b28}), we get the effective Lagrangian 
\begin{equation}
\mathcal{L}_{\rm eff}= \int d^{3}x\,  \mathcal{L}_{2}
= \tfrac{1}{2}\,g_{mn} \dot z^{m}\dot z^{n} +\tfrac{i}{2} \,g_{mn} \,\xi^{m}_{\alpha}\nabla_{t}\xi_{\alpha}^{n}
 +\tfrac{1}{8}\, R_{mnpq}  \, (\xi^{m}_{\alpha}\xi^{n}_{\alpha})(\xi^{p}_{\beta}\xi^{q}_{\beta})  - 2 \pi \tr  \left(  \Phi_{0}G_{0}\right) . 
\end{equation}

\end{appendix}





\vspace{3mm}


\begin{thebibliography}{10}

\bibitem{GNO}
P.~Goddard, J.~Nuyts and D.~I.~Olive,
``Gauge Theories and Magnetic Charge,''
Nucl.\ Phys.\ B \textbf{125}, 1--28 (1977).

\bibitem{Montonen:1977}
C.~Montonen and D.~I.~Olive,
``Magnetic monopoles as gauge particles?,''
Phys.\ Lett.\ B \textbf{72}, 117--120 (1977).

\bibitem{Osborn:1979}
H.~Osborn,
``Topological charges for $N=4$ supersymmetric gauge theories and monopoles of spin 1,''
Phys.\ Lett.\ B \textbf{83}, 321--326 (1979).

\bibitem{Sen:1994}
A.~Sen,
``Dyon--monopole bound states, self-dual harmonic forms on the multi-monopole moduli space, and $SL(2,\mathbb{Z})$ invariance in string theory,''
Phys.\ Lett.\ B \textbf{329}, 217--221 (1994),
hep-th/9402032.



\bibitem{Dorey:1996}
N.~Dorey, C.~Fraser, T.~J.~Hollowood and M.~A.~C.~Kneipp,
``S-duality in $N{=}4$ supersymmetric gauge theories with arbitrary gauge group,''
Phys.\ Lett.\ B \textbf{383}, 422--428 (1996),
hep-th/9605069.


\bibitem{Gauntlett}
J.~P.~Gauntlett,
``Low energy dynamics of N=2 supersymmetric monopoles,''
Nucl.\ Phys.\ B \textbf{411}, 443--460 (1994),
hep-th/9305068.

\bibitem{Blum}
J.~D.~Blum,
``Supersymmetric quantum mechanics of monopoles in $N=4$ Yang--Mills theory,''
Phys.\ Lett.\ B \textbf{333}, 92--97 (1994),
hep-th/9401133.





\bibitem{Witten:1982}
E.~Witten,
``Constraints on supersymmetry breaking,''
Nucl.\ Phys.\ B \textbf{202}, 253--316 (1982).






\bibitem{GauntlettLowe1996}
J.~P.~Gauntlett and D.~A.~Lowe,
``Dyons and S-duality in $N{=}4$ supersymmetric gauge theory,''
Nucl.\ Phys.\ B \textbf{472}, 194--206 (1996),
hep-th/9601085.

\bibitem{LWY:SU3}
K.~Lee, E.~J.~Weinberg and P.~Yi,
``Electromagnetic duality and $SU(3)$ monopoles,''
Phys.\ Lett.\ B \textbf{376}, 97--102 (1996),
hep-th/9601097.

\bibitem{LWY:ManyBPS}
K.~Lee, E.~J.~Weinberg and P.~Yi,
``The moduli space of many BPS monopoles for arbitrary gauge groups,''
Phys.\ Rev.\ D \textbf{54}, 1633--1643 (1996),
hep-th/9602167.

\bibitem{Gibbons:1996}
G.~W.~Gibbons,
``The Sen conjecture for fundamental monopoles of distinct types,''
Phys.\ Lett.\ B \textbf{382}, 53--59 (1996),
hep-th/9603176.

\bibitem{SegalSelby1996}
G.~Segal and A.~Selby,
``The cohomology of the space of magnetic monopoles,''
Commun.\ Math.\ Phys.\ \textbf{177}, 775--787 (1996).





\bibitem{Balachandran1971}
A.~P.~Balachandran, G.~Marmo, N.~Mukunda, J.~S.~Nilsson,
E.~C.~G.~Sudarshan and F.~Zaccaria,
``Monopole topology and the problem of color,''
Phys.\ Rev.\ D \textbf{3}, 1867--1880 (1971).

\bibitem{Balachandran1984a}
A.~P.~Balachandran, G.~Marmo, N.~Mukunda, J.~S.~Nilsson,
E.~C.~G.~Sudarshan and F.~Zaccaria,
``Non-Abelian monopoles break color. I. Classical mechanics,''
Phys.\ Rev.\ D \textbf{29}, 2919--2935 (1984).

\bibitem{Balachandran1984b}
A.~P.~Balachandran, G.~Marmo, N.~Mukunda, J.~S.~Nilsson,
E.~C.~G.~Sudarshan and F.~Zaccaria,
``Non-Abelian monopoles break color. II. Field theory and quantum mechanics,''
Phys.\ Rev.\ D \textbf{29}, 2936--2949 (1984).




\bibitem{Abouelsaood1983a}
A.~Abouelsaood,
``Are there chromodyons?,''
Nucl.\ Phys.\ B \textbf{226}, 309--322 (1983).



\bibitem{Abouelsaood1984b}
A.~Abouelsaood,
``Chromodyons and equivariant gauge transformations,''
Nucl.\ Phys.\ B \textbf{242}, 377--392 (1984).



\bibitem{Abouelsaood1984a}
A.~Abouelsaood,
``Zero modes in the Prasad--Sommerfield limit and SO(5) monopoles,''
Phys.\ Lett.\ B \textbf{137}, 77--82 (1984).



\bibitem{NelsonManohar1983}
P.~Nelson and A.~Manohar,
``Global color is not always defined,''
Phys.\ Rev.\ Lett.\ \textbf{50}, 943--946 (1983).






\bibitem{Weinberg2}
E.~J.~Weinberg,
``Fundamental monopoles and multimonopole solutions for arbitrary simple gauge groups,'' Nucl.\ Phys.\ B \textbf{167}, 500--524 (1980).



\bibitem{Weinberg1}
E.~J.~Weinberg,
``Fundamental monopoles in theories with arbitrary symmetry breaking,''
Nucl.\ Phys.\ B \textbf{203}, 445--471 (1982).

\bibitem{Weinberg1982}
E.~J.~Weinberg,
``A continuous family of magnetic monopole solutions,''
Phys.\ Lett.\ B \textbf{119}, 151--154 (1982).






\bibitem{1}
K.~M.~Lee, E.~J.~Weinberg and P.~Yi,
``Massive and massless monopoles with non-Abelian magnetic charges,''
Phys.\ Rev.\ D \textbf{54}, 6351--6371 (1996),
hep-th/9605229.



\bibitem{Dorey1995a}
N.~Dorey, C.~Fraser, T.~J.~Hollowood and M.~A.~C.~Kneipp,
``Non-abelian duality in $N=4$ supersymmetric gauge theories,''
hep-th/9512116.



\bibitem{BaisSchroers1998}
F.~A.~Bais and B.~J.~Schroers,
``Quantisation of monopoles with non-abelian magnetic charge,''
Nucl.\ Phys.\ B \textbf{512}, 250--294 (1998),
hep-th/9708004.

\bibitem{SchroersBais1998}
B.~J.~Schroers and F.~A.~Bais,
``S-duality in SU(3) Yang-Mills theory with non-abelian unbroken gauge group,''
Nucl.\ Phys.\ B \textbf{535}, 197--218 (1998),
hep-th/9805163.


\bibitem{Bais2009}
L.~Kampmeijer, J.~K.~Slingerland, B.~J.~Schroers and F.~A.~Bais,
``Magnetic charge lattices, moduli spaces and fusion rules,''
Nucl.\ Phys.\ B \textbf{806}, 386--435 (2009),
arXiv:0803.3376 [hep-th].

\bibitem{Bais2010}
L.~Kampmeijer, F.~A.~Bais, B.~J.~Schroers and J.~K.~Slingerland,
``Towards a non-abelian electric-magnetic symmetry: The skeleton group,''
JHEP \textbf{01}, 095 (2010),
arXiv:0812.1230 [hep-th].








\bibitem{BolognesiKonishi2002}
S.~Bolognesi and K.~Konishi,
``Non-Abelian magnetic monopoles and dynamics of confinement,''
Nucl.\ Phys.\ B \textbf{645}, 337--348 (2002),
hep-th/0207161.

\bibitem{Konishi2002}
K.~Konishi,
``Non-Abelian monopoles, vortices and confinement,''
hep-th/0208222.


\bibitem{Auzzi2004b}
R.~Auzzi, S.~Bolognesi, J.~Evslin and K.~Konishi,
``Nonabelian monopoles and the vortices that confine them,''
Nucl.\ Phys.\ B \textbf{686}, 119--134 (2004),
hep-th/0312233.


\bibitem{Auzzi2004a}
R.~Auzzi, S.~Bolognesi, J.~Evslin, K.~Konishi and A.~Yung,
``NonAbelian monopoles,''
Nucl.\ Phys.\ B \textbf{701}, 207--246 (2004),
hep-th/0405070.



\bibitem{Konishi2004}
K.~Konishi,
``Quantum nonabelian monopoles,''
hep-th/0407272.

\bibitem{Bolognesi2005}
S.~Bolognesi, K.~Konishi and G.~Marmorini,
``Light nonAbelian monopoles and generalized r-vacua in supersymmetric gauge theories,''
Nucl.\ Phys.\ B \textbf{718}, 134--152 (2005),
hep-th/0502004.

\bibitem{Konishi2008}
K.~Konishi,
``The magnetic monopole seventy-five years later,''
Lect.\ Notes Phys.\ \textbf{737}, 471--521 (2008),
hep-th/0702102.





\bibitem{Konishi2009}
K.~Konishi,
``Advent of Non-Abelian Vortices and Monopoles: Further Thoughts about Duality and Confinement,''
Prog.\ Theor.\ Phys.\ Suppl.\ \textbf{177}, 83--98 (2009),
arXiv:0809.1370 [hep-th].



\bibitem{Murray1989}
M.~Murray,
``Stratifying monopoles and rational maps,''
Commun.\ Math.\ Phys.\ \textbf{125}, 661--674 (1989).



\bibitem{MurraySinger2003}
M.~K.~Murray and M.~A.~Singer,
``A note on monopole moduli spaces,''
J.\ Math.\ Phys.\ \textbf{44}, 3517--3531 (2003),
math-ph/0302020.




\bibitem{Yi:1996}
P.~Yi,
``Duality, Multi-Monopole Dynamics \& Quantum Threshold Bound States,''
hep-th/9608114.





\bibitem{KW}
A.~Kapustin and E.~Witten,
``Electric-magnetic duality and the geometric Langlands program,''
Commun.\ Number Theory Phys.\ \textbf{1}, no.~1, 1--236 (2007), hep-th/0604151.


\bibitem{Bagger:1984ge}
J.~A.~Bagger,
``Supersymmetric sigma models,''
in \textit{1984 NATO ASI on Supersymmetry},
SLAC-PUB-3461 (1984), pp.~45--87.









\bibitem{Index}
L.~S.~Brown, R.~D.~Carlitz and C.~Lee,
``Massless excitations in pseudoparticle fields,''
Phys.\ Rev.\ D \textbf{16}, 417--422 (1977).


\bibitem{Callias1978}
C.~Callias,
``Axial anomalies and index theorems on open spaces,''
Commun.\ Math.\ Phys.\ \textbf{62}, no.~3, 213--234 (1978).





\bibitem{Dancer1992}
A.~S.~Dancer,
``Nahm data and $SU(3)$ monopoles,''
Nonlinearity \textbf{5}, no.~6, 1355--1373 (1992);
Cambridge preprint DAMTP-91-44.

\bibitem{DancerLeese1997}
A.~S.~Dancer and R.~A.~Leese,
``A numerical study of $SU(3)$ charge-two monopoles with minimal symmetry breaking,''
Phys.\ Lett.\ B \textbf{390}, 252--256 (1997).

\bibitem{Irwin1997}
P.~Irwin,
``$SU(3)$ monopoles and their fields,''
Phys.\ Rev.\ D \textbf{56}, 5200--5208 (1997), hep-th/9704153.








\bibitem{GriffithsHarris}
P.~Griffiths and J.~Harris,
\textit{Principles of Algebraic Geometry},
Wiley--Interscience, New York, 1978.

\bibitem{Humphreys}
J.~E.~Humphreys,
\textit{Introduction to Lie Algebras and Representation Theory},
Graduate Texts in Mathematics, vol.~9,
Springer--Verlag, New York, 1972.

\bibitem{HopfSamelson}
H.~Hopf and H.~Samelson,
``Ein Satz \"uber die Wirkungsr\"aume geschlossener Liescher Gruppen,''
\textit{Comment.\ Math.\ Helv.} \textbf{13} (1941), 240--251.


\bibitem{BottTu}
R.~Bott and L.~W.~Tu,
\textit{Differential Forms in Algebraic Topology},
Graduate Texts in Mathematics, vol.~82,
Springer--Verlag, New York, 1982.

\bibitem{Hatcher}
A.~Hatcher,
\textit{Algebraic Topology},
Cambridge University Press, Cambridge, 2002.



\bibitem{WittenOlive:1978}
E.~Witten and D.~I.~Olive,
``Supersymmetry algebras that include topological charges,''
Phys.\ Lett.\ B \textbf{78}, 97--101 (1978).

\bibitem{GauntlettKimLeeYi2001}
J.~P.~Gauntlett, C.~Kim, K.~Lee and P.~Yi,
``General Low Energy Dynamics of Supersymmetric Monopoles,''
Phys.\ Rev.\ D \textbf{63}, 065020 (2001),
hep-th/0008031.



\bibitem{HarveyStrominger1993}
J.~A.~Harvey and A.~Strominger,
``String Theory and the Donaldson Polynomial,''
Commun.\ Math.\ Phys.\ \textbf{151}, 221--232 (1993),
hep-th/9108020.





\bibitem{WeinbergYi2007}
E.~J.~Weinberg and P.~Yi,
``Magnetic Monopole Dynamics, Supersymmetry, and Duality,''
Phys.\ Rept.\ \textbf{438}, 65--236 (2007),
hep-th/0609055.









\bibitem{Helgason:1978}
S.~Helgason,
\textit{Differential Geometry, Lie Groups, and Symmetric Spaces},
Academic Press, New York, 1978.

\bibitem{FultonHarris:1991}
W.~Fulton and J.~Harris,
\textit{Representation Theory: A First Course},
Springer--Verlag, New York, 1991.



\bibitem{KuennemannTamvakis:2002}
K.~K\"unnemann and H.~Tamvakis,
``The Hodge star operator on Schubert forms,''
Topology \textbf{41}, no.\ 5, 945--960 (2002),
math/0306412.


\bibitem{Kostant:1963}
B.~Kostant,
``Lie Algebra Cohomology and Generalized Schubert Cells,''
Annals of Mathematics \textbf{77}, no.\ 1, 72--144 (1963).


\bibitem{Fulton:YoungTableaux}
W.~Fulton,
\textit{Young Tableaux: With Applications to Representation Theory and Geometry},
London Mathematical Society Student Texts \textbf{35}, Cambridge University Press, Cambridge, 1997.




\bibitem{Cederwall1996}
M.~Cederwall, G.~Ferretti, B.~E.~W.~Nilsson and P.~Salomonson,
``Low energy dynamics of monopoles in $N=2$ SYM with matter,''
Mod.\ Phys.\ Lett.\ A \textbf{11}, 367--376 (1996),
hep-th/9508124.





\end{thebibliography}
\end{document}